\documentclass[aps,pre,preprint,showpacs,groupedaddress]{revtex4-2}

\usepackage{epstopdf}
\usepackage{amsmath}
\usepackage{amssymb}
\usepackage{graphics}
\usepackage{wrapfig}
\usepackage[dvips]{graphicx}
\graphicspath{{./},{./pictures/}}
\usepackage{color}
\begin{document}

\title{Stretching and Lyapunov Exponents of Polymers in Ultra-Dilute Turbulent Solutions}

\author{Demosthenes Kivotides}
\email{demosthenes.kivotides@strath.ac.uk}
\thanks{Corresponding author: demosthenes.kivotides@strath.ac.uk}

\affiliation{Strathclyde University, Glasgow, UK}

\date{\today}

\begin{abstract}
We analyse bead--spring polymers coupled to Navier--Stokes turbulence in
ultra--dilute solutions at Weissenberg number \(Wi\approx 80\). The polymers do not alter the
large-scale turbulent structure, but hydrodynamic interactions generate
sub--Kolmogorov solvent motion, so the mesoscopic coupling remains two--way.
The chains stretch predominantly as material line elements, with measurable
deviations caused by the full mesoscopic bead--spring dynamics. 
Their end-to-end distance exhibits intermediate-range \(1/2\)-power-law
scaling.
Polymer trajectories preferentially sample axisymmetric biaxial extension: the largest
extensions and stretching rates occur in high-strain regions, whereas small
extensions and relaxation events are concentrated in high-enstrophy regions.
The chains align strongly with the intermediate strain-rate eigenvector and
avoid the most compressive direction; together with the positive bias of the
intermediate strain-rate eigenvalue, this gives the intermediate direction a significant
role in stretching. Vorticity sampled along polymer paths aligns with both the
first and second strain-rate eigenvectors, differing from analogous Eulerian and
vortex-stretching statistics. 
We also develop a singular-value-decomposition (SVD)-normalised algorithm for
the tangent-flow equations, enabling finite-time Lyapunov numbers to be computed
along polymer trajectories.
Their late-time statistics become stable after about ten large-eddy turnover
times and, together with ergodic Lyapunov theory, provide estimates of asymptotic
stretching rates.
The intermediate finite-time exponent is positive for all computed trajectories,
with \(E[\lambda_2]/E[\lambda_1]\approx 4/17\), larger than the corresponding
material-line value; the strongest dependence occurs between the largest and
smallest finite-time exponents.
\end{abstract}


\maketitle

\section{Introduction}

Turbulence in simple fluids is a central problem of nonlinear and statistical physics \cite[][]{davidson}.
Important variations of the turbulence problem arise when fluids are coupled
to additional structure. In superfluids this structure is associated with
spontaneous symmetry breaking; in polymer solutions it is the deformable
chain microstructure. Such couplings lead to open problems of significant
conceptual depth and broad applications \cite[][]{larson_book,eringen1,eringen2}. There are 
different types of microstructural motions 
in these so-called complex fluids, which correspond to micropolar (associated with rigid geometrical motions
of microstructure), microstretch (similarity transformations) and micromorphic (affine transformations) fluid media.
A particularly important example of the latter category is polymer fluids
and the associated turbulence phenomena.
The physics of polymer turbulence depend on the volume fraction
of the polymers in the solution, and in this work we will only consider dilute solution cases. Despite their relative simplicity compared to semi-dilute, 
entangled or concentrated solutions, flow structure in dilute solutions could be affected by the elastic effects of polymer
deformations and this, in turn, can have nontrivial effects on flow stability and turbulence structure, including the 
important phenomena of lower drag levels exerted by polymer turbulence on solid walls \cite[][]{saeed,bird,graham,white,crawford}. In this paper, we shall only
present results for ultra-dilute turbulent solutions, when the back-reaction of the polymers to the flow does not alter the 
large scale structure of the latter, although there are hydrodynamic interactions between the chains which induce a microscopic flow field in the solvent.

The complexity of micromorphic continua makes a top--down, rational--mechanics formulation 
as a statistical field theory difficult; accordingly, viscoelastic--fluid models often use bottom-up closures
grounded in microscopic polymer physics
\cite[][]{dimitropoulos,boffetta,likhtman,vincenzi1,beris,angelis}.
More advanced treatments still employ a single continuum velocity field for
the solvent-polymer system, but instead of a statistical
closure they include a Fokker-Planck equation for the probability of polymer-chain conformations, 
which most often neglects bead-level hydrodynamic interactions within and between chains \cite[][]{barrett}. Our 
approach differs from such methodologies in that we have a clear
separation between the solvent fluid and the microstructure, and in that we employ the Langevin equation for polymer conformations
in conjunction with the Navier-Stokes equation for the solvent.
This fully mesoscopic approach provides explicit trajectories of (coarse-grained) polymers, 
and enables the investigation of intricate physics, such as the alignments
between polymer conformations and flow strain-rates, and the finite-time
Lyapunov exponents along chain trajectories.

Our mesoscopic viewpoint is now well established, as a result of work
by many microhydrodynamics and polymer-physics
researchers whose work developed and integrated numerous basic methodologies
\cite[][]{grassia,hsieh,jedrejack,banchio} that enabled this powerful approach to polymeric flows. A major 
motivation for polymer turbulence mesoscopics was provided by the availability 
of experimental methods for the control \cite[][]{liphardt} and observation of individual polymer chains \cite[][]{dnacollins}. When similar capabilities
were extended to polymers in a flow, the need for physics of individual chains became a practical rather than purely theoretical
objective \cite[][]{larson,quake,kivotides_knot,doyle}.
Indeed, starting from simple stagnation flows \cite[][]{juang}, there have been observations of single chains in
fluctuating (albeit not fully developed turbulent) flows \cite[][]{liu}, accompanied by theoretical studies of passive polymer chains in such flows \cite[][]{chertkov,
balkovsky,fouxon,kivotides_random}. The experimental studies focused, among other things, on polymer conformations during coil-stretching transition and elastic stress levels,
while the theoretical studies (with the exception of \cite[][]{kivotides_random}) employed simplified polymer models --- for example, without 
hydrodynamic interactions, excluded volume effects
or multimode conformation structure --- to provide insightful analyses of the probability density function (PDF) of polymer extensions.
The geometry of strain-chain interactions was also tackled \cite[][]{kivotides_random}.

By continuing along these lines, we investigate here chain stretching in ultra-dilute turbulent polymer solutions
with Kolmogorov scalings. We aim to examine the detailed flow-polymer interactions and 
their statistical measures 
by employing a bead--spring polymer model that includes hydrodynamic
interactions, excluded volume, elasticity, Brownian forcing, and finite
chain extensibility
and by tracking individual chain trajectories and the finite-time Lyapunov numbers
along them. 
A special focus of this investigation is the correlation between polymer dynamics 
and vorticity structure. 
Our viewpoint is similar to that of \cite[][]{watanabe1}, although their study 
emphasised the coil--stretch transition, which is not our concern here, and---in 
contrast to our approach---neglected multi--body effects in the hydrodynamic 
interactions and in the Brownian forcing. 
Such effects can alter polymer stretching by modifying the drag forces
acting on the chains.
Indeed, as indicated by Zimm's analysis the inclusion of hydrodynamic interactions
results in the polymer effectively dragging the solvent within its pervaded volume as it moves. So instead of having 
the Stokes drag at each bead location (freely draining Rouse approximation), we have the viscous drag on 
a pseudo-solid object of the size of the 
polymer's pervaded volume \cite[][]{rubinstein}. These important physics are expected to affect polymer 
orientation with respect to the flow, hence also polymer stretching.

Our work adds to a body of important contributions, while differing in scope.
In particular, our approach differs from the mesoscopic frameworks of
\cite[][]{watanabe2,watanabe3,serafini} which, in seeking to quantify the
effect of polymers on turbulence structure \cite[][]{vonlanthen}, employ
computationally tractable reduced polymer models, typically dumbbells
which exclude excluded--volume interactions and
inter--chain multibody effects.
Our calculation has a narrower scope (we do not address alterations of the 
turbulence structure) but employs a sophisticated polymer model that has been 
successfully tested against key rheological experiments 
\cite[][]{kivotides_knot,kivotides_stretch,kivotides_dense}. 
It is important to clarify that, by accounting for both intrachain and 
interchain hydrodynamic interactions, our modelling includes the 
microhydrodynamic flow induced in the fluid by polymer drag. 
The key difference from the aforementioned works is that, in our case, the 
chains exert only weak feedback on the turbulence, so the turbulent structure 
remains unaltered. 
In this respect, the calculation retains the relevant mesoscopic physics
of chain elasticity, excluded volume, Brownian forcing, and hydrodynamic
interactions, while remaining in the ultra-dilute regime.

Investigations closely aligned in scope include \cite[][]{jin}, who employed a
bead--spring model without hydrodynamic interactions to assess the ability of
FENE elasticity to reproduce elastic stress levels in viscoelastic flows.
Their study did not examine the geometry of strain--polymer interactions or the
PDFs of polymer extensions.
In related work, \cite[][]{picardo1,picardo2} studied polymer dynamics in
turbulence using dumbbells. Notably, \cite[][]{picardo2} included the
hydrodynamic interaction between the two beads and thus accounted for two--body
effects in evaluating the Brownian force.
By computing PDFs of polymer extension in the Batchelor regime of the Kraichnan
flow and in Navier--Stokes isotropic turbulence, they concluded that
hydrodynamic interactions do not strongly affect polymer stretching; however,
these conclusions are conditioned by the use of dumbbells.
There was no geometrical analysis of the stretching process or of finite-time
Lyapunov exponents along the trajectories.
The latter point is important, because there is no \emph{a priori} reason for
polymers to move along material trajectories and therefore to experience the
same finite-time Lyapunov exponents as fluid elements.
Indeed, as additional polymer physics are included, polymer motion departs
increasingly from the idealization of material transport by the carrier flow.

A further aim is to characterise the turbulent environment sampled by the
polymers themselves. The strain-rate, vorticity, enstrophy, helicity, and
alignment statistics measured along polymer trajectories need not coincide
with volume-averaged Eulerian statistics, nor with statistics measured along
material trajectories. Their comparison allows us to quantify the deviations
of polymer stretching from material-line stretching, and to relate polymer
extension and stretching rates to the enstrophy, total strain intensity, and
strain-rate eigenvector geometry sampled along polymer trajectories.

The purpose of the present work is therefore to quantify, in an ultra--dilute
turbulent solution with hydrodynamically interacting bead--spring chains, how
polymer stretching, trajectory-conditioned strain and vorticity statistics,
strain--alignment geometry, and finite-time Lyapunov statistics depart from
their material-line and volume-averaged counterparts.

\section{Mathematical formulation}

We now define the mesoscopic system used to study polymer stretching and
trajectory-conditioned turbulent statistics. The governing equations are a
subset of the general formulation of polymer turbulence at all volume
fractions developed in \cite[][]{kivotides_method}. That formulation included
entanglement effects, hydrodynamic interactions, wall and periodic boundary
conditions, and identified a subgrid stress originating in the
microhydrodynamic flow field. As shown by the nondimensional analysis below,
the present ultra-dilute regime does not require the full physical complexity
of that formulation: since the polymer volume fraction \(\phi\ll 1\),
entanglement effects may be neglected, and since 
the polymer--turbulence interaction parameter \(I\ll 1\), the polymer
feedback is too weak to modify the turbulent structure, although polymer
forcing is retained in the linear-response microhydrodynamic motion of the
solvent.

The resulting system contains four coupled components: the Navier--Stokes
equations for the turbulent solvent, the overdamped Langevin equation for the
polymers, the Stokes problem for intrachain and interchain hydrodynamic
interactions, and the tangent-flow equation used to compute finite-time
Lyapunov exponents along polymer trajectories.

For clarity, the main text focuses on the physical assumptions, limitations and
algorithmic role of the formulation, while the governing equations, force
terms, mobility relations, Brownian covariance and tangent-flow variables are
collected in Appendix~\ref{app:governing-system} as a compact mathematical
reference.

\subsubsection{Navier-Stokes equations}

In this work, we consider homogeneous, statistically stationary turbulent
carrier flows. The solvent motion is modelled by the incompressible
Navier--Stokes equations for the fluctuating fluid velocity \(u(x,t)\), with
the Lundgren forcing term used to maintain the statistically stationary
turbulent state:
\begin{equation}
    \frac{\partial u_i}{\partial x_i}=0,
    \label{eq:ns-incompressibility}
\end{equation}
\begin{equation}
    \rho_f \frac{\partial u_i}{\partial t}
    + \rho_f \frac{\partial (u_i u_j)}{\partial x_j}
    +\frac{\partial p}{\partial x_i}
    -\mu_f \frac{\partial^2 u_i}{\partial x_j \partial x_j}
    -\rho_f \frac{\epsilon}{\langle u_i u_i \rangle}u_i=0,
    \label{eq:ns-lundgren}
\end{equation}
where \(i\) is the spatial direction index, \(\mu_f=\rho_f \nu_f\) is the
fluid's dynamic viscosity, \(p\) is the fluid pressure, and
\(\epsilon\equiv -\nu_f \left\langle u_i \partial^2 u_i/
(\partial x_j \partial x_j) \right\rangle\) is the kinetic energy dissipation
rate. The same carrier-flow equations and the definition of \(\epsilon\) are
collected in Appendix~\ref{app:carrier-flow},
Eqs.~\eqref{eq:app-incompressibility}--\eqref{eq:app-dissipation-rate}.
The last term in Eq.~\eqref{eq:ns-lundgren} is Lundgren's turbulence stirring
force; it compensates viscous dissipation and maintains the statistically
stationary turbulent carrier flow
\cite[][]{lundgren,meneveau,blanquart}.
We used Lundgren forcing rather than large-scale Gaussian forcing because our
carrier-flow solver is a projection finite-volume code formulated in physical
space, not a spectral code based on forced Fourier modes.
Moreover, the equation above does not include the feedback of the polymers on
the turbulent carrier-flow field. In ultra-dilute solutions, this feedback is
retained at the level of the linear Stokes response to the bead forces,
represented below through the RPY hydrodynamic-interaction tensor.

It is important to note that the viscous stress may also be viewed as the
dissipative part of a fluctuation--dissipation pair. In fluctuating
hydrodynamics this pairing gives rise to the Landau--Lifshitz stochastic
stress \cite[][]{landau}, associated with microscopic molecular fluctuations
of the solvent. This stochastic stress is distinct from the fluctuating
reaction on the solvent associated with the Brownian forcing of the polymer
beads, which originates in solvent--polymer collisions. Both contributions
are usually omitted in Navier--Stokes turbulence calculations because of their
comparatively small magnitude. However, the Landau--Lifshitz stochastic stress
plays an important role in lattice--Boltzmann approaches to polymer flows
\cite[][]{ladd}.

\subsubsection{Langevin equation}

\paragraph{Bead--spring force balance.}
The polymer physics are taken into account via a bead-spring model that was developed in a series of
papers \cite[][]{kivotides_stretch,kivotides_knot,kivotides_dense}, which should be consulted for
detailed expositions of both the employed formulae and their physical content.
It has two key features. First, the model contains no adjustable physical constants once the coarse-graining
level has been chosen.
To set up the full polymer dynamics, it requires only material properties as input: the \(\theta\) temperature, the chain molecular molar mass \(M\),
the Kuhn monomer molar mass \(M_K\), the Kuhn monomer length \(b\) and the number of monomers in polymer strands in between entanglements in the melt state \(P_e\).
\(P_e\) is only required when the solution is entangled. It has been tested with very good results in the context of single-chain dynamics, where it has accurately predicted the
longest relaxation time \cite[][]{kivotides_dense}, and rheological experiments, where it has reliably predicted elastic stress levels in extensional flows
\cite[][]{kivotides_stretch}.
Second, its predictions are robust under variation of the number of beads
(i.e. of the coarse-graining level).
This is because the model implements a heuristic renormalization-group-like coarse graining
of the polymer physics, so that changing the bead resolution does not introduce
new fitted physical constants. So, one specifies
the number of beads per chain \(N_b\) and the model's equations automatically coarse-grain the dynamics
at the corresponding spring scale. We take advantage of these features and
reduce the computational complexity by choosing \(N_b=5\). The Langevin equation
governing bead motion reads
\begin{equation}
    F^{\mathrm{in},k}_i
    +
    F^{\mathrm{el},k}_i
    +
    F^{\mathrm{ev},k}_i
    +
    F^{\mathrm{vd},k}_i
    +
    F^{\mathrm{th},k}_i
    =
    0 .
    \label{eq:langevin-force-balance}
\end{equation}
where \(k\) is the bead index, \(i\) is the spatial component, and
\(F^{\mathrm{in},k}_i\), \(F^{\mathrm{el},k}_i\),
\(F^{\mathrm{ev},k}_i\), \(F^{\mathrm{vd},k}_i\), and
\(F^{\mathrm{th},k}_i\) are the inertial, elastic, excluded-volume,
viscous-drag, and thermal-fluctuation forces, respectively. 
The force terms entering Eq.~\eqref{eq:langevin-force-balance} are defined
explicitly in Appendix~\ref{app:governing-system}. The bead-level force
balance is restated in Eq.~\eqref{eq:app-langevin-balance}; the FENE elastic
force is specified in
Eqs.~\eqref{eq:app-fene-potential}--\eqref{eq:app-elastic-gradient-form};
and the excluded-volume force and solvent-quality parameters are specified in
Eqs.~\eqref{eq:app-ev-potential-bead}--\eqref{eq:app-kuhn-excluded-volume}.
The hydrodynamic drag, Stokes velocity, mobility and diffusion tensor are
defined in
Eqs.~\eqref{eq:app-carrier-velocity-at-bead}--\eqref{eq:app-diffusion-mobility},
and the RPY diffusion tensor is given in
Eqs.~\eqref{eq:app-bead-separation}--\eqref{eq:app-rpy-near}.

The inertial term is retained here to display the full force balance; in the
computations below we use the overdamped, or Aristotelian, limit. In this
regime the bead inertial relaxation time associated with the small bead mass
is much shorter than the turbulent time scales relevant to polymer stretching,
so bead velocities are determined by the instantaneous balance of drag,
elastic, excluded-volume, and thermal forces.

Elasticity is taken into account via the FENE model, whose nonlinearity prevents
the polymer from stretching beyond its maximum length. A key feature of our
approach is that we exactly enforce this constraint with our numerical methods
\cite[][]{kivotides_stretch} (to be discussed in the next section). The
corresponding elastic potential and spring force are given in
Eqs.~\eqref{eq:app-fene-potential}--\eqref{eq:app-fene-spring-force}.

The excluded-volume force is an effective intermolecular force field,
including steric repulsion, coarse-grained at the bead level with an
exponential cut-off \cite[][]{kivotides_dense}; its bead-level form is given
in Eqs.~\eqref{eq:app-ev-potential-bead}--\eqref{eq:app-ev-force-explicit}.
The strengths of the viscous-drag and thermal-fluctuation forces are related
by the fluctuation--dissipation theorem.

Notably, we treat the Langevin equation as a vector stochastic differential
equation handling all degrees of freedom as one system. Thus, the mobility
includes hydrodynamic interactions between all beads in the system, represented
by the RPY tensor in Eqs.~\eqref{eq:app-bead-separation}--\eqref{eq:app-rpy-near}.

\paragraph{Relation to tracer-reduced hydrodynamic-interaction studies.}
Our physics setting differs from that of Ganesh et al.~\cite{ganesh2026hydrodynamic},
who recently studied how intrachain hydrodynamic interactions alter polymer
stretching and the coil--stretch transition in homogeneous isotropic turbulence.
Their bead--spring formulation contains internal separation variables and
hydrodynamic interactions between beads. However, in the turbulent-flow
implementation an additional simplifying reduction is made: the chain centre
of mass is assumed to move as a tracer of the carrier flow, and the internal
chain dynamics is then evolved using the velocity-gradient history sampled
along that imposed tracer trajectory. Thus the polymer trajectory is not itself
an output of the full bead dynamics.

The present calculation does not make this assumption. We advance the bead
positions directly under the carrier velocity, the hydrodynamic-interaction
contribution, the elastic and excluded-volume forces, and the Brownian
increments. The polymer midpoint and end-to-end vector are then constructed
from the resulting bead positions. Thus the strain, vorticity, alignment and
finite-time Lyapunov statistics reported below are sampled along
polymer-generated trajectories, not along imposed tracer trajectories. This
distinction is important because the validity of replacing polymer trajectories
by material or tracer trajectories is one of the issues examined in the present
work.

\subsubsection{Stokes equations}

Hydrodynamic interactions between beads are represented by the
linear-response microflow generated by the bead drag forces.
This flow satisfies the Stokes equations
\begin{eqnarray}\label{stoma}
\frac{\partial u^{\mathcal S}_i}{\partial x_i}=0,
\end{eqnarray}
\begin{eqnarray}\label{stome}
\frac{\partial p^{\mathcal S}}{\partial x_i}
-\mu_f \frac{\partial^2 u^{\mathcal S}_i}{\partial x_j \partial x_j}
-F^{\mathrm{vd},k}_i \delta(x-r^k)=0,
\end{eqnarray}
where the superscript \(\mathcal S\) denotes the Stokes-flow field, and
summation over the bead index \(k\) is assumed in the forcing term.

In the ultra-dilute regime considered here, the polymer system is sparse
enough that the long-range hydrodynamic interaction can be represented by
Rotne--Prager--Yamakawa (RPY) hydrodynamics in an unbounded domain. RPY
regularises the Stokeslet by replacing point forces with finite-size force
blobs, and gives an analytical expression for the mobility matrix, denoted
here by \(\mathsf{M}\).
This matrix maps bead forces to bead velocities and
is the inverse of the resistance matrix \(\mathsf{Z}\), so that
\(\mathsf{M}=\mathsf{Z}^{-1}\). The corresponding bead-level Stokes
velocity, mobility and diffusion tensor are summarized in
Eqs.~\eqref{eq:app-stokes-relative-velocity}--\eqref{eq:app-diffusion-mobility},
and the RPY diffusion tensor is given explicitly in
Eqs.~\eqref{eq:app-bead-separation}--\eqref{eq:app-rpy-near}.

The same hydrodynamic coupling also enters the Brownian forcing: by the
fluctuation--dissipation theorem, the Brownian increments have covariance
proportional to the mobility matrix and therefore require the action of its
square root on independent Wiener increments.
Thus both the viscous-drag response and
the Brownian covariance contain the multibody hydrodynamic interactions
between beads. The RPY approximation is appropriate for polymer beads,
which are effective, interpenetrating coarse-grained objects rather than
rigid particles requiring lubrication corrections as in Stokesian Dynamics
\cite[]{kivotides_knot,kivotides_dense}.

\subsubsection{Tangent flow equation}

We apply finite-time Lyapunov analysis to better understand polymer stretching processes.
The details of the Lyapunov
theory and its numerical implementation are described in
\cite[][]{kivotides_stringy,kivotides_lagrangian}, which should be
consulted for the physical interpretation and algorithmic details. Here
we mention only the polymer-specific aspects.

Let \(\mathcal{L}(t)\) denote the configuration of all
polymer chains at time \(t\), and let \(\mathcal{L}(0)\) be the
material reference configuration.
For each polymer chain, we denote by \(Z\) the midpoint of the chain
end-to-end vector in the reference configuration, and by \(z(Z,t)\) its
trajectory.
We also introduce an infinitesimal reference vector
\(f_R(Z)\), associated with a perturbation of the chain end-to-end
vector, and denote by \(f(Z,t)\) its image under the tangent-flow
evolution. Along the trajectory \(z(Z,t)\), obtained from the combined
Navier--Stokes, Stokes and Langevin system, we evolve \(f_R(Z)\) into
\(f(Z,t)\) using the tangent linear system
\begin{equation}
  \frac{df(Z,t)}{dt}
  =
  \frac{\partial u(x,t)}{\partial x}
  \bigg\rvert_{x=z(Z,t)} f(Z,t),
  \label{linsys}
\end{equation}
where
\begin{equation}
L_{ij}(x,t)\stackrel{\text{def}}{=}
\frac{\partial u_i(x,t)}{\partial x_j}
\end{equation}
is the continuum-mechanical velocity-gradient tensor.
Here \(u\) denotes the carrier Navier--Stokes velocity field.
Equivalently,
\begin{equation}
        \frac{df(Z,t)}{dt}=L(z(Z,t),t) f(Z,t).
        \label{temp_stretch}
\end{equation}
Introducing the deformation-gradient tensor \(F(Z,t)\), defined by
\(f(Z,t)=F(Z,t)f_R(Z)\), gives
\begin{equation}
	\frac{dF(Z,t)}{dt}=L(z(Z,t),t)F(Z,t),
        \qquad F(Z,0)=I .
        \label{defdyn}
\end{equation}
Formally,
\begin{equation}
        F(Z,t)=
        \mathcal{T}\exp\left[
        \int_0^t L(z(Z,t'),t')\,dt'
        \right],
        \label{solution}
\end{equation}
where \(\mathcal{T}\) denotes time ordering.
More explicitly, with
\(L_Z(t)\equiv L(z(Z,t),t)\),
\[
\mathcal{T}\exp\left[\int_0^t L_Z(t')\,dt'\right]
=
I+\int_0^t L_Z(t_1)\,dt_1
+\int_0^t dt_1\int_0^{t_1}dt_2\,L_Z(t_1)L_Z(t_2)+\cdots ,
\]
so that velocity-gradient tensors evaluated at later times appear to the
left of those evaluated at earlier times. This ordering is required because
\(L_Z(t_1)L_Z(t_2)\neq L_Z(t_2)L_Z(t_1)\) in general along a turbulent trajectory.

We solve Eq.~\eqref{defdyn} for \(F(Z,t)\) along each polymer end-to-end
midpoint trajectory \(z(Z,t)\).
This is a hybrid diagnostic: the
velocity-gradient tensor is sampled along polymer trajectories, while
\(f_R\) is evolved as an infinitesimal material vector of the carrier flow.
The distinction is important because polymer end-to-end midpoints are not
material fluid markers.
As indicated by the Langevin equation, elastic, excluded-volume, Brownian,
and hydrodynamic-interaction terms can make the bead motion differ from that
of material fluid points.

We also perform, along the chain trajectories, the singular value
decomposition (SVD) \cite[][]{matrix_analysis}
\begin{equation}
F=W\Sigma V^T ,
\label{eq:svd-decomposition}
\end{equation}
where \(W\) and \(V\) are orthogonal tensors and \(\Sigma\) is diagonal
with positive entries \(\sigma_i\),
the singular values of \(F\). Since \(F\) is a square matrix,
\(VV^T=V^TV=I\), and therefore
\[
F=W\Sigma V^T
  =W I\Sigma V^T
  =W V^T V\Sigma V^T
  =(WV^T)(V\Sigma V^T).
\]
Thus
\begin{equation}
F=\mathcal{R}U,\qquad
\mathcal{R}=WV^T,\qquad
U=V\Sigma V^T ,
\label{eq:right-polar-decomposition}
\end{equation}
which is the right polar decomposition of \(F\). Here \(\mathcal{R}\) is the
rotational part and \(U\) is the right stretch tensor. We also have
\[
U^2=F^TF,
\]
and the eigenvalues of \(U\), the principal stretches, are the singular
values \(\sigma_i\) of \(F\).
By the spectral decomposition theorem,
\[
U=\sum_{i=1}^3 \sigma_i r_i r_i^T ,
\]
where the \(U\) eigenvectors \(r_i\) are the right principal directions, equivalently the right
singular vectors of \(F\).

Because the carrier flow is incompressible, the tangent evolution is
isochoric and the product of the principal stretches is unity. In the
calculations, this condition was satisfied with typical accuracy
\(10^{-6}\).
The finite-time Lyapunov exponents are then defined by
\begin{equation}
\lambda_i(t;Z)=\frac{\ln \sigma_i(t;Z)}{t}
\label{eq:finite-time-lyapunov}
\end{equation}
following \cite[][]{kivotides_stringy,kivotides_lagrangian}.
These are finite-time quantities; asymptotic Lyapunov exponents would require
an additional \(t\to\infty\) limit.
Hence the
finite-time Lyapunov exponents contain the same stretching information
as the singular values of \(F\).
The polymer midpoint,
sampled velocity-gradient tensor, tangent-flow equation,
SVD and accumulated finite-time Lyapunov exponents are also collected in
Appendix~\ref{app:governing-system},
Eqs.~\eqref{eq:app-chain-midpoint}--\eqref{eq:app-svd-normalised-lyapunov}.

\subsection{Polymer stretching}

We examine several informative metrics of polymer stretching. 
Metrics derived from the tangent (linearized) system are reliable only when
\(|f(Z,t)|\) is small --- the regime where Lyapunov linearization holds.
When \(|f(Z,t)|\) is large, the polymer state is no longer a small 
perturbation of the reference trajectory, so propagating it with the instantaneous
\(L(z(t),t)\) is not a valid approximation --- the dynamics 
are dominated by finite-amplitude, state-dependent nonlinearities instead.
For this reason, computing long-time statistics of the alignment between
\(f(Z,t)\) and the strain-rate eigenvectors at \(z(t)\)
is not meaningful: the two sets of vectors eventually decorrelate. 
Since we solve the full nonlinear problem here, we evaluate these key quantities directly.

\subsubsection{Polymer stretching in the nonlinear system}

We can also characterise polymer stretching via fully nonlinear measures.
For this purpose, we employ, for each polymer chain, the end-to-end vector
\(\Delta z=z_e-z_s=r^e(t)-r^s(t)\), where \(z_e\) and \(z_s\) are spatial
vectors belonging to \(\mathcal{L}(t)\). The corresponding midpoint is
\(z=(z_e+z_s)/2\). These configuration variables are collected in
Appendix~\ref{app:governing-system},
Eqs.~\eqref{eq:app-chain-midpoint}--\eqref{eq:app-chain-end-to-end}.

Flow effects on polymers are depicted by the action of the velocity-gradient
tensor \(L(z,t)\) on fibers \(\Delta z\) \cite[][]{arcady}.
Since \(L=D+W\) and \(W\) is a rigid motion, to investigate polymer stretching,
we only need to consider the action \(D\Delta z\) of the strain-rate tensor.
If \(\mu_j\) and \(s_j\) \((j=1,2,3)\) are, respectively, the eigenvalues and
orthonormal eigenvectors of \(D\), the spectral decomposition theorem
\cite[][]{matrix_analysis} gives
\begin{equation}
    D
    =
    \sum_{j=1}^{3}\mu_j s_j s_j^T .
    \label{eq:strain-rate-spectral-decomposition}
\end{equation}
Here, positive \(\mu_j\) indicates polymer stretching along \(s_j\), and,
similarly, negative \(\mu_j\) indicates polymer compression, which includes
folding. Incompressibility implies \(\sum_{j=1}^{3}\mu_j=0\). Hence
\begin{equation}
    D\Delta z
    =
    \sum_{j=1}^{3}
    \mu_j
    |\Delta z|
    \cos(s_j,\Delta z)
    s_j,
    \qquad
    |D\Delta z|^2
    =
    |\Delta z|^2
    \sum_{j=1}^{3}
    \mu_j^2
    \cos^2(s_j,\Delta z).
    \label{eq:strain-action-on-polymer}
\end{equation}
Thus, it is necessary to consider both the magnitude of \(\mu_j\) and the
alignment cosines \(\cos(s_j,\Delta z)\) to understand \(D\)'s role in
instantaneous polymer stretching. Similarly,
\begin{equation}
    (D\Delta z)\cdot \Delta z
    =
    |\Delta z|^2
    \sum_{j=1}^{3}
    \mu_j
    \cos^2(s_j,\Delta z).
    \label{eq:strain-projection-on-polymer}
\end{equation}

At time \(t\), we can also define the scalar
\(\delta\stackrel{\text{def}}{=}|z_e-z_s|\), and show
\cite[][]{gurtin} the useful identity
\[
    \delta\dot{\delta}
    =
    (z_e-z_s)\cdot[u(z_e)-u(z_s)],
\]
where \(u\) is the Navier--Stokes velocity. Thus, the scalar
\begin{equation}
    \beta
    =
    \frac{\dot{\delta}}{\delta}
    =
    \frac{(z_e-z_s)\cdot[u(z_e)-u(z_s)]}{\delta^2}
    \label{eq:instantaneous-material-line-stretching-rate}
\end{equation}
measures the instantaneous material-line stretching rate associated with
the current polymer end-to-end segment.

A useful characterisation of the flow field stretching the polymers is provided
by the \(\lambda^*\) parameter of Kerr, Lund and Rogers \cite[][]{kerr,lund},
\begin{equation}
    \lambda^*
    =
    -3\sqrt{6}
    \frac{\mu_1\mu_2\mu_3}
    {(\mu_1^2+\mu_2^2+\mu_3^2)^{3/2}},
    \label{eq:kerr-lund-rogers-lambda-star}
\end{equation}
which ranges from \(-1\) to \(1\). It is easy to verify that
\(\lambda^*=-1\) corresponds to axisymmetric uniaxial extension,
\(\lambda^*=0\) to planar shear and \(\lambda^*=1\) to axisymmetric biaxial
extension. The PDF of \(\lambda^*\) is uniform for a Gaussian velocity field,
whilst, for isotropic turbulence, Lund and Rogers showed that the most
probable strain process is axisymmetric biaxial extension that is associated
with high dissipation flow regions. We therefore examine the distribution of
\(\lambda^*\) sampled along polymer configurations in turbulence.

The flow-type parameter \(\lambda^*\) is closely related to the normalized
strain-eigenvalue parameter \(s\) used by
Kopyev~\cite[][]{kopyev2018} in his analysis of strain-rate tensor statistics
in isotropic incompressible flows. In fact, Kopyev's \(s\) is the same
strain-shape diagnostic as the present \(\lambda^*\). Kopyev
also introduces a second normalized eigenvalue parameter, denoted here by
\(\beta_K\) to distinguish it from the instantaneous stretching rate
\(\beta\) used below. This parameter measures the normalized intermediate
strain-rate eigenvalue. However, \(s\) and \(\beta_K\) are not independent:
for incompressible strain they are related algebraically by
\[
        s=\frac{\beta_K(3-\beta_K^2)}{2}.
\]
Thus Kopyev's \(\beta_K\) is an alternative parametrisation of the same
normalized strain-eigenvalue shape information already represented here by
\(\lambda^*\).

For this reason, introducing \(\beta_K\) would not add independent information
to the present analysis. The relevant eigenvalue-ratio information is already
contained in \(\lambda^*\) and in the separate strain-eigenvalue PDFs. In
particular, we compute directly the PDF of the intermediate strain-rate
eigenvalue, so the information emphasized by \(\beta_K\) is already represented
explicitly in the eigenvalue statistics. The additional information needed for
the present polymer-stretching problem is the orientation of the polymer
end-to-end vector relative to the local strain eigenvectors, which we have also
computed in this work.

\subsubsection{Polymer stretching in the tangent system}

The tangent linear system allows us to track the limiting deformation of
polymer-fibre directions. Appropriate finite candidates for the latter
are the two vectors connecting the middle bead with the two end-beads in the
initial conditions. Indices \([s,m,e]\) denote start bead, middle bead and end
bead correspondingly. In particular, let us consider, for each chain, the line
segment
\[
    \Delta Z_e
    =
    Z_e-Z_m
    =
    r^e(0)-r^m(0)
    =
    L_e g_e ,
\]
where \(g_e\) is the unit vector along \(\Delta Z_e\), \(L_e\) is the finite
length of \(\Delta Z_e\), and \(Z_e,Z_m\in\mathcal{L}(0)\). At time \(t\),
the spatial vector corresponding to \(\Delta Z_e\) is
\[
    \Delta z_e
    =
    z_e-z_m
    =
    r^e(t)-r^m(t),
\]
where \(z_e,z_m\in\mathcal{L}(t)\). Then, taking the limit for small \(L_e\),
one can show \cite[][]{gurtin} that
\begin{equation}
    \zeta_e
    =
    \lim_{L_e\to0}
    \frac{|\Delta z_e|}{L_e}
    =
    |U(Z_m,t)g_e| .
    \label{eq:tangent-fibre-stretch}
\end{equation}
Here \(U\) is the right stretch tensor introduced in
Eq.~\eqref{eq:right-polar-decomposition}. The quantity \(\zeta_e\) represents
the stretch at \(z_m\) along the direction \(g_e\) at \(Z_m\), since it
measures the limiting length of the deformed fibre in units of the length of
the undeformed fibre in the material direction \(g_e\). Thus, we say that
fibres at \(Z_m\) in the direction \(g_e\) are stretched if \(\zeta_e\neq1\).

The quantities \(\beta\) and \(\zeta_e\) are complementary stretching
measures. The former, defined in
Eq.~\eqref{eq:instantaneous-material-line-stretching-rate}, refers to finite
spatial end-to-end segments and gives the material-line stretching rate
associated with the current polymer configuration, whereas the latter, defined
in Eq.~\eqref{eq:tangent-fibre-stretch}, refers to infinitesimal reference
material fibres and is obtained from the \(L_e\to0\) tangent-flow limit.

Finally, we can employ \(F\) to define the Green--St. Venant tensor
\begin{equation}
    E
    =
    \frac{1}{2}(F^T F-I),
    \label{eq:green-st-venant-strain}
\end{equation}
where \(I\) is the identity tensor. This tensor is identically zero for rigid
rotation, since then \(F^T F=I\). Thus \((E:E)^{1/2}\), with
\(E:E=\operatorname{tr}(E^T E)\), is a measure of pure strain.

\section{Methods of analysis}

Before describing the numerical and algorithmic methods used to solve the
governing equations, we first identify the nondimensional parameters that
define the physical regime of the calculation. The analysis is therefore
organised in two levels. The first is dimensional and scaling analysis: it
locates the present calculation within the parameter space of mesoscopic
polymeric fluid dynamics. The second is numerical and algorithmic analysis: it
describes how the corresponding equations are discretised, coupled, and
advanced in time.

\subsection{Dimensional and scaling analysis}

The mesoscopic formulation admits a large space of possible dynamical
regimes, ranging from isolated-chain stretching to polymer-modified
turbulence and, at larger volume fractions, entangled polymer dynamics. To
specify the regime studied here, we introduce nondimensional numbers that
measure concentration, solvent quality, turbulent inertia, polymer feedback
on the carrier flow, elastic stiffness, excluded-volume effects,
hydrodynamic-interaction effects, and trajectory-dependent stretching. Some
of these parameters are needed to classify the present calculation directly;
others are introduced because they organise the broader mesoscopic problem
and provide a guide for future computations in regimes where polymer
feedback, concentration effects, or entanglements become dynamically
important. Thus these quantities are not auxiliary diagnostics: they locate
the present calculation within the parameter space of mesoscopic polymeric
fluid dynamics and determine the relative contribution of different physical
effects to the dynamics.

In particular, the present calculation is designed to realise an
ultra-dilute turbulent solution in a good solvent. The nondimensional
numbers below verify that the polymer volume fraction is far below the
semidilute threshold and that the polymer--turbulence interaction parameter
is small enough for the resolved turbulent structure to remain essentially
unchanged. They also characterise the bead-level polymer dynamics, showing
that the chains are strongly deformed by turbulent drag while their motion
remains governed by elasticity, Brownian forcing, excluded volume, and
hydrodynamic interactions rather than by purely passive material-line
transport. We pay special attention to Weissenberg number \(Wi\), 
Deborah number \(De\), and the Lyapunov number \(L\),
because they characterise distinct aspects of polymer response: the
competition between local turbulent stretching and polymer relaxation, the
liquid-like or solid-like character of the viscoelastic response, and the
accumulated material stretching along polymer trajectories, respectively.

{\renewcommand{\labelenumi}{(\arabic{enumi})}
\begin{enumerate}

\item The most basic nondimensional number is the volume fraction $\phi$, the ratio of the occupied volume of the polymer chains
over the volume of the solution. We have that $0 < \phi < \phi^{*} < \phi_e < \phi^{**} < 1$,
where $\phi^{*}$, $\phi_e$ and $\phi^{**}$ are, correspondingly, the chain overlap, entanglement and concentrated solution thresholds.
It is useful to compare $\phi$ with the semidilute solution threshold $\phi^{*}\approx (N b^3)/R^3$, where 
$b$ is the Kuhn length, $N$ is the number of Kuhn monomers per chain, and $R$ is the chain size.
According to Flory theory, $R \approx b (v/b^3)^{2\nu-1} N^{\nu}$, where the swelling exponent $\nu=0.588$,
and $v$ is the excluded volume. In our solution, we have $\phi/\phi^{*}=7.602\ 10^{-12} \ll 1$, so we are well below
		the semidilute limit.

\item Another basic number is the solvent quality number $\Theta=T/\theta$ (where $T$ is the solution temperature). 
When $\Theta=1$, we have ideal chains
with zero excluded volume, and the solvent is called a $\theta-$solvent.
In our study, $\Theta=1.6$, hence, we have a good solvent, where the excluded volume 
of the chain is positive and effective intrachain interactions cause the polymers to swell under equilibrium conditions.
For consistency, our polymer physics assume $\Theta> 1$, and our equations automatically
incorporate temperature dependencies in this $\Theta$ regime.

\item In the absence of any imposed time scales in the carrier fluid, the Strouhal number is irrelevant, hence
the Taylor Reynolds number $Re_{\lambda}=u^{\prime} \lambda/\nu_f$, 
where \(u'\) is the turbulence intensity, \(\lambda\) is the Taylor scale,
and \(\nu_f\) is the solvent kinematic viscosity, suffices
to characterize the ratio between inertial and viscous effects in turbulence. In our turbulent
		flow, $Re_{\lambda}\approx 82$.

\item The fluid is coupled with the polymer chains via the drag force, whose effects on turbulence structure we quantify 
by means of an {\em interaction parameter} $I$, which is the ratio of the drag force from the polymers to the fluid with turbulent inertial forces. 
For $I> 1$, we expect the polymer microstructure to alter the vortical turbulence structure.
To evaluate $I$, we will consider these forces within a fluid volume of size $\lambda^3$. The choice of the 
Taylor scale is motivated by the fact that most of turbulent strain production takes place there, and strain is what 
stretches the polymers via the mediation of the viscous drag force on the chains.
If $\rho_f$ is the fluid mass density, then the (turbulent) inertial force within this volume
is $\rho_f \lambda^3 u^{\prime} (u^{\prime}/\lambda)$. To evaluate the drag force from the polymers 
we make use, as shown above, of the fact that our solution is dilute and the solvent is good.
Accordingly, the drag force associated with a single polymer chain is $f_d\approx \zeta u^{\prime}$, where 
$\zeta \approx \rho_f \nu_f R$ is the chain friction coefficient.
Notice that we use the Zimm formula for $\zeta$, employing the chain size (rather than a bead size) to scale the drag force.
Indeed, since $\phi\ll \phi^{*}$ here, the hydrodynamic interactions are not screened and 
the polymer appears to a turbulent fluctuation as a pseudo-solid object of size $R$ \cite[][]{rubinstein}.
To find the total reaction from the polymers we need to multiply this value with the
total number of chains within $\lambda^3$, which is $P=(\phi \lambda^3)/N b^3$.
Consequently,
\begin{eqnarray}
I=\frac{P \rho_f \nu_f R u^{\prime}}{\rho_f \lambda^3 u^{\prime} (u^{\prime}/\lambda)}=
\frac{P (\nu_f/\lambda^2)}{(u^{\prime}/R)}=\frac{P \tau_i}{\tau_{vd}},   \nonumber
\end{eqnarray}
where $\tau_{vd}$ is a viscous time scale based on the {\it strain producing volume size}, and $\tau_i$ is an inertial flow time scale
based on the {\it chain size scale}.
Notably, the Zimm/good-solvent estimate gives larger values of \(I\) than
an ideal-chain/Rouse-size estimate, because it gives a larger chain size
\(R\), hence a larger chain-scale inertial time \(\tau_i=R/u'\).
In our setting, \(I=1.5\times10^{-7}\), which indicates an
ultra-dilute turbulent solution, where the turbulence structure is unaffected
by the polymers, although the latter still induce microscopic flow fields via
their hydrodynamic interactions. The value of \(I\) also provides a useful
scale estimate for turbulence-modifying mesoscopic calculations. Since \(I\)
scales linearly with polymer loading, achieving \(I=O(1)\) would require
increasing the polymer population in each Taylor-scale strain-producing
volume by a factor \(I^{-1}\simeq 6.7\times10^{6}\). At a minimal resolution
of three beads per chain, this corresponds to approximately \(2\times10^{7}\)
beads per Taylor-scale volume. For a representative turbulent calculation
with a \(256^3\) grid, \(\Delta x\approx\eta\), and
\(\lambda\approx10\eta\), the number of Taylor-scale volumes is
\((256/10)^3\simeq1.7\times10^4\). The corresponding total bead count is
therefore
\[
N_{\mathrm{beads,total}}
\sim
2\times10^{7}\times1.7\times10^{4}
\simeq 3.4\times10^{11}.
\]
Thus, turbulence-modifying mesoscopic calculations require
\(O(10^{11})\) beads, placing them well beyond the present ultra-dilute
calculation and motivating future work.

\item Noting that in linear elasticity the ideal elastic constant is a measure of the elastic force per unit length, we can 
form its ratio with the drag force per unit length and gauge chain elastic stiffness in viscous drag units.
The corresponding elastic stiffness number $S=(kT/N b^2)/(\rho_f \nu_f u^{\prime})$ is indicative of the {\it propensity} of drag forces to strain the polymer
in the linear (small-deformation) regime.
Notably, $S$ is not indicative
of elastic stress levels in the fluid, since these depend on accumulated chain deformation which is not what $S$ encodes. 
In other words, elastic stiffness and elastic force (or energy) are distinct: it is possible (depending on the specific elasticity law) to have large 
elastic energy with small elastic stiffness. 
Although related to relaxation physics, \(S\) expresses the balance in force
units rather than time units: it compares the small-deformation elastic
stiffness with the viscous-drag force scale. Thus \(S\gg1\) implies that
viscous drag is too weak to significantly deform the polymers, whereas
\(S\ll1\) indicates chains that are readily deformed by the flow.
In our polymer solution, \(S=1.5\times10^{-3}\), so the small-deformation
elastic stiffness is far below the viscous-drag scale. The chains are
therefore nonstiff and readily deformed by the turbulent flow.

\item Excluded volume effects are expected to be much weaker than drag and elastic force effects in turbulent flows, but they can be important in the initial stages of 
coil-stretch transition \cite[][]{kivotides_knot}, so for completeness, we define here the excluded volume force number $O$, which is the ratio of excluded volume forces 
to viscous drag forces. To scale the former, we use the formulae of \cite[][]{kivotides_dense} for the strength of their potential
$k T N^2 v/R^3$ divided by their range $\delta\approx R$. The excluded volume $v$ is given by $v=b^3 (T-\theta)/T$ \cite[][]{rubinstein}.
Notably, it is important to define here the equilibrium chain length via the {\it dilute solution} formula
$R=b {(v/b^3)}^{2\nu-1} N^{\nu}$, that is via Flory mean-field theory and its renormalization group refinement of the swelling exponent $\nu$ (i.e. $\nu=0.588$).
This gives $O=k T N^2 v/(\rho_f \nu_f u^{\prime} R^5)$.
In our present study, \(O=2.2\times10^{-4}\), so excluded-volume forces are
small compared with the viscous-drag scale and are not expected to
significantly affect the turbulent stretching dynamics directly. Their
possible importance is instead indirect: by influencing the early stages of
coil--stretch deformation, they can affect the subsequent distribution of
chain extensions and hence the observed polymer-stretching statistics.

\item It is also important to estimate the ratio $\Delta$ of thermal-fluctuations forces to the viscous drag. 
It is easy to show that $\Delta$ is the polymer fluid analog of the inverse P\'eclet number \(\mathrm{Pe}^{-1}\)
of granular fluids. To evaluate $\Delta$, we scale the Brownian forces with $k T/R$ and the viscous drag force with $\zeta u^{\prime}$, hence
\begin{eqnarray}
\Delta=\frac{(k T/R)}{\zeta u^{\prime}}=\frac{k T}{\zeta u^{\prime} R}=\frac{D}{u^{\prime} R}, \nonumber
\end{eqnarray}
where the Einstein relation $D=k T/\zeta$ was used for the diffusion coefficient $D$, i.e. $D=k T/(\rho_f \nu_f (b N^{\nu}))$, with $\nu=0.588$.
Noting that the Brownian diffusion time-scale $\tau_d$ is $\tau_{d}\approx R^2/D$, and the convective time scale for polymer motion is 
$\tau_c\approx R/u^{\prime}$, we can write
\begin{eqnarray}
\Delta=\frac{D}{u^{\prime} R}=\frac{D R}{u^{\prime} R^2}=\left(\frac{D}{R^2}\right) \left(\frac{R}{u^{\prime}}\right)=\frac{\tau_c}{\tau_{d}}=\mathrm{Pe}^{-1}. \nonumber
\end{eqnarray}
In our flow, \(\Delta=3\times10^{-5}\), so the Brownian force scale is much
smaller than the viscous-drag scale. Comparing with the excluded-volume
number \(O=2.2\times10^{-4}\), the excluded-volume force scale is larger
than the Brownian force scale by approximately one order of magnitude.
Although both are small compared with viscous drag, their relative magnitude
is physically informative: when elastic forces become weak, Brownian
diffusion relaxes the chain through a conformational landscape shaped by
excluded-volume effects. This provides a possible mechanism for slow
late-stage relaxation after the dominant elastic response has decayed.

\item As already mentioned, the polymer stiffness number $S$ does not represent a measure of elastic effects in the flow.
Indeed, a direct indicator of the linear or nonlinear character of the dynamics 
is the magnitude of polymer deformation or strain. It is not easy to a priori gauge the extent of chain strain accumulation
in a turbulent flow. An alternative but not equivalent measure is the Weissenberg number $Wi=\tau_R \dot{\gamma}$,
where \(\dot{\gamma}\) is the fastest characteristic turbulent strain rate, associated here with the Kolmogorov scale,
and $\tau_R$ is the longest 
chain relaxation time scale. For $\phi \ll 1$ here, $\tau_R$ is well approximated by Zimm theory \cite[][]{rubinstein,kivotides_knot}, i.e. 
\begin{equation}
    \tau_R
    \approx
    \frac{\rho_f \nu_f}{k_B T} R^3 .
    \label{eq:zimm-relaxation-time}
\end{equation}
The heuristic idea is that since the fluid motions
are faster than the polymer relaxation time, the polymer ought to accumulate strain faster than release it, thus, the level of elastic 
stresses in the flow will grow. Although this argument is plausible, it is important to note that chain deformation also depends on 
the geometry of polymer-flow interactions, quantified by the alignments between the strain rate eigenvectors and the polymer end-to-end vector.
One of our aims here is to investigate in depth these kinematical features.
Moreover, the heuristic interpretation assumes 
that polymer deformation is well approximated by considering it to be
a material line --- it is in this latter case that the flow strain rate can directly inform about chain deformation tendencies.
However polymers do not move with the flow, since the drag force in the Langevin equation is 
balanced by elastic and excluded volume forces, and elastic forces can easily match viscous drag at high polymer extensions.
It is for this reason that polymers have an upper extensibility limit, and
sufficiently large tensions can ultimately lead to chain scission
\cite[][]{vincenzi2}.
With these qualifications in mind, we proceed by noting that in our homogeneous turbulence
we can identify $\dot{\gamma}$ with the shear rate $\dot{\gamma}_{\eta}$
within a Kolmogorov eddy,
and use it as a global characteristic shear rate scale. Since \cite[][]{frisch}, $\dot{\gamma}_{\eta}=\epsilon^{1/3} \eta^{-2/3}$ where $\epsilon$ is the 
rate of kinetic energy dissipation in the turbulent flow and $\eta$ is the Kolmogorov length scale,
we can define $Wi=\tau_R \epsilon^{1/3} \eta^{-2/3}$. In our study, $Wi=81.174$, which implies that the smallest turbulent
scales are much faster than polymer relaxation.

\item We can improve upon \(Wi\) physics, by characterising polymer deformation by means of
material-line deformation sampled along the chain trajectory.
Material stretching is well quantified by the Lyapunov exponents which describe the evolution of material length perturbations in the linearised (tangent) system
employing the velocity gradient tensor along polymer trajectories.
The calculation of the exponents requires the consideration of the $t \to \infty$ limit, although, in practice, convergence occurs over much smaller time scales.
In the asymptotic stretching regime, the largest Lyapunov exponent
\(\Lambda_1\) gives the leading estimate
\(\gamma(t)\sim e^{2\Lambda_1 t}-1\).
Here, \(\gamma\) measures the relative increase in squared length of a
material element over time \(t\), given the average stretching rate
\(\Lambda_1\). 
Since the polymers considered here remain smaller than the
Kolmogorov scale, such a material element may be identified kinematically
with the initial polymer end-to-end vector when estimating the stretching
imposed by the carrier-flow gradient.
More explicitly,
$\gamma=|f(t)|^2/|f(0)|^2 - 1$, where $f(t)$ signifies a sufficiently small material vector at time $t$, hence $\gamma\equiv 0$ at $t=0$.
Notably, at $t=\tau_R$, a value of the non-dimensional number $L$, $L\equiv \tau_R \Lambda_1 \approx 0.5$ indicates 
strong deformations and could be associated with coil-stretch transitions in the  
polymer system. Our aim here is to compute the Lyapunov exponents along polymer trajectories, and under the plausible assumption
of the statistical universality of stretching processes in ultra-dilute turbulent solutions,
to employ these Lyapunov exponents to characterise similar processes in experiments and applications.
Remarkably, if the $L$ numbers prove useful, it would seem to
indicate that material stretching is a good approximation of polymer stretching.

The usefulness of stretching--relaxation measures is standard in rheology and
polymer turbulence. The basic idea that polymer stretching is governed by the
deformation accumulated over a polymer-chain relaxation time goes back to
classical rheological and time-scale arguments, including Lumley's argument
for polymer stretching in turbulence~\cite{lumley1969drag,lumley1973drag} and
Tanner's interpretation of the Weissenberg number in terms of dumbbell
stretching accumulated over one relaxation time~\cite{tanner1976test}.
Balkovsky, Fouxon and Lebedev~\cite{balkovsky} and
Chertkov~\cite{chertkov} then developed Lyapunov-based theories of
polymer stretching in random and turbulent flows, in which the relevant
stretching statistics are those of the carrier-flow or material deformation.

The role of \(L\) in the present work is different. It is not introduced as a
new general coil--stretch criterion, nor is it an application of a simplified
random-flow theory based only on material-flow Lyapunov statistics. It is used
as a diagnostic for testing whether pure material stretching, evaluated along
the polymer-specific trajectories generated by the bead--spring dynamics,
approximates the full stretching dynamics of the same polymers. The formulation
is therefore agnostic about the relation between material stretching and
polymer stretching. It allows the computed solutions to determine whether the
polymer behaves approximately as a material line along its own
polymer-generated trajectory, or whether its stretching differs because of
elasticity, excluded volume, Brownian forcing, finite extensibility and
hydrodynamic interactions.

\item Although \(Wi\) and \(L\) provide useful measures of stretching tendency and
material deformation along polymer trajectories, 
it is important to also characterize the quality of these effects. Is the viscoelastic continuum's behaviour closer to
that of a purely viscous fluid or closer to that of a purely elastic solid? This question is independent 
of elastic and strain levels in the flow, as it is concerned with 
the nature of the viscoelastic response to strain or stress loads over a wide range of magnitudes.
The appropriate nondimensional number here is the Deborah number $De$. In a Maxwell fluid of relaxation time 
$\tau_R=\eta_M/G_M$ (where $\eta_M$ is the shear viscosity and $G_M$ the shear modulus of the fluid)
subject to sinusoidal shear strain with frequency $\omega$, there are two well defined limits.
$De=\omega \tau_R \gg 1$ indicates a purely solid response,
with the storage modulus equal to the shear modulus and loss modulus equal to zero,
and $De=\omega \tau_R \ll 1$ purely viscous response, with storage modulus zero and
loss modulus equal to $\omega$ times the (zero shear rate) viscosity.
In between these extremes there is mixed viscoelastic behaviour \cite[][]{tanner}.
It is {\it assumed} that these rigorous conclusions could apply in more general viscoelastic media under an appropriate
generalization of the applied strain frequency $\omega$. In actual flows \cite[][]{tanner,leal}, the variation of 
strain rates applied to a moving polymer is Lagrangian in nature and the analog of $De$ could be 
$De=\tau_R \left|\frac{D}{Dt}\ln |\nabla u|\right|$, where $u$ is the instantaneous velocity field. In a
homogeneous turbulent flow, most of the strain takes place
in scales of the order of the Taylor scale $\lambda$. Since the usefulness of $Re_{\lambda}$
indicates the physical relevance of turbulence intensity $u^{\prime}$ at the Taylor scale, we can use the strain rate $u^{\prime}/\lambda$
or the flow characteristic time $\lambda/u^{\prime}$, to define $De=\tau_R u^{\prime}/\lambda$. Notably, this time scale can be interpreted as representative of the polymer residence
time in space regions of size $\lambda$ with homogeneous significant strain. After that time, the polymer finds itself within a realization of
``different" vigorously straining eddy motions. Similarly to our discussion of the usefulness of the Lyapunov exponents,
our turbulent $De$ number substitutes the material residence time for the actual polymer residence time (which is difficult
to know a priori). Certainly, such simplifications can only be empirically justified. 
In our turbulent flow, \(De=21\). 
Hence, the polymers exhibit a strongly
elastic response, which, for appropriately large values of the interaction
parameter \(I\), is compatible with an effective viscoelastic-fluid continuum
description.
Although the present ultra-dilute calculation does not generate a
polymer elastic stress large enough to feed back into the carrier-flow
equation, a corresponding dilute-solution elastic stress can be constructed by
employing Kramer's polymer stress formula
\cite[][]{kivotides_stretch} and scaling it to a finite dilute polymer
concentration. The computed Deborah number would then characterize the nature
of the elastic response of such a dilute solution.
\end{enumerate}

The nondimensional numbers introduced in this subsection are collected in
Table~\ref{tab:nondimensional-parameters}. This table is included to make the
computed parameter regime explicit.
\begin{table}[!htbp]
\centering
\caption{
Nondimensional parameters introduced in the dimensional and scaling analysis.
The table collects the values used to identify the physical regime of the
present calculation.
}
\label{tab:nondimensional-parameters}
\begin{tabular}{lll}
\hline
Quantity & Meaning & Value \\
\hline
\(\phi/\phi^\ast\) & Diluteness ratio relative to overlap threshold & \(7.602\times 10^{-12}\) \\
\(\Theta=T/\theta\) & Solvent-quality number & \(1.6\) \\
\(Re_\lambda=u'\lambda/\nu_f\) & Taylor-scale Reynolds number & \(82\) \\
\(I\) & Polymer--turbulence interaction parameter & \(1.5\times 10^{-7}\) \\
\(S\) & Polymer stiffness number & \(1.5\times 10^{-3}\) \\
\(O\) & Excluded-volume force number & \(2.2\times 10^{-4}\) \\
\(\Delta=Pe^{-1}\) & Brownian-force number & \(3.0\times 10^{-5}\) \\
\(Wi\) & Weissenberg number & \(80\) \\
\(L=\tau_R\Lambda_1\) & Lyapunov stretching--relaxation number & \(20.6\) \\
\(De\) & Deborah number & \(21\) \\
\hline
\end{tabular}
\end{table}
\subsection{Numerics}

The numerical methods employed in this work have been discussed in detail elsewhere. Here, we provide only a brief overview, 
emphasizing the mathematical properties that make them particularly suitable for the present physical problem. However, we 
describe in detail the SVD normalization method developed in this study, highlighting its geometric 
characteristics and presenting the corresponding mathematical formulation.

\subsubsection{The Navier-Stokes solver}

We apply a staggered grid, fractional step, finite volume method \cite[][]{taira,kivotides_cloud}. 
All spatial partial derivatives are computed with second
order accurate schemes. The method employs an explicit,
third order accurate, low storage Runge-Kutta (RK) method for the computation of the advective terms.
In addition, an implicit, second order accurate Crank-Nicolson (CN) scheme
is applied to the viscous terms. Since this scheme is implicit
and the boundary conditions are periodic, the method requires the 
solution of cyclic-tridiagonal, linear algebraic
equation systems. We have used the Sherman-Morrison formula to 
reduce the latter to much more easily solved tridiagonal systems.
The CN scheme is incorporated into the
RK steps and the method becomes a hybrid RK/CN scheme.
The pressure gradient is
incorporated explicitly into the RK steps, but in order to
satisfy incompressibility, an additional calculation after each
RK substep projects the velocity field onto the space of
divergence-free vector fields (Hodge projection). This latter
computation is equivalent to a separate velocity update due
to the pressure gradient. 
The net result of the RK and Hodge projection procedures is that --- depending
on the chosen scheme parameters --- pressure-gradient effects are captured
with first- or second-order temporal accuracy.
We have found the first order accurate scheme to be more stable, 
and we have employed it here.

\subsubsection{The Langevin solver}

The choice of numerical method for solving the Langevin equation is guided
by the smallness of the polymer inertial force on the turbulent time scales
of interest. As the flow field changes, it exerts time-dependent viscous drag
forces on the polymer beads. Because the bead masses are very small, these
forces would produce very large accelerations over very short inertial time
scales. Consequently, bead velocities relax much faster than the viscous
diffusion time scales of the turbulent flow or the Brownian diffusion time
scales of the chains. On the latter time scales, the inertial force is
therefore negligible, and bead motion is determined by the instantaneous
balance of hydrodynamic drag, elastic, excluded-volume, and random Brownian
forces.

By choice, we do not resolve this rapid inertial relaxation, since it does
not control the polymer stretching processes studied here. 
We therefore neglect chain inertia and solve the Langevin equation in the
diffusive limit. In this limit it takes the form of the overdamped Langevin
equation, or Brownian dynamics, and the dynamics become Aristotelian.
In doing so, we discard the high-frequency inertial components of
the bead response, while retaining the conformational dynamics generated by
drag, elasticity, excluded volume, hydrodynamic interactions, and Brownian
forcing. Methods for integrating the Langevin equation in the inertial regime
are discussed in \cite[][]{coffey}.

The overdamped Langevin equation represents an algebraic stochastic force balance for the bead velocity. To integrate it, 
one must select the appropriate stochastic calculus interpretation, since two definitions of the stochastic integral are possible. 

The Ermak--McCammon scheme~\cite[][]{ermak,kivotides_knot} is a first-order, It\^o-compatible extension of the Euler--Maruyama method that 
includes a drift correction for position-dependent diffusion. The Fixman scheme~\cite[][]{fixman,kivotides_dense} is 
a stochastic midpoint (Heun-type) integrator consistent with the Stratonovich interpretation. The former is 
computationally efficient but requires explicit drift evaluation, while the latter avoids this correction at 
the cost of roughly doubling the computational work per step. For sufficiently small timesteps, both 
yield equivalent statistical results; however, Fixman's method produces smoother, Stratonovich-type trajectories, whereas 
the Ermak--McCammon trajectories reflect the characteristic roughness of It\^o processes owing to their strictly non-anticipative character.

We employ the \emph{Ermak--McCammon scheme} for two reasons. First, our use of Rotne--Prager--Yamakawa (RPY) hydrodynamics 
eliminates the It\^o drift by construction, making implementation straightforward and efficient. 
Second, we have adapted the predictor--corrector algorithm of~\cite[][]{somasi} to the Ermak--McCammon scheme, ensuring that its time-accuracy properties 
are preserved. A significant advantage of this combined approach is that
the It\^o character of the dynamics is maintained, and the \emph{maximum chain-length constraint} is enforced exactly.
This is crucial for polymer dynamics in turbulent flows, where material lines can stretch indefinitely but polymer 
chains cannot. The resulting method improves numerical stability and allows larger timesteps~\cite[][]{kivotides_stretch}.

It is important to note that, when the It\^o drift vanishes, the overdamped
Brownian dynamics take the schematic form
\[
    dr^k
    =
    \left[
    u(r^k)+u^S(r^k)
    \right]dt
    +
    {\rm thermal\ effects}.
\]
Since \(u^S\) is generally much smaller than \(u\), it is reasonable to expect
that polymer stretching is predominantly controlled by the turbulent carrier
flow. This does not, however, imply that polymer trajectories are statistically
equivalent to material-fluid trajectories.

Hydrodynamic interactions encoded in \(u^S\) can become important when a
polymer is stretched near its maximum length. A material line can continue to
stretch indefinitely, whereas a polymer develops a strong elastic force that
limits its extension and increases the bead velocities relative to the local
carrier flow. Brownian displacements and excluded-volume interactions provide
additional sources of relative bead motion. The polymer midpoint trajectory is
therefore generated jointly with the evolving polymer conformation by the full
bead--spring stochastic dynamics.

The relevant mathematical distinction is between the one-time spatial
distribution of a trajectory and its complete stochastic path law. Let
\(X_{\rm p}(t)\) and \(X_{\rm m}(t)\) denote polymer-midpoint and
material-fluid trajectories, respectively. Even if both have the same uniform
one-time spatial distribution,
\[
    {\mathbb P}\{X_{\rm p}(t)\in dx\}
    =
    {\mathbb P}\{X_{\rm m}(t)\in dx\}
    =
    \frac{dx}{|V|},
\]
their multi-time distributions need not coincide:
\[
    {\cal L}\!\left(
    X_{\rm p}(t_1),\ldots,X_{\rm p}(t_n)
    \right)
    \neq
    {\cal L}\!\left(
    X_{\rm m}(t_1),\ldots,X_{\rm m}(t_n)
    \right).
\]
Thus two trajectory processes may visit the same spatial regions with the same
long-time frequencies while possessing different transition probabilities,
residence times, temporal correlations, and ordered histories of sampled
velocity gradients. Uniform spatial sampling therefore does not imply
equivalence of polymer and material trajectories.

Motivated by this stochastic-process distinction, we use the full
bead--spring dynamics to determine jointly the polymer conformation and the
sequence of midpoint positions at which the turbulent velocity-gradient tensor
is sampled. The resulting accumulated stretching, finite-time Lyapunov
exponents, and alignment statistics depend on the complete ordered path rather
than only on the spatial distribution of the polymer midpoints. The purpose of
the present calculation is therefore to determine which properties of
material-line stretching are preserved along polymer-generated trajectories
and which are modified by elasticity, finite extensibility, excluded-volume
interactions, Brownian forcing, and hydrodynamic interactions.

In ultra-dilute turbulent solutions, these questions are expected to have a
degree of universality arising from the combination of universal small-scale
Navier--Stokes turbulence and coarse-grained polymer dynamics. This
universality statement should, however, be understood as applying only to the
high-\(Wi\), nonlinear stretching regime considered here. If \(Wi\) is reduced
towards the coil--stretch transition, or if mildly stretched polymers are
considered, the balance between turbulent stretching, Brownian fluctuations,
hydrodynamic interactions, and elastic relaxation changes qualitatively.

\subsubsection{A method for the computation of finite-time Lyapunov exponents in the tangent system}

\paragraph{Finite-precision difficulty.}

Equation~\eqref{defdyn} for the dynamics of the deformation-gradient tensor
\(F\) is solved with a third-order accurate Runge--Kutta (RK) method. Although stable
from the numerical-analysis viewpoint, large accumulated material
deformations lead to finite-precision arithmetic issues. In particular, the
Lyapunov exponents are computed via the logarithms of the singular values of
\(F\).
The latter can be computed via SVD but also much more easily via the
spectral decomposition of the \(U^2=F^TF\) matrix, since its eigenvalues
are the squares of the singular values of \(F\).

However, as more and more strain is accumulated in \(F\) for sufficiently large
times, its singular values \(\sigma_i\) can become numerically extreme under
finite-precision arithmetic. For example, very small \(\sigma_i\) values may
lead to roundoff errors that cause unphysical effects, such as slightly
negative eigenvalues of
\[
F^TF .
\]
We developed a method to avoid such issues by ensuring that only
well-behaved \(\sigma_i\) values appear in the computation, while preserving
all physical information carried by \(F\). The method is based on the SVD of
\(F\), Eq.~\eqref{eq:svd-decomposition}.

\paragraph{Geometry of the SVD.}

To understand how this works, we reflect on the geometry of the SVD. The
matrix \(F\) acts on a material reference vector \(f_R\) as follows: the matrix
\(V\) is a rotation matrix whose columns are the eigenvectors \(r_i\) of
\[
F^TF .
\]
The initial action
\[
V^T f_R
\]
rotates \(f_R\) and expresses its Cartesian components in the basis defined by
the \(r_i\). Next, \(\Sigma\), a diagonal matrix with entries \(\sigma_i\),
performs a pure stretch by scaling each of these components. Finally, the
matrix \(W\), whose columns are the eigenvectors of
\[
FF^T ,
\]
rotates the stretched vector from the \(r_i\) basis to the Cartesian
components of the spatial vector \(f\).

\paragraph{SVD normalisation.}

Accordingly, the SVD shows that shear is a coordinate-dependent effect, and
that any general deformation can be decomposed into a pure stretch and two
rotations. To avoid extreme singular values in \(\Sigma\), we periodically
replace \(F\) with
\begin{equation}
    F
    \leftarrow
    WV^T ,
    \label{eq:svd-reset-main}
\end{equation}
i.e. we remove \(\Sigma\) while storing the logarithms of the singular values
in three corresponding cumulative sums
\begin{equation}
    G_i
    =
    \sum_{k=1}^{n}\ln\sigma_i^{(k)},
    \label{eq:svd-stored-log-stretches}
\end{equation}
where the sum runs over the \(n\) normalisations of \(F\). In this way,
the new \(F\) remains well-conditioned, while the cumulative stretching up to
each normalisation is preserved. Since we retain \(W\) and \(V^T\), the
singular-vector geometry of \(F\) is also preserved.

\paragraph{Continuation after normalisation.}

At the next time step, \(F\) evolves again according to Eq.~\eqref{defdyn}.
The new \(F\) captures the incremental deformation generated by the current
velocity-gradient history, while the previously accumulated stretch is stored
in the variables \(G_i\). Importantly, stretching arises from the instantaneous
velocity gradient and does not depend on the accumulated magnitude of \(F\);
therefore the normalisation procedure introduces no physical inconsistency.

At any time \(t\), we can then compute the Lyapunov exponents using
\begin{equation}
    \lambda_i(t)
    =
    \frac{\ln\sigma_i(t)+G_i}{t},
    \label{eq:svd-normalised-lyapunov-main}
\end{equation}
where \(\sigma_i(t)\) are the current singular values of the normalised
\(F\). Moreover, because the singular-vector geometry of the SVD has been
preserved, the \(F\) matrix at time \(t\) can still be used to deduce
directional information, including the angles between deformed fibres.

\paragraph{Recovery of strain diagnostics.}

In this context, we have also developed formulae for the recovery, after
multiple \(F\) normalisations, of other important strain-related quantities.
They are based on the semigroup property of deformation gradients: for
times \(t_1<t_2<t_3\),
\begin{equation}
    F(t_1,t_3)
    =
    F(t_2,t_3)F(t_1,t_2),
    \label{eq:deformation-gradient-semigroup}
\end{equation}
and on an SVD-based reorientation/renormalisation at each normalisation step,
which aligns successive blocks so that principal-stretch factors are
multiplicative, i.e. log-stretches add. This additivity does not hold for
general, unaligned products of \(F\).

Accordingly, the pure strain measure \((E:E)^{1/2}\) is based on the
Green--St. Venant tensor in Eq.~\eqref{eq:green-st-venant-strain}, which
requires \(F^TF\). After simple matrix algebra, and using the SVD, we obtain
\begin{equation}
    F^TF
    =
    V\Sigma^T\Sigma V^T,
    \label{eq:svd-recovered-cauchy-green}
\end{equation}
where the true, unnormalised matrix \(\Sigma\) needs to be used. This is a
diagonal matrix with entries
\begin{equation}
    \sigma_i^{\rm true}
    =
    e^{G_i}\sigma_i .
    \label{eq:svd-true-singular-values}
\end{equation}
The true singular values also need to be used in place of \(\sigma_j\) in the
previously mentioned expression for the deformed fibres,
\begin{equation}
    U(Z_m)g_e
    =
    \left(
    \sum_{j=1}^{3}
    e^{G_j}\sigma_j r_j r_j^T
    \right)g_e .
    \label{eq:svd-recovered-tangent-fibre}
\end{equation}

\paragraph{Evaluation of the left singular vectors.}

Finally, note that whilst \(V\) is a matrix with columns the eigenvectors of
\(U^2\), hence available from the spectral decomposition of the latter, the
evaluation of \(W\) --- which is necessary to perform the replacement of \(F\)
by \(WV^T\) --- is not directly available. One can easily show that
\begin{equation}
    W
    =
    FV\Sigma^{-1},
    \label{eq:left-singular-vectors-from-f}
\end{equation}
where the entries of \(\Sigma^{-1}\) are equal to \(1/\sigma_i\). Notably,
\(\sigma_i\) are the singular values of the currently evolved \(F\), and we
need not take into account the singular values stored in \(G_i\) to evaluate
\(W\).

The SVD-normalised tangent-flow formulation developed here is not merely an
alternative way of accumulating scalar finite-time Lyapunov exponents.
QR-based methods also act on the deformation gradient and can compute
Lyapunov exponents and, with additional procedures, Lyapunov-vector
information. However, Lyapunov-vector information is not the same as the
finite-time singular-vector geometry of the deformation gradient. QR is based
on orthonormalisation and triangular growth factors, whereas SVD gives the
principal finite-time stretches and their associated singular directions.

The special feature of the present method is that this finite-time singular
geometry is not discarded after stabilisation. In SVD implementations used only
for exponent accumulation, the retained output is essentially scalar
stretching information. In the present formulation, by contrast, the
accumulated singular stretches are stored while the singular-vector geometry
is preserved. This allows singular values, Hencky-type strain measures, and
principal-stretch/alignment diagnostics to be reconstructed along
polymer-generated trajectories. Such information is essential for a physical
investigation of polymer stretching, not only for the numerical calculation of
Lyapunov exponents. The corresponding SVD, stored-stretch, reset and
normalised finite-time Lyapunov formulae are summarized in
Appendix~\ref{app:governing-system},
Eqs.~\eqref{eq:app-svd}--\eqref{eq:app-svd-normalised-lyapunov}.

The principal ingredients of the present procedure have been tested in our
earlier work. The evolution of the deformation-gradient tensor and the
extraction of the Lyapunov exponents were developed and validated by Kivotides
and Leonard in \cite[][]{kivotides_stringy}, while the SVD-based
calculation of the Lyapunov spectrum was implemented and tested by Kivotides
in \cite[][]{kivotides_lagrangian}. The periodic
normalisation introduced here is not an additional approximation, but an exact
algebraic reformulation of the same evolution. Its numerical implementation
preserves the continuity of the three Lyapunov-number trajectories across every
normalisation event, as shown in Figs.~16--18 (left), while the
incompressibility condition is satisfied at every time step with a typical
accuracy of \(10^{-6}\).

\subsection{Algorithmics}

As already mentioned, round-off errors play a key role in Lyapunov dynamics.
Within the employed finite-precision floating-point number set \(\mathbb{F}\),
the distance between \(1\) and the next larger floating-point number is
\(\epsilon_m=0.222 \times 10^{-15}\).
The smallest and largest numbers that can be represented are
\(2.2 \times 10^{-308}\) and \(1.8 \times 10^{308}\), respectively.
The algorithm's arithmetic employs the round-to-nearest-even rounding method.

In the Langevin solver, a key bottleneck is the computation of the
square root of the diffusion matrix for use in the evaluation of the
Wiener process. The standard Cholesky procedure has \(O(N^3)\) complexity,
so we have employed the method of Chebyshev-polynomial approximation, which
has \(O(N^{2.25})\) complexity \cite[][]{kivotides_stretch}.
To achieve convergence with error tolerance \(10^{-6}\), the calculations
used a minimum Chebyshev order of \(6\) and allowed increments by \(6\) up
to a maximum order of \(400\).
The method requires estimates of the extremal eigenvalues of the diffusion
matrix, and we used for this purpose the well-known Arnoldi algorithm of
\(O(N^2)\) complexity, allowing up to \(400\) iterations to achieve
convergence with error tolerance \(10^{-6}\). There were no convergence
issues under these specifications.

We have nondimensionalised the various physical variables with combinations
of the characteristic scales \([l^*,t^*,m^*]\) for length, time and mass,
respectively. We have
\begin{equation}
    l^*=b,\qquad
    t^*=\frac{6 \pi \mu r_b b^2}{k_B T},
    \qquad
    m^*=\frac{(6 \pi \mu r_b b)^2}{k_B T},
    \label{eq:nondimensional-scales}
\end{equation}
where \(r_b\) is the bead radius \cite[][]{kivotides_knot}.
The corresponding cgs values of these scales are
\(l^*=7.370 \times 10^{-7}\ {\rm cm}\),
\(t^*=1.051 \times 10^{-4}\ {\rm s}\), and
\(m^*=8.233 \times 10^{-10}\ {\rm g}\).
With this nondimensionalisation, the main DNS, turbulence and polymer
parameters used in the calculation are collected in
Table~\ref{tab:calculation-parameters}. The table also gives the Kolmogorov
scales, the large-eddy turnover time, the polymer relaxation time, and the
length- and time-scale ratios used to identify the computed regime.
\begin{table}[!htbp]
\centering
\caption{
Computational parameters and characteristic scale ratios. \(N\) is the number
of grid points per spatial direction, \(\Delta x\) and \(\Delta t\) are the
spatial and temporal discretisations, \(\nu\) is the kinematic viscosity, and
\(\epsilon\) is the mean energy-dissipation rate. The Kolmogorov length and
time scales are \(\eta=(\nu^3/\epsilon)^{1/4}\) and
\(\tau_\eta=(\nu/\epsilon)^{1/2}\). The quantities \(u'\), \(\lambda\),
\(\tau_e=l_e/u'\), and \(Re_\lambda=u'\lambda/\nu\) are the turbulence
intensity, Taylor microscale, large-eddy turnover time and Taylor-scale
Reynolds number. \(N_c\) is the number of polymer chains, \(N_b\) is the
number of beads per chain, \(L_{\max}\) is the maximum polymer length,
\(\tau_R\) is the longest polymer relaxation time, \(\tau_s\) is the spring
relaxation time, \(\tau_m\) is the monomer relaxation time, and
\(Wi=\tau_R/\tau_\eta\) is the Weissenberg number.
}
\label{tab:calculation-parameters}
\begin{tabular}{lll}
\hline
Quantity & Definition & Value \\
\hline
Grid resolution & \(N^3\) & \(256^3\) \\
Grid spacing & \(\Delta x\) & \(33921.302\) \\
Time step & \(\Delta t\) & \(1.610\) \\
Kinematic viscosity & \(\nu\) & \(1668605.211\) \\
Mean dissipation rate & \(\epsilon\) & \(44.096\) \\
Kolmogorov length scale & \(\eta=(\nu^3/\epsilon)^{1/4}\) & \(18016.304\) \\
Kolmogorov time scale & \(\tau_\eta=(\nu/\epsilon)^{1/2}\) & \(194.526\) \\
Turbulence intensity & \(u'\) & \(504.312\) \\
Taylor microscale & \(\lambda\) & \(271311.464\) \\
Taylor Reynolds number & \(Re_\lambda=u'\lambda/\nu\) & \(82\) \\
Large-eddy turnover time & \(\tau_e=l_e/u'\) & \(8609.604\) \\
Number of chains & \(N_c\) & \(301\) \\
Beads per chain & \(N_b\) & \(5\) \\
Maximum polymer length & \(L_{\max}\) & \(16575.000\) \\
Longest polymer Zimm relaxation time & \(\tau_R\) & \(15562.086\) \\
Spring Zimm relaxation time & \(\tau_s\) & \(1349.212\) \\
Monomer relaxation time & \(\tau_m\) & \(9.409 \times 10^{-4}\) \\
Weissenberg number & \(Wi=\tau_R/\tau_\eta\) & \(80.00\) \\
Polymer/Kolmogorov length ratio & \(L_{\max}/\eta\) & \(0.92\) \\
Polymer/grid-spacing ratio & \(L_{\max}/\Delta x\) & \(0.49\) \\
Polymer/Kolmogorov time ratio & \(\tau_R/\tau_\eta\) & \(80.00\) \\
\hline
\end{tabular}
\end{table}
\paragraph{Carrier-flow calculation.}
The carrier-flow boundary conditions are periodic. 
Descriptions of our implementation of the linear forcing used to maintain
statistically stationary turbulence are available in several publications
\cite[][]{kivotides_cloud,kivotides_sftur,kivotides_gravity}, which should
be consulted for further details.
The Navier--Stokes equations are solved on a
\(256^3\) grid, and the resulting Taylor Reynolds number is
\(Re_\lambda=82\). The spatial resolution is
\[
    \frac{\Delta x}{\eta}=1.88.
\]
Using the scalar-wavenumber convention
\[
    k=\frac{i}{l_b},
\]
where \(i\) is the Fourier-mode index and \(l_b\) is the computational-box
size, the largest resolved wavenumber is
\[
    k_{\max}=\frac{N}{2l_b}=\frac{1}{2\Delta x}.
\]
Consequently,
\[
    k_{\max}\eta=\frac{\eta}{2\Delta x}=0.266.
\]
This spatial resolution is comparable with that employed in established
physical-space DNS calculations
\cite[][]{vandine_dns,watanabe_grid}.
Consistent with this resolution,
Fig.~\ref{fig:energy-spectrum} shows that the dimensionless carrier-flow
kinetic-energy spectrum exhibits an interval compatible with the
\(k^{-5/3}\) scaling, followed by a well resolved dissipative range. The
smooth spectral decay at the highest wavenumbers confirms that the smallest
dynamically active scales of the carrier flow are adequately resolved. This
is particularly important for the present study because the polymers remain
at sub-Kolmogorov scales.

The computational time step satisfies
\[
    \frac{\Delta t}{\tau_\eta}=8.28\times10^{-3},
\]
corresponding to approximately \(121\) time steps per Kolmogorov time.
As described below, the time step is selected
according to the considerably shorter polymer time scales and is
much smaller than that required to resolve the viscous processes in the fluid.
The maximum CFL number is \(0.75\). The carrier-flow velocity at the bead
positions is evaluated by trilinear interpolation, which is linear in each
spatial coordinate. The velocity-gradient tensor at the polymer end-to-end
midpoint is evaluated using second-order central finite differences of the
staggered-grid variables in the cells surrounding the polymer.
\begin{figure*}
\includegraphics[width=0.49\linewidth]{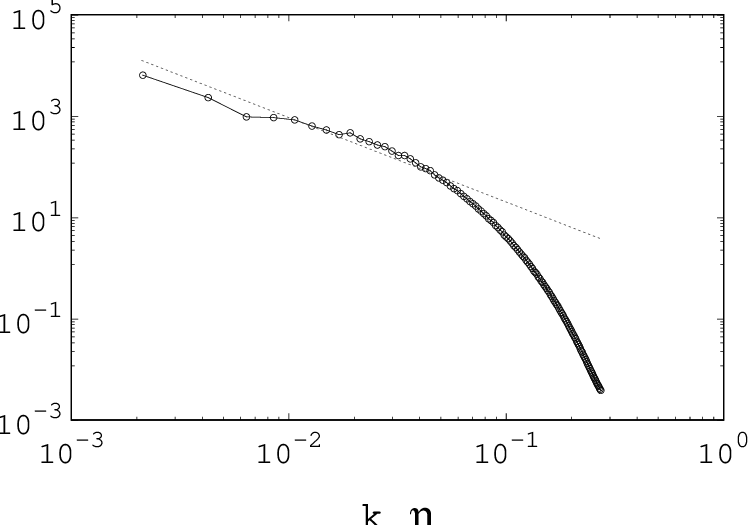}
\caption{\label{fig:energy-spectrum}
Dimensionless carrier-flow kinetic-energy spectrum. The horizontal
axis shows \(k\eta\), where \(k=i/L\) is the scalar-wavenumber convention
used here, and the vertical axis shows
\(E(k)/(\epsilon\nu^5)^{1/4}\). The dotted line indicates the
\((k\eta)^{-5/3}\) scaling for reference.}
\end{figure*}

\paragraph{Polymer calculation and statistical sampling.}
The polymer boundary conditions are periodic. The maximum chain length is
\(0.92\eta\) and \(0.49\Delta x\). The calculation uses \(N_c=301\) chains
with \(N_b=5\) beads per chain. These values were chosen to obtain converged
polymer statistics with minimal computational complexity, while retaining the
complete mesoscopic bead--spring force balance described above and summarized
in Appendix~A.

The polymer time step is one hundred times smaller than the spring relaxation
time. This choice is not imposed by numerical-stability requirements, since
the spring forces are treated implicitly. Rather, the small time step is used
to improve temporal accuracy because the Ermak--McCammon scheme employed for
the stochastic differential equations is first-order accurate in time.

The reported probabilities and statistics are based on a sample size of
\(10^7\). Sampling started after the turbulence had reached a statistically
stationary state. Unless stated otherwise, statistics reported at time \(t\)
are computed by combining an ensemble average over the \(301\) chains with a
time average over the preceding sliding window
\begin{equation}
    [t-20\tau_e,t],
\end{equation}
where
\begin{equation}
    \tau_e=\frac{l_e}{u'}
\end{equation}
is the large-eddy turnover time, \(u'\) is the turbulence intensity, and
\(l_e\) is half the system size. Thus, all reported statistics use the same
averaging-window duration, while the sampled interval shifts with the reporting
time. For times \(t<20\tau_e\), the available averaging interval is
correspondingly shorter, \([0,t]\), and the associated sample size is smaller.

The calculation is designed to isolate the sub-Kolmogorov, high-\(Wi\)
polymer-stretching regime in homogeneous isotropic Navier--Stokes
turbulence. The Taylor Reynolds number \(Re_{\lambda}=82\) provides a
turbulent flow with the characteristic small-scale strain geometry and
intermittency of Navier--Stokes turbulence. Although the finite value of
\(Re_{\lambda}\) limits the extent of the inertial range, this is not the
controlling scale range for the present polymers: their maximum length
remains smaller than the Kolmogorov scale, so they sample the locally smooth
velocity-gradient field rather than inertial-range velocity differences.

At the same time, the value \(Wi\simeq 80\) places the chains well beyond
the coil--stretch transition, in a strongly stretched nonlinear elastic
regime. Moreover, the polymer model has a complementary universality
property. Its renormalisation-group-like coarse graining preserves the
mesoscopic bead--spring force balance under changes of bead resolution; in
particular, Table~II of Ref.~\cite[][]{kivotides_dense} reports
bead-resolution comparisons showing that changing \(N_b\) does not require
new fitted physical constants and does not alter the predicted dynamics in
any essential way. Thus, the computed statistics characterise the interaction
between two universal structures: the small-scale strain field of
Navier--Stokes turbulence and the coarse-grained nonlinear dynamics of
bead--spring polymers in the high-\(Wi\), sub-Kolmogorov regime.

\subsubsection{Algorithmic procedure}

Based on the above formulation, the algorithmic steps are as follows:
{\renewcommand{\labelenumi}{(\arabic{enumi})}
\begin{enumerate}
    \item Specify initial conditions for the turbulent velocity field \(u\),
    the bead positions \(r^k\), and the deformation-gradient tensor \(F\) of
    each chain.

    \item Increment time by the computational time step.

    \item Update the turbulent velocity field \(u\) with the Navier--Stokes
    solver, Eq.~\eqref{eq:ns-lundgren}.

    \item Using the current bead positions \(r^k\), compute the elastic and
    excluded-volume forces.

    \item Insert the above forces in the overdamped Langevin force balance,
    Eq.~\eqref{eq:langevin-force-balance}, to evaluate the hydrodynamic drag
    forces at \(r^k\).

     \item Using the RPY diffusion matrix
    (Eqs.~\eqref{eq:app-bead-separation}--\eqref{eq:app-rpy-near})
    and the hydrodynamic drag forces
    (Eq.~\eqref{eq:app-viscous-drag}), calculate the Stokes velocity
    \(u^{\mathcal S}(r^k)\)
    (Eq.~\eqref{eq:app-stokes-relative-velocity}).

    \item Use the square root of the RPY diffusion matrix to calculate the
    Wiener process.

    \item Employ \(u(r^k)\), \(u^{\mathcal S}(r^k)\), and the Wiener process
    to update the bead positions \(r^k\).

    \item Employing the updated bead positions \(r^k\), evaluate the velocity
    gradient tensor \(L\) at the midpoint of the end-to-end distance of each
    chain, and update \(F\) according to Eq.~\eqref{defdyn}.

    \item On prescribed time steps, perform the singular value decomposition
    of the deformation-gradient matrices and normalise \(F\) according to
    Eq.~\eqref{eq:svd-reset-main}, while storing the accumulated logarithmic
    stretches as in Eq.~\eqref{eq:svd-stored-log-stretches}.

    \item Repeat step (2) until the final time is reached.
\end{enumerate}
}

\section{Results}

We organise the results in two parts. First, we analyse the full nonlinear
bead--spring dynamics, including chain extension, sampled strain states,
alignment statistics, and correlations with turbulent invariants. These results
show how actual polymer stretching differs from material-line, Eulerian
vorticity, and vortex-filament reference statistics. Second, we analyse a mixed
linear material-stretching problem: the initial polymer end-to-end vector is
treated as an infinitesimal material element and evolved by the linearized
carrier-flow dynamics, but with the velocity-gradient tensor sampled along the
actual polymer trajectory. This separates nonlinear polymer deformation from
polymer-conditioned material-element stretching.

\subsection{Polymer stretching in the nonlinear system}

\begin{figure*}
\begin{tabular}{cc}
\hspace{-3mm}\includegraphics[width=0.49\linewidth]{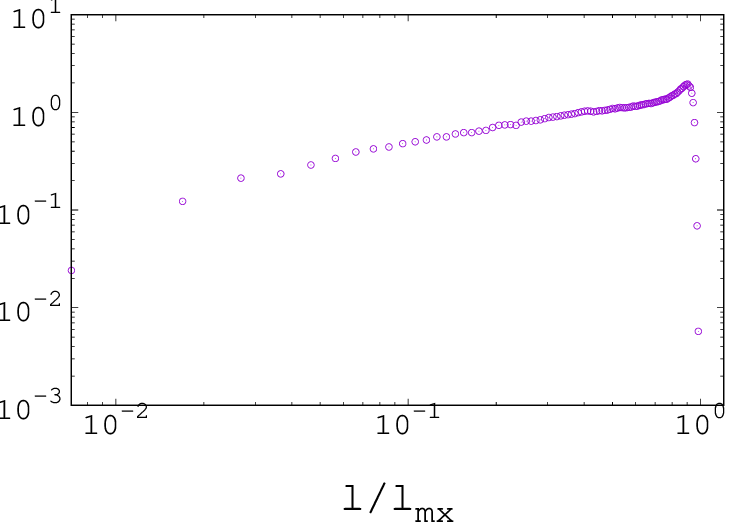}&
\hspace{2mm}\includegraphics[width=0.49\linewidth]{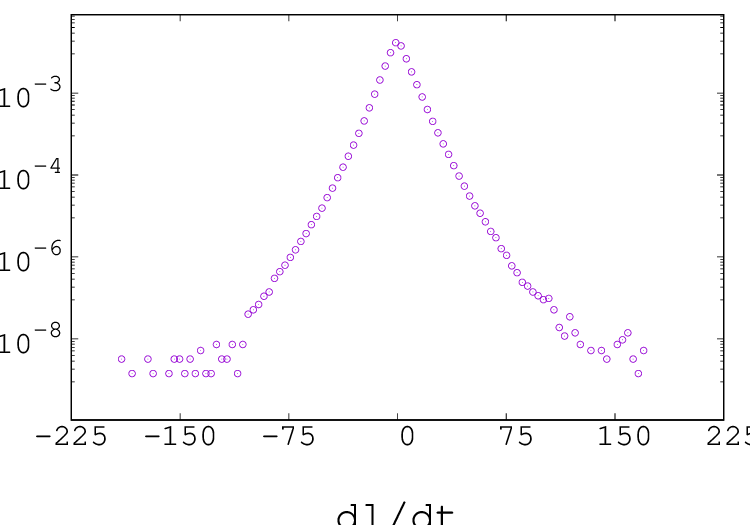}\\
\end{tabular}
\caption{\label{lengths}
(Left) PDF of the polymer end-to-end distance, normalized by the fixed maximum
chain length \(l_{\max}\). A log--log fit of
\(P(x)=C x^\alpha\), with \(x=l/l_{\max}\), over the interval
\(4.5\times10^{-2}\leq x\leq0.8\), gives
\(\alpha=0.506\pm0.005\), with rms logarithmic residual \(0.03279\).
A constrained fit with \(\alpha=1/2\) gives essentially the same rms
logarithmic residual, \(0.03282\). The data are therefore consistent with the
\(x^{1/2}\) reference scaling over this interval. 
(Right) PDF of polymer end-to-end length
derivatives.
}
\end{figure*}
\begin{figure*}
\begin{tabular}{cc}
\hspace{-3mm}\includegraphics[width=0.49\linewidth]{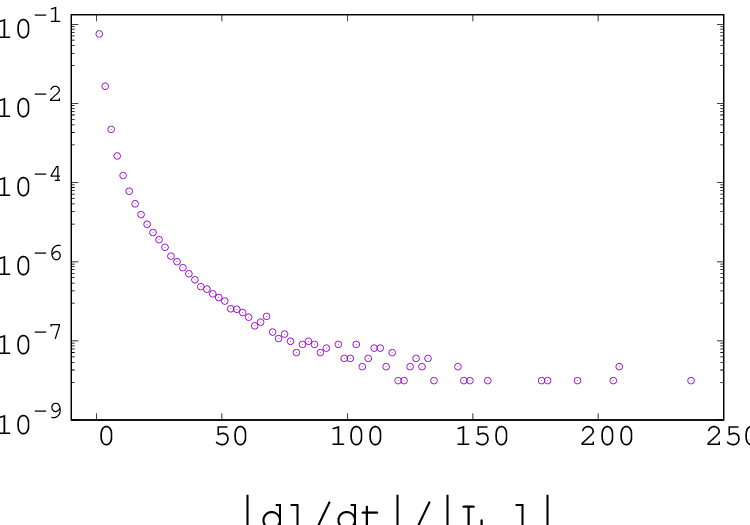}&
\hspace{2mm}\includegraphics[width=0.49\linewidth]{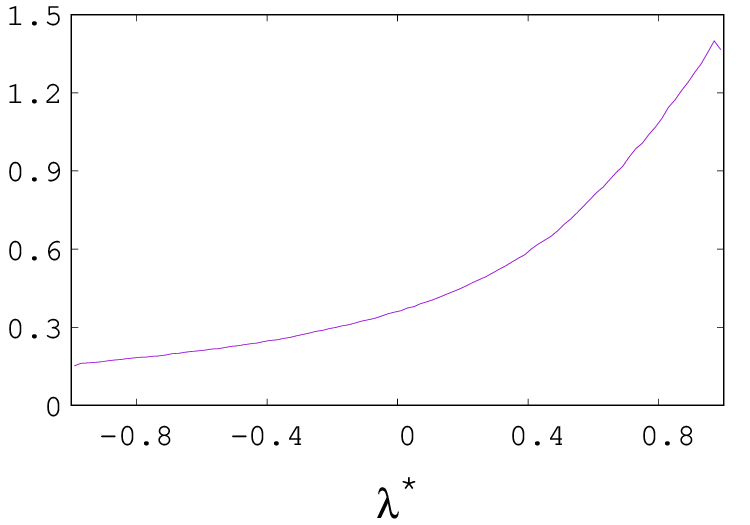}\\
\end{tabular}
\caption{\label{stretchchar}
(Left) PDF of \(R\), the ratio of the magnitude of the actual polymer
end-to-end length change to the magnitude of the corresponding length change
produced by material deformation. A value \(R=1\) denotes exact agreement
between the two stretching measures. The histogram bin containing \(R=1\),
centred at \(R=1.074\), contains \(91.85\%\) of the total probability; over
the complete histogram range,
\(\langle R\rangle=1.319\) and
\({\rm Var}(R)=1.073\). Here \(l\) denotes the polymer end-to-end length and
\(L\) denotes the sampled velocity-gradient tensor, defined in
Eq.~\eqref{eq:app-chain-sampled-gradient}. The plotted quantity is
dimensionless because its numerator and denominator have the same units.
(Right) PDF of the flow-type parameter \(\lambda^*\), defined in
Eq.~\eqref{eq:kerr-lund-rogers-lambda-star}.
}
\end{figure*}
\begin{figure*}
\begin{tabular}{cc}
\hspace{-3mm}\includegraphics[width=0.49\linewidth]{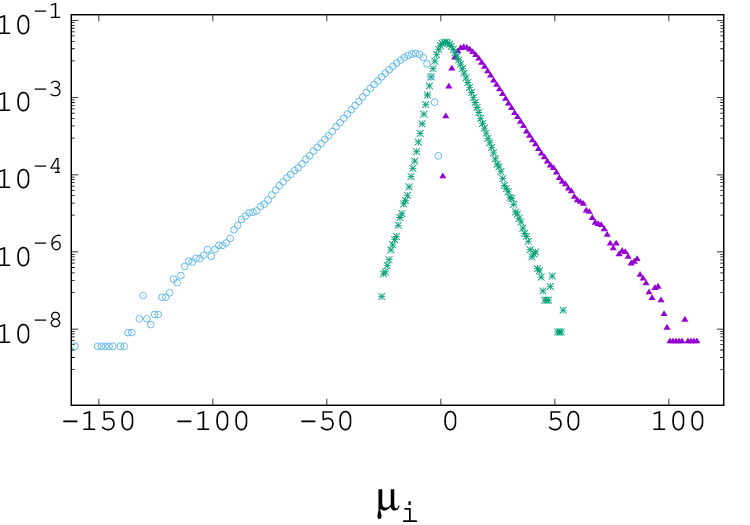}&
\hspace{2mm}\includegraphics[width=0.49\linewidth]{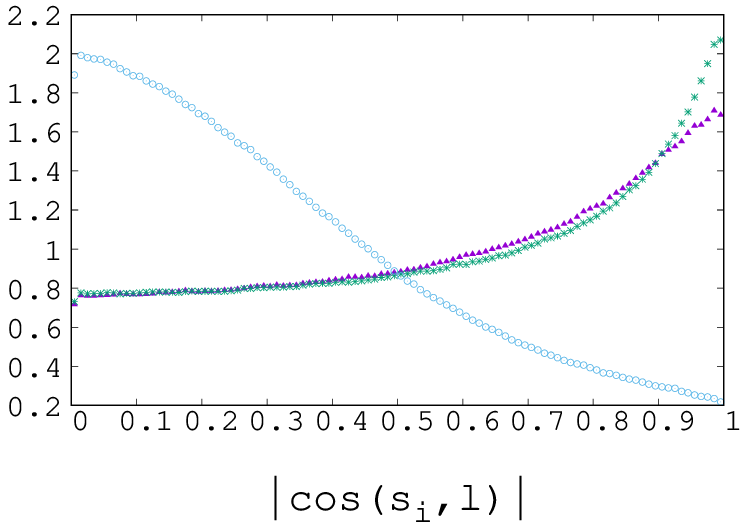}\\
\end{tabular}
\caption{\label{eigenvals}
        (Left) PDFs of eigenvalues of strain rate tensor along polymer trajectories.
	Triangles, stars and circles indicate $\mu_1$, $\mu_2$ and $\mu_3$ values respectively.
	(Right) PDF of cosines $cos(s_i,l)$ between strain rate eigenvectors $s_i$ and chain
	end-to-end vector $l$. Symbols convention as with $\mu_i$.
}
\end{figure*}
\begin{figure*}
\begin{tabular}{cc}
\hspace{-3mm}\includegraphics[width=0.49\linewidth]{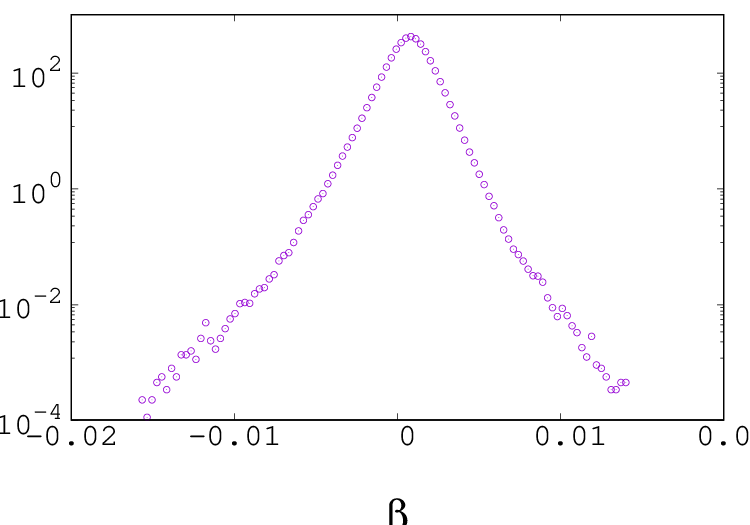}&
	\hspace{2mm}\includegraphics[width=0.49\linewidth]{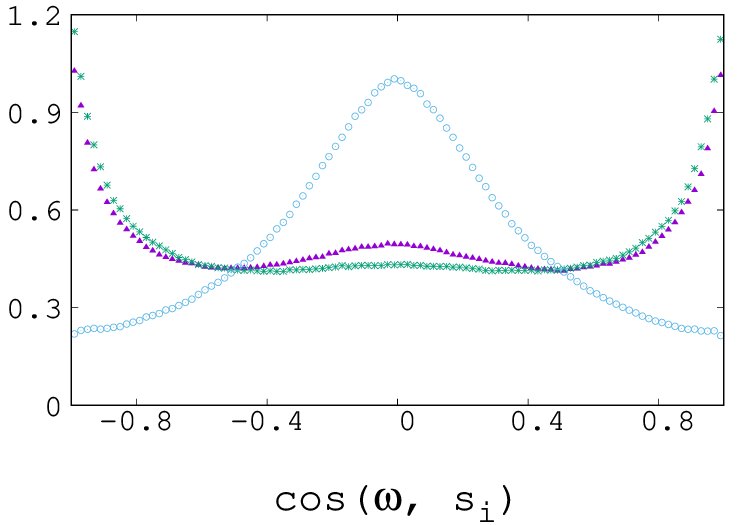}\\
\end{tabular}
\caption{\label{growths}
(Left) PDF of the instantaneous material stretching rate \(\beta\) of the
polymer end-to-end distance. Here \(\beta\) is defined in
Eq.~\eqref{eq:instantaneous-material-line-stretching-rate}. In dimensional
variables \(\beta\) has units of inverse time; the plotted values are
nondimensional, according to the scales introduced in the algorithmic section.
(Right) PDF of the cosines \(\cos(\omega,s_i)\) between the vorticity vector
\(\omega\) and the strain-rate eigenvectors \(s_i\). Triangles, stars and
circles indicate alignments with \(s_1\), \(s_2\) and \(s_3\), respectively.
Although the signs of the strain-rate eigenvectors, and hence of the
corresponding signed alignments, have no physical meaning, we retain the signed
cosines here to display the symmetry about zero and confirm the absence of a
spurious sign bias in the sampling.
}
\end{figure*}

All stochastic results presented here are conditioned, since they were
measured along polymer trajectories. They therefore describe the flow
structures and polymer configurations sampled by the chains, rather than
Eulerian statistics of the turbulent field. The nonlinear results are
obtained from the full bead--spring dynamics, including viscous drag,
Brownian forcing, elasticity, excluded volume and hydrodynamic interactions.
Our purpose is to understand the flow encountered along turbulent polymer
trajectories, the mechanisms that keep chains strongly out of equilibrium,
and the statistics of the stretched polymer states themselves in
ultra-dilute turbulent polymer solutions with strongly nonlinear chain
elasticity.

\paragraph{Chain extension and nonlinear stretching.}

The PDF of the end-to-end distance, normalised by the maximum chain length
(Fig.~\ref{lengths}, left), exhibits a broad power-law-like scaling range,
approximately consistent with an \(l^{1/2}\) behaviour, over
\(3\times 10^{-2}\lesssim l/l_{mx}\lesssim 0.9\).
The important point is the existence of this extended scaling regime: the
polymers do not fluctuate around a single typical extension, but sample a
broad hierarchy of conformations, from weakly stretched states to states close
to the finite-extensibility limit. These results are consistent with the
\(Wi=50\) Navier--Stokes calculations of \cite[][]{watanabe1}, although here
the observed scaling law extends over a wider range. This agreement suggests
that the broad extension distribution is a robust feature of polymer
stretching in Navier--Stokes turbulence in the sub-Kolmogorov regime.

By contrast, the present PDF is qualitatively different from the high-\(Wi\)
results of \cite[][]{kivotides_random}. That comparison is useful because it
marks the distinction between two polymer-stretching settings: the present
case, where sub-Kolmogorov polymers are stretched mainly by the locally smooth
Navier--Stokes velocity-gradient field, and the synthetic random-phase case,
where stretching was dominated by \emph{Gaussian} inertial-range turbulent
motions. Thus the extension statistics are not merely a function of \(Wi\),
but reflect the statistical structure of the turbulent stretching process
sampled by the polymer and the relation between polymer length and the length
scale of the stretching motions.

The PDF of the time derivative of the end-to-end distance
(Fig.~\ref{lengths}, right) is nearly symmetric about zero, reflecting the
strongly oscillatory behaviour of chain lengths. Thus, highly stretched
states are not produced by monotonic extension, but by repeated stretching
and relaxation events along polymer trajectories. 
By comparing the magnitude of the polymer length change with that of the
corresponding material deformation (Fig.~\ref{stretchchar}, left), we confirm
that polymers predominantly stretch like material-line elements. Denoting this
ratio by \(R\), the value \(R=1\) represents exact agreement between the actual
polymer length change and the material-deformation estimate. The histogram bin
containing \(R=1\), centred at \(R=1.074\), contains \(91.85\%\) of the total
probability, whereas only \(8.15\%\) is distributed over the remaining
large-\(R\) tail. Over the complete histogram range,
\[
    \langle R\rangle=1.319,
    \qquad
    {\rm Var}(R)=1.073.
\]
Material-line-like stretching is therefore overwhelmingly the predominant
behaviour. Finite deviations nevertheless occur because of elasticity and
excluded-volume forces and have measurable probability. These deviations
quantify the part of the nonlinear polymer dynamics that cannot be reduced to
material-line stretching by the carrier flow.

\paragraph{Rate-of-strain state sampled by polymer trajectories.}

The PDF of \(\lambda^*\) (Fig.~\ref{stretchchar}, right) demonstrates
that polymers most frequently reside in regions with predominant
axisymmetric biaxial extension. Remarkably, our result is strikingly similar
to that of \cite[][]{lund}, where the sampling was performed over the whole
flow domain. This comparison shows that, at the level of the
rate-of-strain eigenvalue state, polymer trajectories do not preferentially
select a class of strain-rate configurations different from those already
dominant in homogeneous isotropic turbulence. In other words, the frequent
sampling of axisymmetric biaxial extension reflects a robust feature of the
Navier--Stokes strain field rather than a special bias of the polymer
trajectories.

This observation does not by itself determine polymer stretching, because
\(\lambda^*\) classifies the eigenvalue state of the strain-rate tensor but
does not contain information about polymer orientation relative to the
corresponding eigenvectors. Polymer stretching in steady axisymmetric
biaxial extensional flows at very high \(Wi\) numbers was carefully studied
in \cite[][]{kivotides_stretch}, which can be consulted for detailed insight
into the particular stretching processes.

\paragraph{Strain eigenvalues and polymer--strain alignment.}

We now turn to results that characterize the flow structure encountered by the
polymers and the geometry of polymer--flow interactions. The strain-rate
eigenvalues sampled along polymer trajectories (Fig.~\ref{eigenvals}, left)
closely resemble the Eulerian Navier--Stokes statistics, exhibiting the
characteristic skewness of the intermediate eigenvalue towards positive
values and the slower decay of the largest eigenvalue PDF compared with that
of the smallest one \cite[][]{arcady}. These features are also present in the
Lagrangian statistics along material particle paths in
\cite[][]{kivotides_stringy}. Thus, at the level of strain-rate eigenvalue
statistics, polymer trajectories sample the familiar structure of
Navier--Stokes turbulence rather than a qualitatively different strain-rate
environment.

The PDFs of the cosines \(\cos(s_i,l)\) between the strain-rate eigenvectors
\(s_i\) and the polymer end-to-end vector \(l\) (Fig.~\ref{eigenvals}, right)
differ markedly from analogous results for material lines
\cite[][]{kivotides_stringy}, Eulerian vorticity \cite[][]{ashurst,arcady},
and Lagrangian vortex filaments \cite[][]{kivotides_lagrangian}. None of
these cases exhibit a tendency for simultaneous alignment of the
corresponding vector with both \(s_1\) and \(s_2\). For example, in the
material-line case --- which provides the most natural point of comparison
--- the intermediate eigenvector \(s_2\) shows no particular correlation with
the line direction \cite[][]{kivotides_stringy}.

By contrast, our polymers display preferential orientation of \(l\) in the
\(s_1\)-\(s_2\) plane, with stronger alignment toward \(s_2\) than toward
\(s_1\), and with \(l\) tending to be nearly perpendicular to \(s_3\).
The results
show that \(s_2\) aligns better with the polymer end-to-end vector than
\(s_1\). Interestingly, this pattern closely matches the result reported by
Koide and Goto \cite[][]{koide} at \(Wi=20\) and \(Re_{\lambda}=220\) for
dumbbells without hydrodynamic interactions. That our calculations, performed
at \(Wi=81\) and \(Re_{\lambda}=82\) with full multi-bead chains including
excluded-volume and hydrodynamic interactions, recover the same behavior
strongly suggests that these alignment tendencies are generic and robust
features of polymer--turbulence interaction within the nondimensional
parameter range considered here.

By combining the \(\lambda^*\), \(\mu_i\), and alignment statistics, we
obtain a more precise picture of the stretching mechanism. The computed PDF
of \(\mu_2\) along polymer trajectories (Fig.~\ref{eigenvals}, left) shows
that the intermediate eigenvalue takes both positive and negative values, but
is biased towards positive values. The PDF of \(\lambda^*\)
(Fig.~\ref{stretchchar}, right) further shows that polymers most frequently
sample biaxial extensional strain states. Thus, along polymer trajectories,
the intermediate eigendirection is often part of the extensional subspace of
the rate-of-strain tensor.

The alignment statistics provide the essential additional information. The
polymer end-to-end vector \(l\) aligns strongly with both \(s_1\) and
\(s_2\), but the alignment with \(s_2\) is stronger than that with \(s_1\).
Since the material stretching rate associated with the
end-to-end direction contains the contributions
\[
\frac{l\cdot D l}{|l|^2}
=
\sum_{j=1}^3 \mu_j \cos^2(s_j,l),
\]
polymer stretching is not determined by the largest eigenvalue alone. The
\(s_1\) direction carries the strongest extensional rate, but the polymer
orientation gives exceptional weight to the \(s_2\) direction. 
Consequently, positive-\(\mu_2\) events provide a secondary extensional
channel that is geometrically weighted by the strong \(s_2\)-alignment.
Conversely, the strong alignment with \(s_2\) does not imply an analogous
compressive role for the intermediate direction. Compression through \(s_2\)
would require \(\mu_2<0\), but such events correspond to the uniaxial side
of the strain statistics and are comparatively rare. Moreover, when
\(\mu_2<0\), its magnitude is typically small compared with \(|\mu_3|\).
Thus the intermediate direction can provide a significant secondary
stretching channel when \(\mu_2>0\), but it is not expected to provide a
comparable compression channel. 
Compression is associated primarily with the eigendirection \(s_3\), which
corresponds to the negative eigenvalue \(\mu_3\), but its contribution to the
end-to-end stretching rate is weighted by the polymer orientation through
\(\cos^2(s_3,l)\).

\paragraph{Material-segment stretching and vorticity--strain alignment.}

The PDF of the instantaneous stretching rate \(\beta\) that the polymer
end-to-end segment would have if it were a material line element of the
carrier flow (Fig.~\ref{growths}, left) peaks at positive \(\beta\),
consistent with the behaviour of material lines in turbulence
\cite[][]{kivotides_stringy}. In contrast, the PDF of the actual polymer
end-to-end length derivative (Fig.~\ref{lengths}, right) is nearly symmetric
with its maximum at zero. Thus, even when evaluated at the same polymer
positions, the material-line stretching tendency and the actual nonlinear
polymer stretching rate are statistically different. This comparison gives a
direct measure of the extent to which polymer elasticity, excluded volume,
Brownian forcing, and hydrodynamic interactions modify the stretching that
would be inferred from the carrier-flow velocity gradient alone.

The distinction between polymer-conditioned sampling and standard turbulent
sampling is not limited to polymer orientation; it also appears in the
vorticity--strain geometry of the flow encountered by the chains. The
polymer-conditioned character of the sampled trajectories becomes even
clearer in the alignments between the vorticity vector
\({\omega}\) and the strain-rate eigenvectors \(s_i\)
(Fig.~\ref{growths}, right). Along polymer paths, vorticity tends to align
with both \(s_1\) and \(s_2\), a behaviour that differs from both
corresponding Eulerian statistics \cite[][]{ashurst,jimenez,arcady} and
Lagrangian vortex-stretching phenomenology \cite[][]{kivotides_lagrangian}.
In the Eulerian case, there is only weak correlation between
\({\omega}\) and \(s_1\), while the alignment with \(s_2\) is
dominant. In contrast, for Lagrangian vortex-stretching of vortex filaments,
\({\omega}\) tends to align with \(s_1\) and shows only weak
correlation with \(s_2\). The present polymer-conditioned statistics
therefore define a distinct sampling regime: along polymer trajectories,
vorticity has significant alignment with both \(s_1\) and \(s_2\), 
while the Eulerian tendency for \({\omega}\) to be nearly
orthogonal to \(s_3\) is preserved.

\begin{figure*}
\begin{tabular}{cc}
	\hspace{2mm}\includegraphics[width=0.49\linewidth]{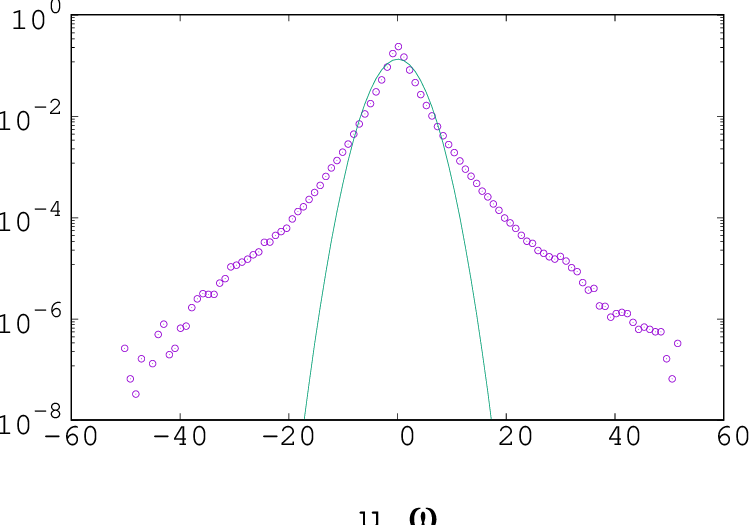}&
\hspace{2mm}\includegraphics[width=0.49\linewidth]{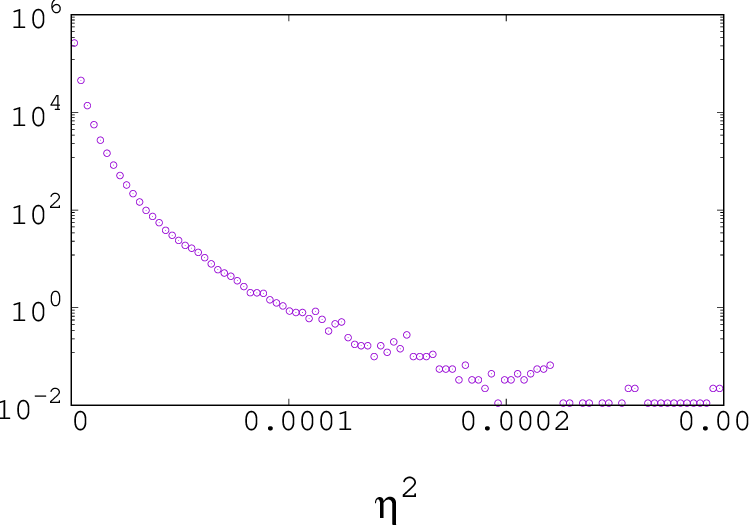}\\
\end{tabular}
\caption{\label{aligns}
        (Left) PDF of helicity $h=u \cdot \omega$.
        (Right) PDF of the square of the rate of change of vorticity direction $\eta^2$.
}
\end{figure*}
\begin{figure*}
\begin{tabular}{cc}
\hspace{-3mm}\includegraphics[width=0.49\linewidth]{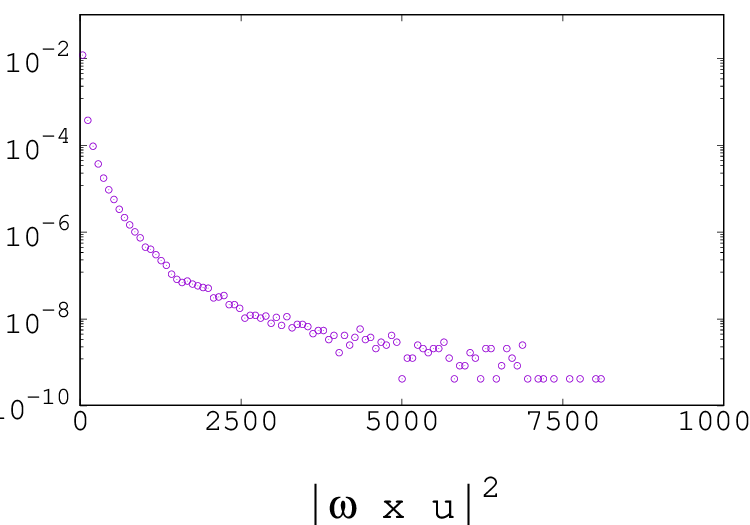}&
\hspace{3mm}\includegraphics[width=0.49\linewidth]{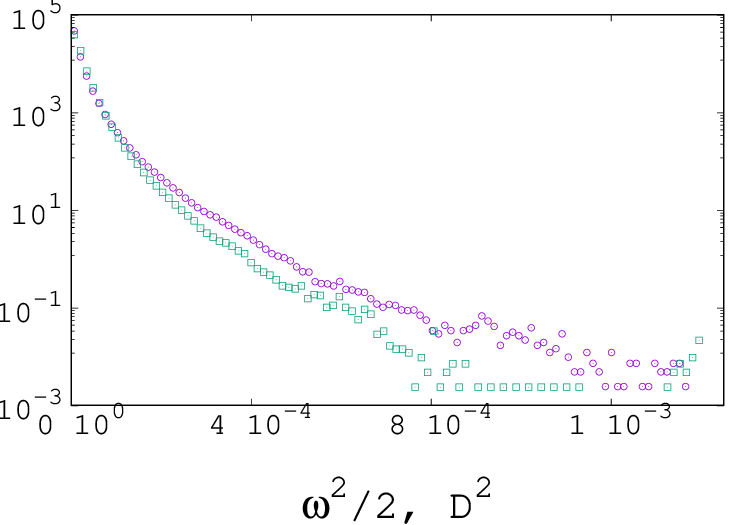}\\
\end{tabular}
\caption{\label{lamb}
        (Left) PDF of the square of the Lamb vector $\omega \times u$.
	(Right) PDFs of enstrophy $\omega^2/2$ (top curve, circles) and total strain $D^2$ (bottom curve, squares). 
}
\end{figure*}
\begin{figure*}
\begin{tabular}{cc}
\hspace{-3mm}\includegraphics[width=0.49\linewidth]{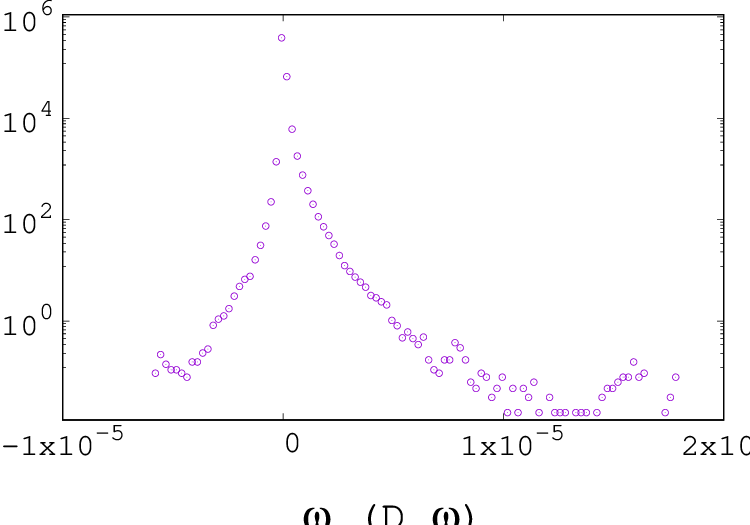}&
	\hspace{3mm}\includegraphics[width=0.49\linewidth]{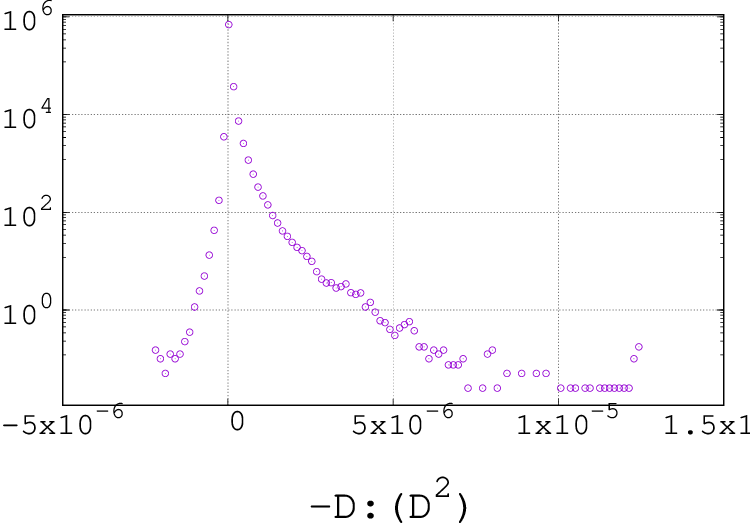}\\
\end{tabular}
\caption{\label{enstrampl}
	(Left) PDF of enstrophy amplification $\omega \cdot (D \omega)$.
        (Right) PDF of the total strain amplification $-D:(D^2)$.
}
\end{figure*}
\begin{figure*}
\begin{tabular}{cc}
	\hspace{-3mm}\includegraphics[width=0.49\linewidth]{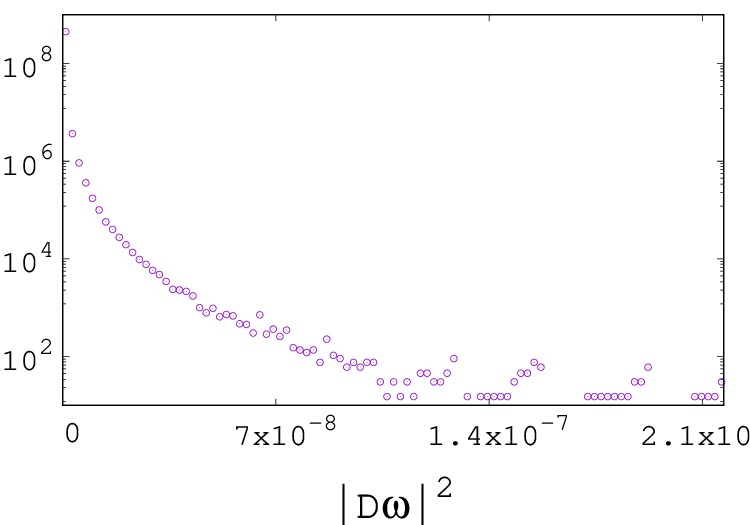}&
\hspace{5mm}\includegraphics[width=0.49\linewidth]{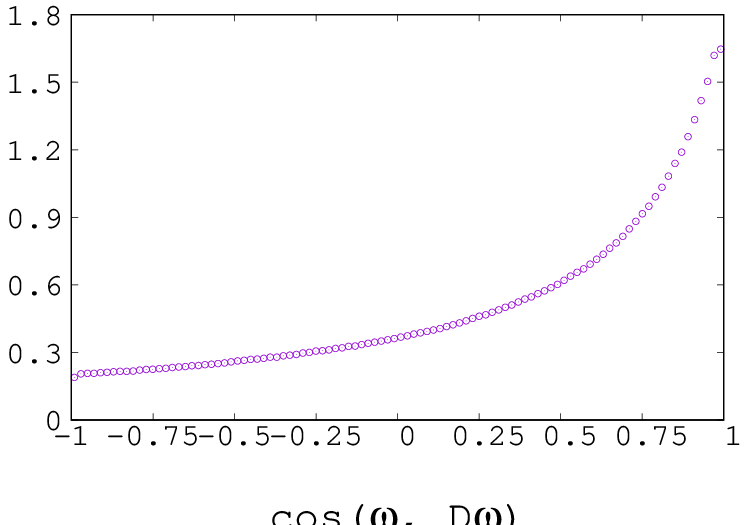}\\
\end{tabular}
\caption{\label{enstrostr}
	(Left) PDF of the square of the vorticity stretching vector $|D\omega|^2$.
	(Right) PDF of $cos(\omega, D\omega)$.
}
\end{figure*}
\begin{figure*}
\begin{tabular}{cc}
\hspace{-3mm}\includegraphics[width=0.49\linewidth]{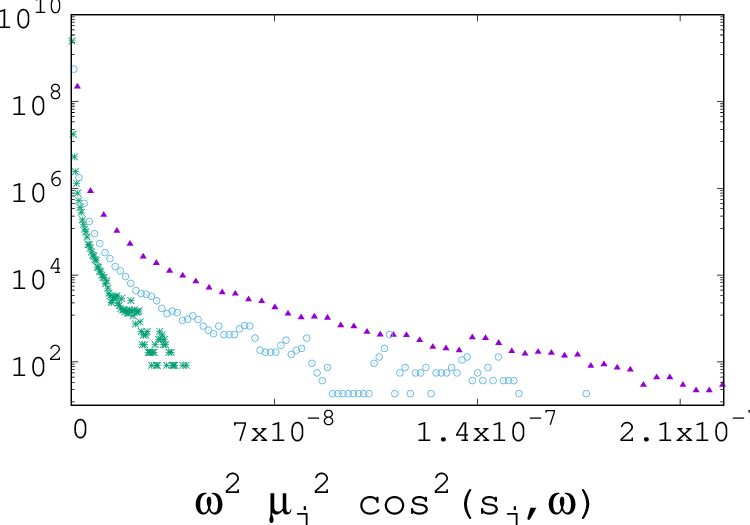}&
\hspace{2mm}\includegraphics[width=0.49\linewidth]{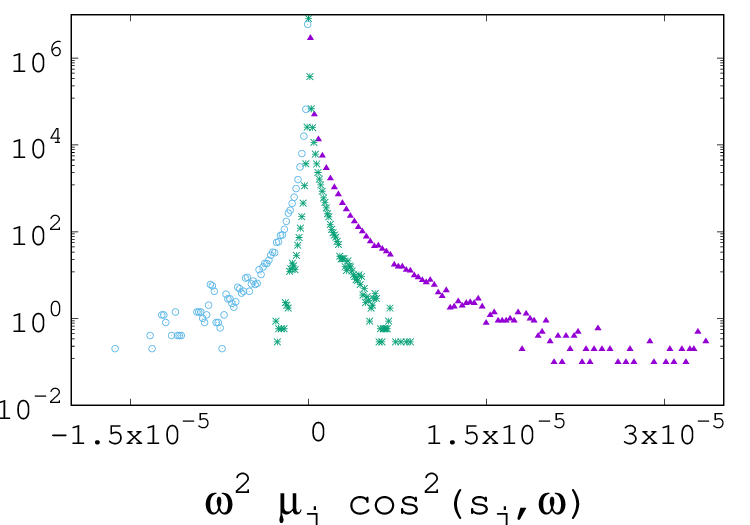}\\
\end{tabular}
\caption{\label{contributions}
(Left) PDFs of the individual strain-eigendirection contributions
$\phi_j=\omega^2\mu_j^2\cos^2(s_j,\omega)$ to the squared magnitude of the
vorticity-stretching vector,
$|D\omega|^2=\sum_{j=1}^{3}\phi_j$. The upper, lowest, and intermediate
curves correspond, respectively, to the $s_1$, $s_2$, and $s_3$
contributions.
(Right) PDFs of the individual strain-eigendirection contributions
$\psi_j=\omega^2\mu_j\cos^2(s_j,\omega)$ to enstrophy amplification,
$(D\omega)\cdot\omega=\sum_{j=1}^{3}\psi_j$. The positive, negative, and
mixed-sign curves correspond, respectively, to the $s_1$, $s_3$, and $s_2$
contributions.
}
\end{figure*}

\paragraph{Turbulent invariants and amplification mechanisms along polymer trajectories.}

The preceding results show that polymer trajectories define a nontrivial
conditioned sampling of the turbulent field: the strain--polymer and
vorticity--strain alignments measured along polymer paths differ from
standard Eulerian and material-line statistics. It is therefore important to
ask whether this conditioning also alters more basic one-point turbulent
signatures, or whether these remain close to their homogeneous-isotropic
form.

To address this question, we examine the PDFs of three quantities along the
polymer trajectories: the helicity density
\(h={u}\cdot{\omega}\)
(Fig.~\ref{aligns}, left), the squared magnitude of the vorticity tilting
vector \({\eta}^2\) (Fig.~\ref{aligns}, right), and the squared magnitude of
the Lamb vector
\(|{\omega}\times{u}|^2\)
(Fig.~\ref{lamb}, left). Here
\({\eta}^2=H_1-H_2^2\), with
\[
H_1=\sum_{i=1}^3 \mu_i^2 \cos^2(\omega,s_i),
\qquad
H_2=\sum_{i=1}^3 \mu_i \cos^2(\omega,s_i)
\]
\cite[][]{arcady,kivotides_filaments}.
If \(W=D{\omega}\) is the vortex-straining vector and
\(\phi={\omega}\cdot D{\omega}\) is the enstrophy
production, then
\begin{equation}
    {\eta}^2
    =
    \frac{W^2}{\omega^2}
    -
    \frac{\phi^2}{\omega^4}.
    \label{eq:vorticity-direction-change-rate}
\end{equation}

For the helicity density, the polymer-conditioned PDF remains symmetric
about zero, indicating that sampling along polymer trajectories does not
select a preferred sign of helical motion. The quantities \({\eta}^2\) and
\(|{\omega}\times{u}|^2\) are non-negative amplitudes, so
the relevant issue is different. Their PDFs start at zero and attain their
largest values at zero, showing that the most probable state along polymer
trajectories is weak vorticity tilting and weak Lamb-vector activity. Intense
events are present only through the intermittent tails. This is the same
qualitative organisation found in homogeneous isotropic turbulence: weak
events are most frequent, while strong activity events are rare. Thus,
although polymer trajectories define a conditioned sampling of the turbulent
flow, these one-point PDFs preserve the basic HIT-like character of the
corresponding turbulent quantities.

We also examined the PDFs of enstrophy \(\omega^2/2\) and total strain
\(D^2\) along polymer trajectories (Fig.~\ref{lamb}, right). These
quantities measure the rotational and straining intensity of the turbulent
regions sampled by the polymers. Both PDFs attain their largest values at
zero, showing that zero-enstrophy and zero-strain events are the most
probable states in these one-point statistics. Their tails, however, are very
different: the enstrophy distribution is considerably broader than the strain
distribution. Thus polymer trajectories retain the classical HIT feature that
vorticity, or enstrophy, is more intermittent than strain
\cite[][]{yeung,arcady}. The result is useful because it shows that, although
polymer paths are conditioned trajectories, their sampling preserves this
basic hierarchy of turbulent intermittency.

It is important to examine whether the same is
true of the corresponding dynamical amplification mechanisms. This is an
important control on the interpretation of the polymer statistics. If polymer
trajectories preferentially sampled regions with non-standard enstrophy or
strain production, then the stretching results above could simply reflect a
bias towards an anomalous subset of turbulent events. The following
diagnostics test whether this is the case.

The enstrophy amplification term
\({\omega}\cdot D{\omega}\)
(Fig.~\ref{enstrampl}, left) gives the stretching contribution to the
evolution of \(\omega^2/2\). The corresponding total-strain amplification
term is
\(-D:(D^2)\)
(Fig.~\ref{enstrampl}, right). Along polymer trajectories, the PDFs of both
quantities are strongly leptokurtic and peak at zero, indicating that weak
amplification or reduction events are the most probable states, while
occasional bursts of strong amplification or reduction populate the tails.
Moreover, the production sides of the PDFs display much longer tails than the
reduction sides, showing that intense growth events are considerably more
probable than equally intense decay events. This asymmetry is consistent with
classical results for HIT \cite[][]{ashurst,jimenez,arcady}, and demonstrates
that the intermittent production of enstrophy and strain persists along
polymer-conditioned trajectories.

Similarly, the PDFs of the squared magnitude of the vorticity-stretching
vector \(|D{\omega}|^2\) (Fig.~\ref{enstrostr}, left) and of the
cosine between vorticity \({\omega}\) and its stretching vector
\(D{\omega}\) (Fig.~\ref{enstrostr}, right) agree very well with
canonical Eulerian results \cite[][]{ashurst,arcady}. The PDF of
\(|D{\omega}|^2\) is strongly leptokurtic and peaks at zero,
reflecting the intermittent character of vorticity amplification along
polymer trajectories. The PDF of
\(\cos({\omega},D{\omega})\) is nearly
indistinguishable from those Eulerian HIT results, exhibiting a strong
preference for alignment.

These results show that the polymer-conditioned trajectories retain the
classical HIT structure of enstrophy production, strain amplification, and
vorticity stretching. Therefore the polymer-specific stretching behaviour
reported above should not be interpreted as a consequence of sampling a
qualitatively anomalous population of amplification events. Rather, the
distinctive polymer physics enters through the orientation of the chain
relative to the local strain field and through the nonlinear bead--spring
response to that field.

To characterize further the flow along polymer trajectories, we decomposed
the vorticity-stretching and enstrophy-amplification mechanisms into their
strain-eigenvector contributions. Since
\[
D\omega
=
\sum_{j=1}^{3}
\mu_j(\omega\cdot s_j)s_j ,
\]
the contributions to the square of the vorticity-stretching vector are
\[
\phi_j=\omega^2\mu_j^2\cos^2(\omega,s_j),
\]
so that \(|D\omega|^2=\sum_{j=1}^{3}\phi_j\)
(Fig.~\ref{contributions}, left). Similarly, the contributions to
enstrophy amplification are
\[
\psi_j=\omega^2\mu_j\cos^2(\omega,s_j),
\]
so that
\[
(D\omega)\cdot\omega
=
\sum_{j=1}^{3}\psi_j
\]
(Fig.~\ref{contributions}, right).

For enstrophy amplification, the PDFs show a clear ordering of tails, with
the slowest decay for \(\psi_1\) and the fastest for \(\psi_2\). They
indicate that the \(\psi_1\) production effect is larger than the
\(\psi_3\) destruction effect. This follows from the combined eigenvalue and
alignment structure: \(s_1\) tends to align with \({\omega}\),
whereas \(s_3\) tends to be nearly perpendicular to
\({\omega}\) (Fig.~\ref{growths}, right). Thus the negative
\(\mu_3\) contribution is geometrically weakened by the small projection of
\({\omega}\) on \(s_3\). Interestingly, although \(s_2\) often
aligns with \({\omega}\), its contribution to enstrophy reduction
is smaller, albeit non-negligible, owing to the much smaller magnitude of the
intermediate eigenvalue. Moreover, the intermediate eigendirection shows a
tendency to favour enstrophy intensification rather than damping. These
results agree with Eulerian isotropic-turbulence studies
\cite[][]{arcady2}.

The decomposition of the squared vorticity-stretching magnitude
\(|D{\omega}|^2\) shows similar features, both in the hierarchy of
tails and in the relative strengths of the eigenvector contributions.

\begin{figure*}
\begin{tabular}{cc}
\hspace{-3mm}\includegraphics[width=0.49\linewidth]{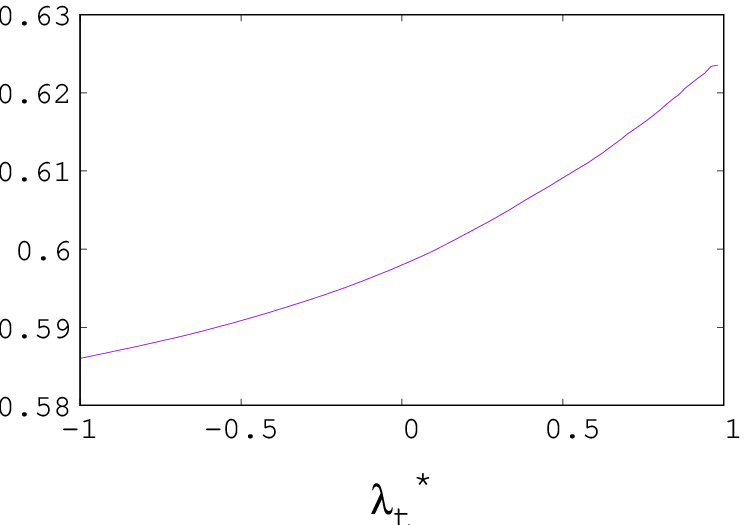}&
\hspace{2mm}\includegraphics[width=0.49\linewidth]{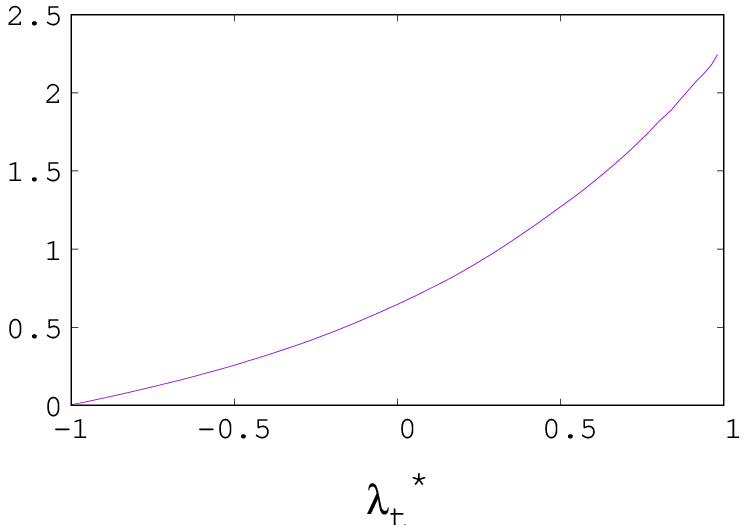}\\
\end{tabular}
\caption{\label{lamlencond}
	(Left) Conditional averages of polymer end-to-end distance, conditioned on ${\lambda}^*$.
        (Right) Conditional averages of the derivative of polymer end-to-end distance, conditioned on ${\lambda}^*$.
}
\end{figure*}
\begin{figure*}
\begin{tabular}{cc}
\hspace{-3.4mm}\includegraphics[width=0.49\linewidth]{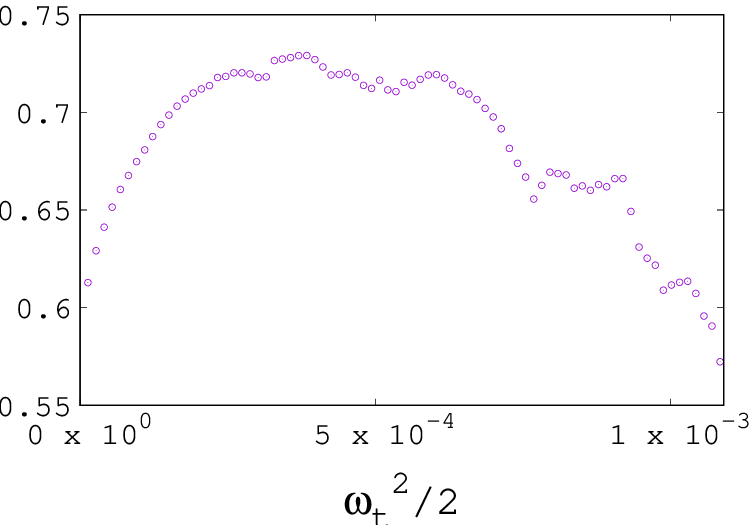}&
\hspace{2.4mm}\includegraphics[width=0.49\linewidth]{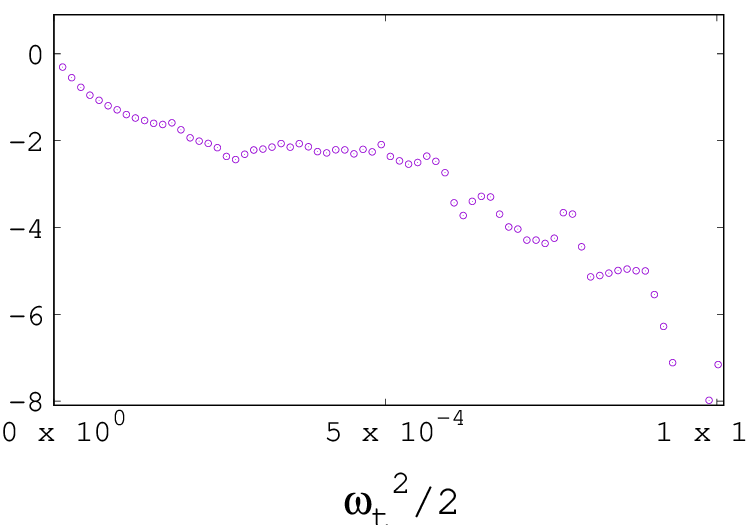}\\
\end{tabular}
\caption{\label{enstrolencond}
	(Left) Conditional averages of polymer end-to-end distance, conditioned on enstrophy ${\omega}^2/2$.
        (Right) Conditional averages of the derivative of polymer end-to-end distance, conditioned on ${\omega}^2/2$.
}
\end{figure*}
\begin{figure*}
\begin{tabular}{cc}
\hspace{-3mm}\includegraphics[width=0.49\linewidth]{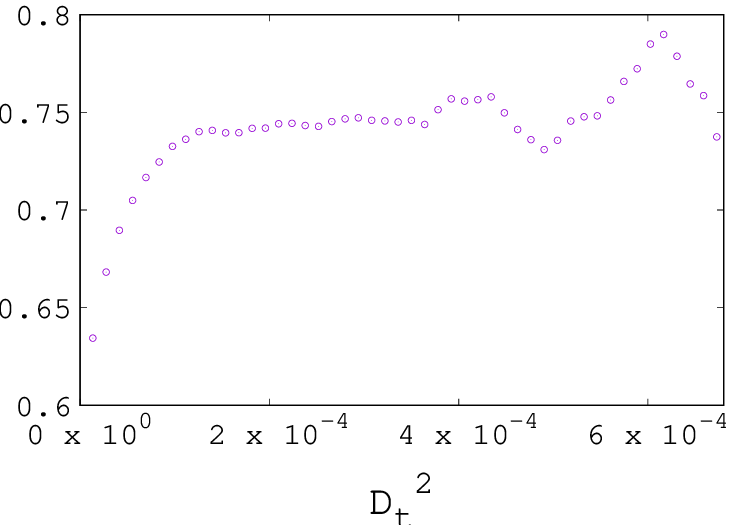}&
\hspace{2mm}\includegraphics[width=0.49\linewidth]{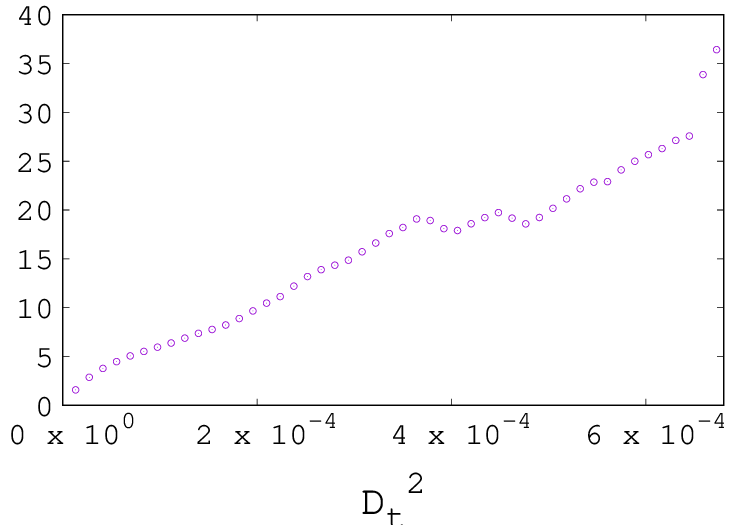}\\
\end{tabular}
\caption{\label{strainlencond}
        (Left) Conditional averages of polymer end-to-end distance, conditioned on total strain $D^2$.
	(Right) Conditional averages of the derivative of polymer end-to-end distance, conditioned on total strain $D^2$.
}
\end{figure*}

\paragraph{Conditional stretching statistics.}

A useful statistical measure for analysing the relation between flow
features and polymer stretching is the tail conditional average
\[
F_t(g_t)=\mathbb{E}[f \mid g>g_t],
\]
i.e. the expected value of a stochastic variable \(f\) conditioned on another
stochastic variable \(g\) exceeding a threshold \(g_t\). This statistic is
particularly useful here because it asks a direct physical question: how do
polymer length \(l\) and length-change rate \(dl/dt\) behave when the
polymer samples increasingly intense or increasingly specialised turbulent
events?

The functions \(F_t(g_t)\) admit a local interpretation when the threshold is
moved from \(g_t\) to \(g_{t+1}\), i.e. when the slice
\((g_t,g_{t+1}]\) is removed from the retained tail. A rising region,
\(F_{t+1}>F_t\), occurs if
\[
        \mathbb{E}[f \mid g \in (g_t, g_{t+1}]]
        <
        \mathbb{E}[f \mid g > g_{t+1}],
\]
so that the discarded slice has a lower \(f\)-mean than the retained
higher-\(g\) tail. In this case, larger values of \(g\) are locally
associated with larger values of \(f\). A falling region, \(F_{t+1}<F_t\),
occurs if
\[
        \mathbb{E}[f \mid g \in (g_t, g_{t+1}]]
        >
        \mathbb{E}[f \mid g > g_{t+1}],
\]
so that larger values of \(g\) are locally associated with smaller values of
\(f\). A flat region, \(F_{t+1}=F_t\), occurs if
\[
        \mathbb{E}[f \mid g \in (g_t, g_{t+1}]]
        =
        \mathbb{E}[f \mid g > g_{t+1}],
\]
so including or excluding the slice does not change the conditional average.
Thus, in this threshold range, increasingly large values of \(g\) do not
systematically select larger or smaller mean values of \(f\). In this
restricted sense, the tail conditional average indicates weak local
conditional association between \(f\) and \(g\). 

We apply this statistic to the polymer end-to-end distance \(l\) and its
time derivative \(dl/dt\), conditioning on \(\lambda^*\), enstrophy
\(\omega^2/2\), and total strain \(D^2\). Conditioning on \(\lambda^*\)
(Fig.~\ref{lamlencond}) shows that both \(l\) and \(dl/dt\) increase with the
tail threshold. Since the PDF of \(\lambda^*\) peaks near \(\lambda^*=1\),
the unconditional statistics already show that polymers frequently sample
biaxial extensional strain states. The conditional averages add the stronger
statement that, within this sampled population, the regions with larger
\(\lambda^*\) are associated with longer polymers and faster instantaneous
stretching. Thus biaxial extensional strain states are not merely frequently
visited; their more intense realisations are also the ones most directly
connected with polymer extension.

The enstrophy-conditioned averages show a different structure. The function
\(\mathbb{E}[l \mid \omega^2/2>\omega_t^2/2]\)
(Fig.~\ref{enstrolencond}, left) indicates that the smallest polymer lengths
are associated with both low and high enstrophy levels, while a broad
intermediate enstrophy range is only weakly correlated with \(l\). By
contrast,
\[
\mathbb{E}[dl/dt \mid \omega^2/2>\omega_t^2/2]
\]
(Fig.~\ref{enstrolencond}, right) shows that intense enstrophy regions are
associated with the strongest polymer relaxation events. Thus enstrophy does
not act like strain intensity in these conditional statistics: high-enstrophy
regions are more clearly connected with polymer relaxation than with polymer
stretching.

The strain-conditioned averages complete this picture. The function
\(\mathbb{E}[l \mid D^2>D_t^2]\)
(Fig.~\ref{strainlencond}, left) shows an extended intermediate range in
which polymer length and strain intensity are weakly correlated, followed by
a strong positive correlation at high \(D^2\). The derivative statistic
\[
\mathbb{E}[dl/dt \mid D^2>D_t^2]
\]
(Fig.~\ref{strainlencond}, right) gives the clearest dynamical signal:
larger strain-rate levels are directly associated with larger polymer
stretching rates. Taken together, the conditional statistics separate the
roles of the main turbulent quantities: biaxial extensional strain states
and high strain intensity promote polymer extension and stretching, whereas
intense enstrophy is associated primarily with relaxation.

In summary, polymers most frequently sample regions of axisymmetric biaxial
extension. The conditional averages further show that, as the threshold in
\(\lambda^*\) is increased, the polymer configurations satisfying
\(\lambda^*>\lambda^*_t\) have larger end-to-end distances and larger
positive values of \(dl/dt\). Thus, within the polymer-conditioned ensemble,
stronger biaxial-extensional strain states are associated with the largest
chain lengths and the fastest stretching events.

Polymer extension is only weakly associated with strain intensity over an
extended intermediate range of \(D^2\), but becomes strongly associated with
\(D^2\) at high strain levels, while \(dl/dt\) shows a direct positive
association with \(D^2\). 
By contrast, intense enstrophy regions are associated with both the smallest
polymer lengths and the strongest polymer relaxation events. The
conditional-statistics framework therefore provides a direct quantitative
connection between turbulent flow structures and nonlinear polymer
conformational dynamics.

\subsection{Polymer stretching in the tangent system}

We next examine stretching in the tangent system. This analysis is distinct from the
nonlinear chain-extension statistics discussed above. The full bead--spring dynamics
gives the actual polymer deformation, whereas the tangent system evolves infinitesimal
material fibres of the carrier flow along the polymer end-to-end midpoint trajectories.
It therefore measures the material-line stretching environment sampled by the polymers,
rather than the nonlinear polymer stretch itself. This distinction is important because
polymer trajectories are not material-fluid trajectories, and because the finite-time
Lyapunov quantities reported below are not asymptotic Lyapunov exponents. We use the
dimensionless finite-time Lyapunov numbers
\[
        L_i(t)=\tau_R \lambda_i(t),
\]
where \(\tau_R\) is the longest polymer relaxation time.
\begin{figure*}
\begin{tabular}{cc}
\hspace{-3mm}\includegraphics[width=0.49\linewidth]{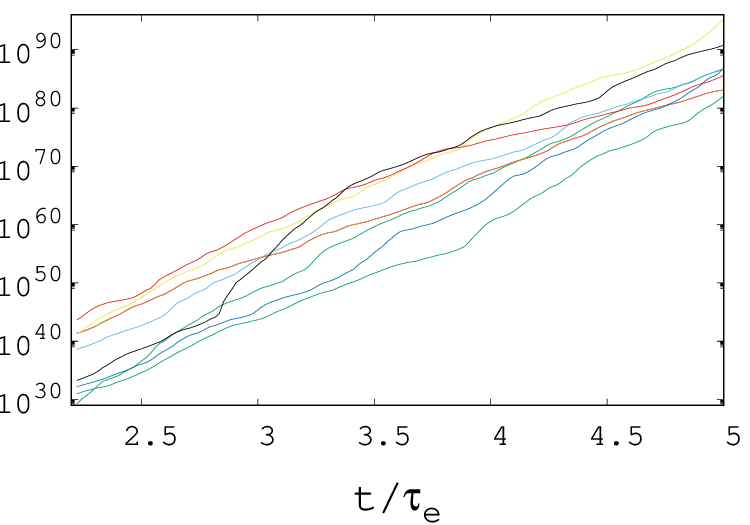}&
\hspace{2mm}\includegraphics[width=0.49\linewidth]{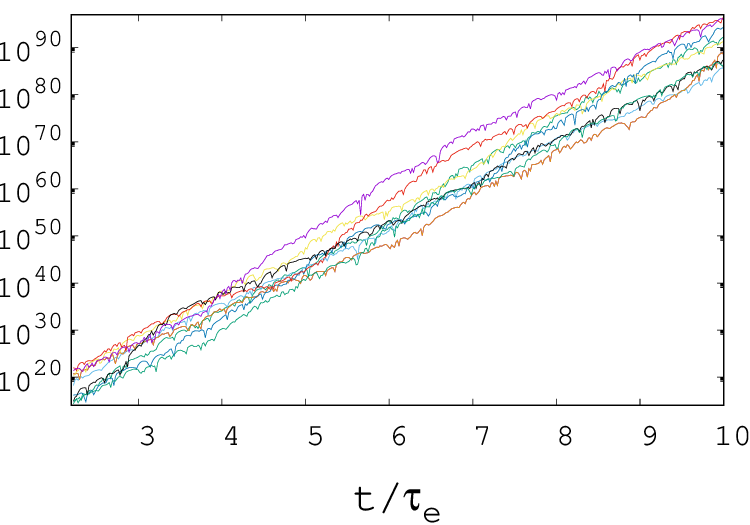}\\
\end{tabular}
\caption{\label{strainstretch}
        (Left) Histories of the pure-strain measure $(E E)^{1/2}$ along polymer
        trajectories. (Right) Histories of the tangent-system stretch $\zeta_e$
        along polymer trajectories.
}
\end{figure*}
\begin{figure*}
\begin{tabular}{cc}
\hspace{-3mm}\includegraphics[width=0.49\linewidth]{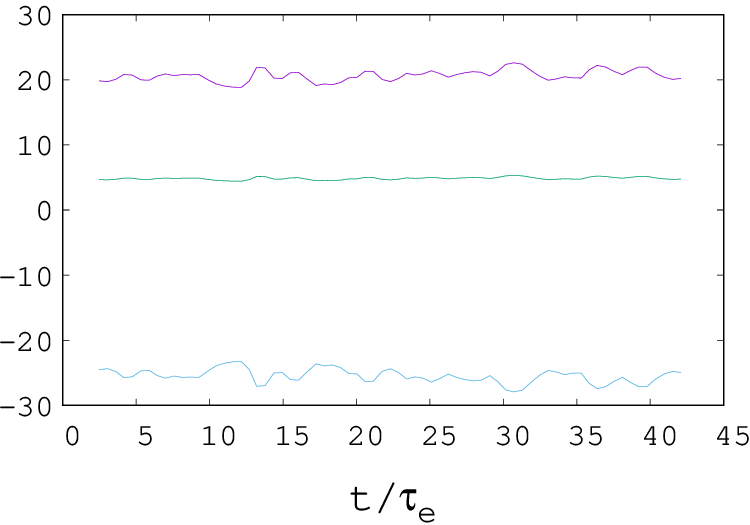}&
\hspace{2mm}\includegraphics[width=0.49\linewidth]{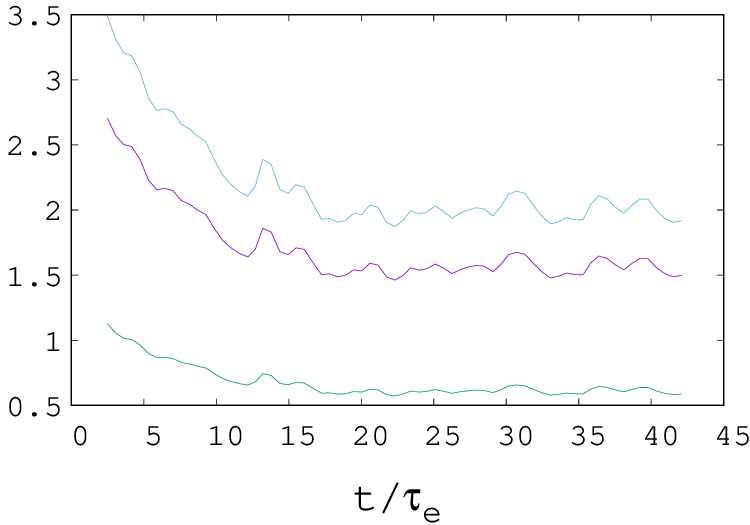}\\
\end{tabular}
\caption{\label{meandevlyap}
        (Left) Ensemble means $E[L_i(t)]$ of the finite-time Lyapunov numbers
        $L_i(t)=\tau_R\lambda_i(t)$. (Right) Standard deviations
        $\sigma[L_i(t)]$ of the same quantities. The three curves correspond to
        $i=1,2,3$. On the right panel, the top line corresponds to $\lambda_3$,
	the middle line to $\lambda_1$ and the bottom line to $\lambda_2$. 
}
\end{figure*}
The histories of the pure-strain measure $(E E)^{1/2}$ (Fig.~\ref{strainstretch},
left) and of the tangent-system stretch $\zeta_e$ (Fig.~\ref{strainstretch}, right)
show the strong accumulated deformation produced by the turbulent velocity-gradient
field sampled along polymer trajectories. The stretch histories grow rapidly, as expected
for material-fibre evolution in a turbulent flow, but they also display small-scale
fluctuations. These fluctuations reflect local relaxation and reorientation events along
the sampled material directions, and show that the tangent-system stretching is not a
smooth monotone process even when the accumulated deformation is large.

The ensemble means \(E[L_i(t)]\) and standard deviations \(\sigma[L_i(t)]\),
where \(L_i(t)=\tau_R\lambda_i(t)\), reach statistically stationary levels
during the late stage of the calculation. 
The corresponding PDFs are also insensitive, within the accuracy of the present
sampling, to replacing the window ending at \(22\tau_e\) by the later window
ending at \(42.2\tau_e\) (Figs.~\ref{lyap1trajpdf}--\ref{lyap3trajpdf}).
These observations
indicate convergence of the finite-time Lyapunov-number statistics over the
computed interval. This statistical convergence should be distinguished from
trajectory-wise collapse of the finite-time exponents to a common asymptotic
Lyapunov spectrum, which is not observed over the present integration time.
Nevertheless, since the finite-time PDFs and low-order moments have reached
a stable late-time regime, the late-time ensemble mean provides a useful
finite-time estimate of the corresponding asymptotic stretching rate.

Guided by the ergodic Lyapunov theory, we therefore use
\[
        \Lambda_1 \simeq E[\lambda_1]
\]
as an estimate of the largest asymptotic Lyapunov exponent, rather than as an
exact finite-time identification. This gives the estimated largest asymptotic
Lyapunov number
\[
        L_1=\tau_R\Lambda_1 \simeq \tau_R E[\lambda_1]\approx 20.6,
\]
consistent with the strongly stretched polymer conformations observed in the
nonlinear bead--spring dynamics.

The ratio of the finite-time mean values,
\[
        \frac{E[\lambda_2]}{E[\lambda_1]}
        \approx \frac{4}{17},
\]
is useful for comparing the character of polymer-sampled stretching with other
stretching processes. It is larger than the corresponding material-line values
reported for Navier--Stokes turbulence,
\(E[\lambda_2]/E[\lambda_1]\approx 1/7\) \cite[][]{yeung,bentkamp2022,johnson_meneveau2015},
and for vortex-dynamics calculations,
\(E[\lambda_2]/E[\lambda_1]\approx 1/8\) \cite[][]{kivotides_stringy}.
This indicates a stronger intermediate stretching contribution, or equivalently
a broader ribbon-like component of the polymer-sampled material deformation.
By contrast, vortex-filament stretching gives
\(E[\lambda_2]/E[\lambda_1]\approx 1/56\) \cite[][]{kivotides_lagrangian},
consistent with the expectation that polymer stretching is closer to material-line
stretching than to vortex-filament stretching.
\begin{figure*}
\begin{tabular}{cc}
\hspace{-3mm}\includegraphics[width=0.49\linewidth]{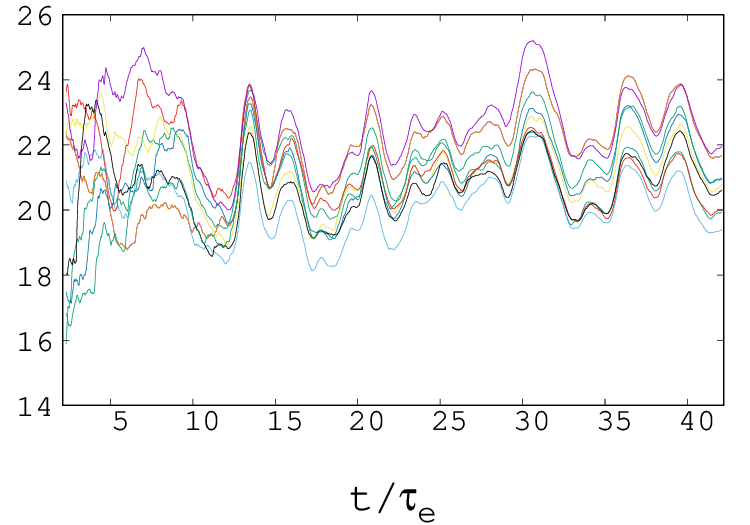}&
\hspace{2mm}\includegraphics[width=0.49\linewidth]{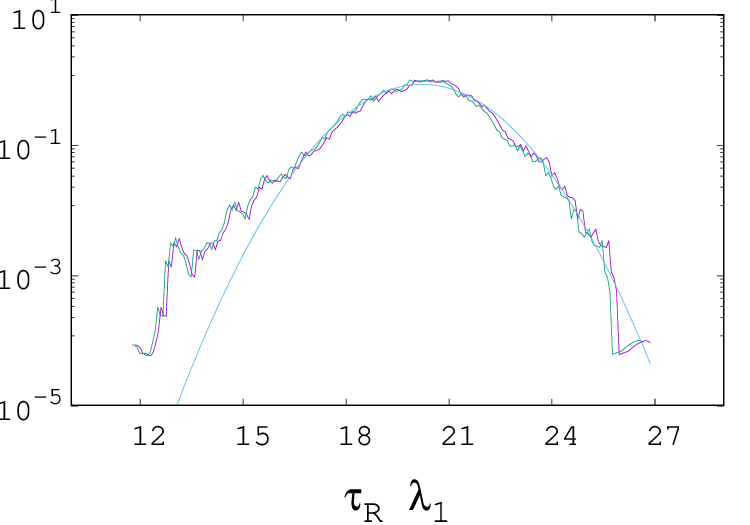}\\
\end{tabular}
\caption{\label{lyap1trajpdf}
        (Left) Histories of the largest finite-time Lyapunov number
        $L_1(t)=\tau_R\lambda_1(t)$ along polymer trajectories.
        (Right) PDFs of $L_1$ obtained at times
        $t_1=22\tau_e$ and $t_2=42.2\tau_e$. The smooth curves are
        Gaussian fits shown as reference distributions.
}
\end{figure*}
\begin{figure*}
\begin{tabular}{cc} 
\hspace{-3mm}\includegraphics[width=0.49\linewidth]{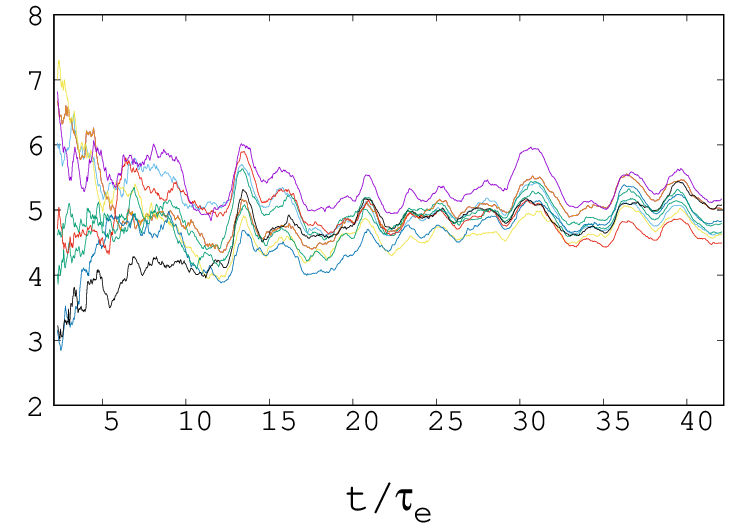}&
\hspace{2mm}\includegraphics[width=0.49\linewidth]{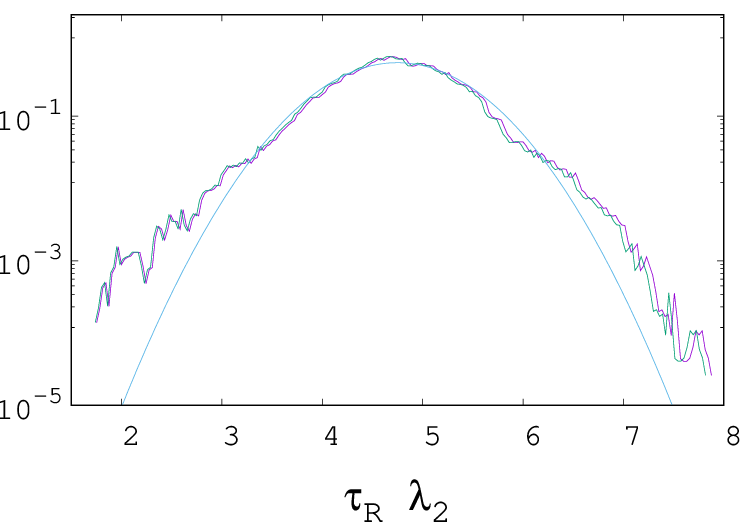}\\
\end{tabular}
\caption{\label{lyap2trajpdf}
        (Left) Histories of the intermediate finite-time Lyapunov number
        $L_2(t)=\tau_R\lambda_2(t)$ along polymer trajectories.
        (Right) PDFs of $L_2$ obtained at times
        $t_1=22\tau_e$ and $t_2=42.2\tau_e$. The smooth curves are
        Gaussian fits shown as reference distributions.
}
\end{figure*}
\begin{figure*}
\begin{tabular}{cc} 
\hspace{-3mm}\includegraphics[width=0.49\linewidth]{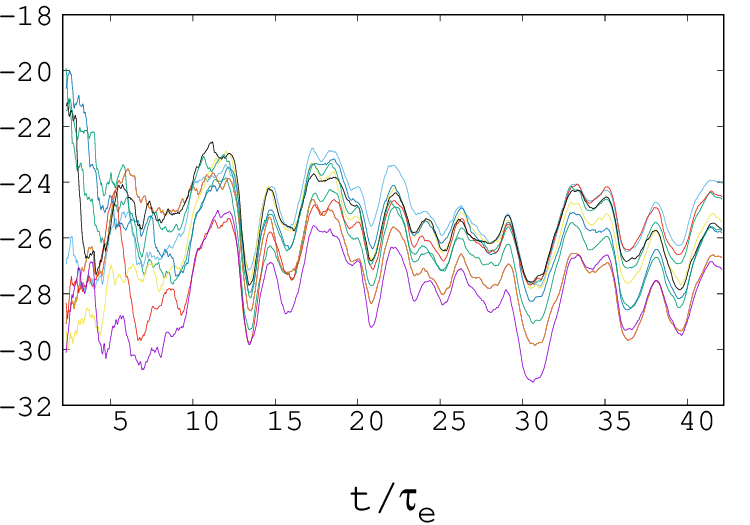}&
\hspace{2mm}\includegraphics[width=0.49\linewidth]{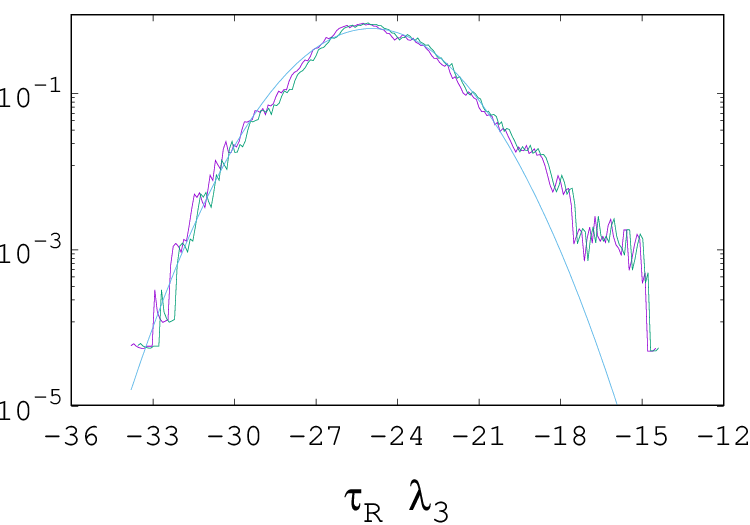}\\
\end{tabular}
\caption{\label{lyap3trajpdf}
        (Left) Histories of the smallest finite-time Lyapunov number
        $L_3(t)=\tau_R\lambda_3(t)$ along polymer trajectories.
        (Right) PDFs of $L_3$ obtained at times
        $t_1=22\tau_e$ and $t_2=42.2\tau_e$. The smooth curves are
        Gaussian fits shown as reference distributions.
}
\end{figure*}
\begin{figure*}
\includegraphics[width=0.49\linewidth]{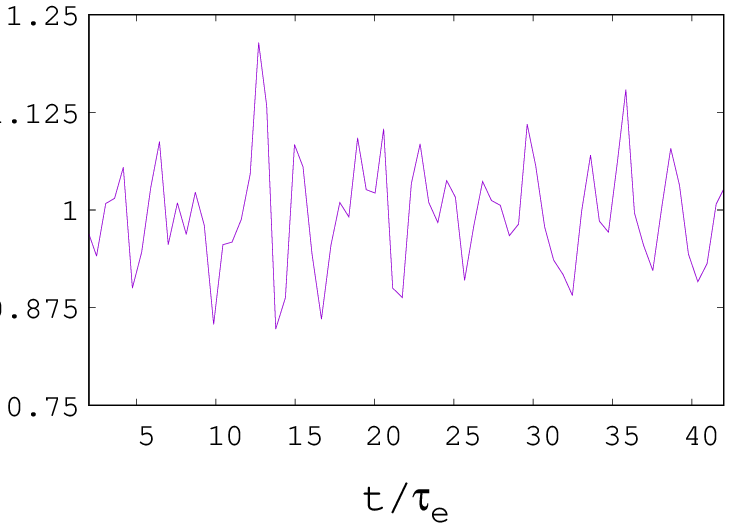}
\caption{\label{dissipation}
        Time series of the domain-averaged energy-dissipation rate
        \(\epsilon(t)\), normalised by its time mean \(\epsilon_a\).
}
\end{figure*}
After approximately \(10\tau_e\),
the finite-time Lyapunov-number histories (Figs.~\ref{lyap1trajpdf}--\ref{lyap3trajpdf}, left)
enter a late-time regime in which their
ensemble means and standard deviations fluctuate around statistically 
stationary levels (Fig.~\ref{meandevlyap}).
The individual histories do not collapse onto trajectory-independent curves. Instead,
they display trajectory-dependent separation together with a common low-frequency
modulation. This common modulation is consistent with the time dependence of the
domain-averaged dissipation statistic shown in Fig.~\ref{dissipation}, which reflects
the residual modulation of the statistically maintained turbulent state. 
The remaining trajectory-to-trajectory spread is interpreted as finite-time variability
of the polymer-sampled velocity-gradient cocycle, not as evidence for distinct
asymptotic Lyapunov spectra.

A notable feature of the finite-time Lyapunov-number PDFs
(Figs.~\ref{lyap1trajpdf}--\ref{lyap3trajpdf}, right) is that the intermediate
finite-time exponent 
has positive support/mean over the sampled statistics.
This is consistent
with the polymers preferentially sampling regions of axisymmetric biaxial extension
(Fig.~\ref{stretchchar}, right), and with the pronounced alignment between the
intermediate strain-rate eigenvector and the chain end-to-end vector
(Fig.~\ref{eigenvals}, right). 
This connection should not be understood as a direct equivalence between local
strain-rate eigenvalues and finite-time Lyapunov exponents. The latter result from
the accumulated, time-ordered product of velocity-gradient tensors along polymer
trajectories. Nevertheless, the local strain geometry provides a physical explanation
for the positive intermediate finite-time stretching statistics.

It is useful to analyse the interdependence of the finite-time Lyapunov numbers by
considering the independence residual function and accompanying statistical measures
\cite[][]{vaart}. To avoid confusing these finite-time random variables with
asymptotic Lyapunov exponents, we denote by \(p_i(\ell_i)\) the PDF of the sampled
finite-time quantity \(\ell_i\), and by \(p_{ij}(\ell_i,\ell_j)\) the corresponding
joint PDF, where \(\ell_i=L_i(t)=\tau_R\lambda_i(t)\) in the figures below.
The independence model density is then \(p_i(\ell_i)p_j(\ell_j)\), and the
independence residual is
\begin{equation}
    \Delta(\ell_i,\ell_j)
    =
    p_{ij}(\ell_i,\ell_j)
    -
    p_i(\ell_i)p_j(\ell_j).
    \nonumber
\end{equation}
Besides plotting \(\Delta(\ell_i,\ell_j)\), we also compute several scalar global
measures of the departure from independence.

\paragraph{Independence energy.}
The squared \(L^2\) distance between the joint distribution and the independence model is
\begin{equation}
    \mathcal{E}_{ij}
    =
    \iint
    \big[
    p_{ij}(\ell_i,\ell_j)
    -
    p_i(\ell_i)p_j(\ell_j)
    \big]^2
    \,d\ell_i\,d\ell_j
    =
    \|\Delta\|_2^2 .
    \nonumber
\end{equation}
Large localized deviations contribute disproportionately because of the quadratic form,
so \(\mathcal{E}_{ij}\) acts as an energy of dependence: it amplifies strong departures
from independence. Unlike the total variation and Hellinger distances, \(\mathcal{E}_{ij}\)
is not bounded from above.

\paragraph{Total variation distance.}
The \(L^1\) distance between the joint distribution and the independence model is
\begin{equation}
    \mathrm{TV}_{ij}
    =
    \frac{1}{2}
    \iint
    \big|
    p_{ij}(\ell_i,\ell_j)
    -
    p_i(\ell_i)p_j(\ell_j)
    \big|
    \,d\ell_i\,d\ell_j
    =
    \frac{1}{2}\|\Delta\|_1 .
    \nonumber
\end{equation}
When the independence model density and the true joint probability density differ,
this measure gives the probability mass that must be redistributed from regions where
\(p_i(\ell_i)p_j(\ell_j)\) overshoots the true joint density to regions where it undershoots it.
The total variation distance is bounded, \(\mathrm{TV}_{ij}\in[0,1]\).

\paragraph{Hellinger distance.}
The Hellinger distance can be written as the \(L^2\) norm of the difference between
the square-root densities of the true joint distribution and the independence model:
\begin{equation}
    H_{ij}
    =
    \frac{1}{\sqrt{2}}
    \left[
        \iint
        \left(
        \sqrt{p_{ij}(\ell_i,\ell_j)}
        -
        \sqrt{p_i(\ell_i)p_j(\ell_j)}
        \right)^2
        \,d\ell_i\,d\ell_j
    \right]^{1/2}.
    \nonumber
\end{equation}
Equivalently,
\begin{equation}
    H_{ij}^2
    =
    1
    -
    \iint
    \sqrt{
    p_{ij}(\ell_i,\ell_j)
    p_i(\ell_i)p_j(\ell_j)}
    \,d\ell_i\,d\ell_j .
    \nonumber
\end{equation}
Thus \(H_{ij}=0\) when
\(p_{ij}(\ell_i,\ell_j)=p_i(\ell_i)p_j(\ell_j)\), corresponding to independence.
At the other extreme, \(H_{ij}\to 1\) when the two probability measures become
mutually singular. Deterministic functional dependence is an example of this limiting
case for continuous variables, since then the joint probability is concentrated on a
lower-dimensional set,
\begin{equation}
    p_{ij}(\ell_i,\ell_j)
    =
    p_i(\ell_i)\delta\!\big(\ell_j-g(\ell_i)\big),
    \nonumber
\end{equation}
rather than being distributed over the full \((\ell_i,\ell_j)\) plane. In the general
case of noisy statistical dependence, \(0 < H_{ij} < 1\).

\begin{table}
\begin{center}
    \begin{tabular}[t]{c|c|c|c}
      \hline
      \hline
            Pair & \(\mathcal{E}_{ij}\) & \(\mathrm{TV}_{ij}\) & \(H_{ij}\)\\
      \hline
            \((L_1,L_2)\)& \(0.3479 \times 10^{-9}\)& \(0.3002\)& \(0.3138\)\\
      \hline
            \((L_2,L_3)\)& \(0.5583 \times 10^{-9}\)& \(0.4134\)& \(0.4127\)\\
      \hline
            \((L_3,L_1)\)& \(1.1284 \times 10^{-9}\)& \(0.6967\)& \(0.6623\)\\
      \hline
      \hline
    \end{tabular}
\end{center}
\vspace{5mm}
\caption{\label{dependence}
Dependence measures for the finite-time Lyapunov-number pairs
\((L_1,L_2)\), \((L_2,L_3)\), and \((L_3,L_1)\), where
\(L_i(t)=\tau_R\lambda_i(t)\). The quantities \(\mathcal{E}_{ij}\),
\(\mathrm{TV}_{ij}\), and \(H_{ij}\) measure departures from the independence
model \(p_i(\ell_i)p_j(\ell_j)\).
}
\end{table}

\begin{figure*}
\begin{tabular}{cc}
\hspace{-3mm}\includegraphics[width=0.49\linewidth]{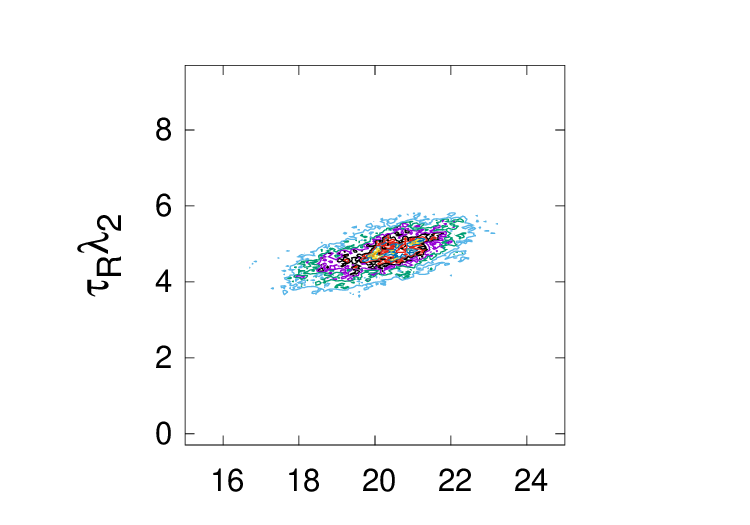}&
\hspace{2mm}\includegraphics[width=0.49\linewidth]{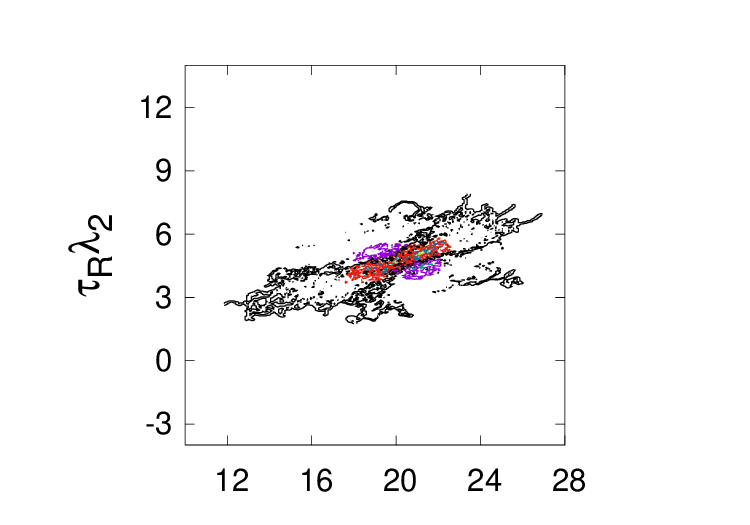}\\
\end{tabular}
\caption{\label{12jntdeltpdf}
        (Left) Isolines of the joint PDF \(p_{12}(\ell_1,\ell_2)\) of the finite-time
        Lyapunov numbers \(L_1(t)=\tau_R\lambda_1(t)\) and
        \(L_2(t)=\tau_R\lambda_2(t)\). (Right) Isolines of the corresponding
        independence residual \(\Delta(\ell_1,\ell_2)\).
}
\end{figure*}
\begin{figure*}
\begin{tabular}{cc}
\hspace{-3mm}\includegraphics[width=0.49\linewidth]{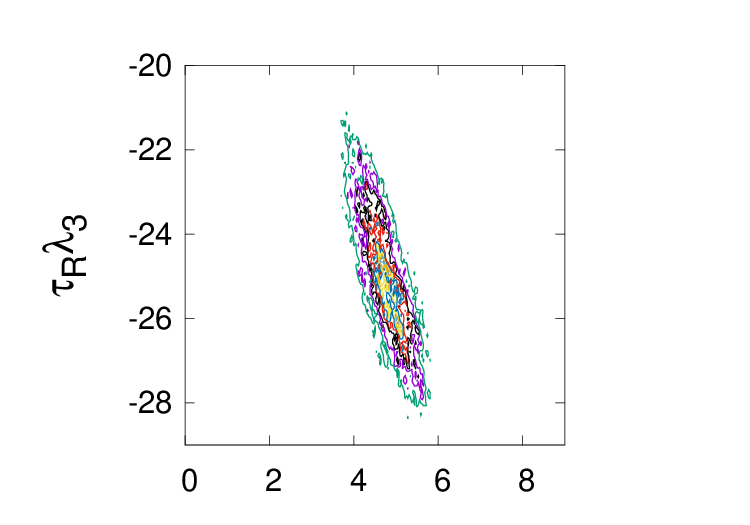}&
\hspace{2mm}\includegraphics[width=0.49\linewidth]{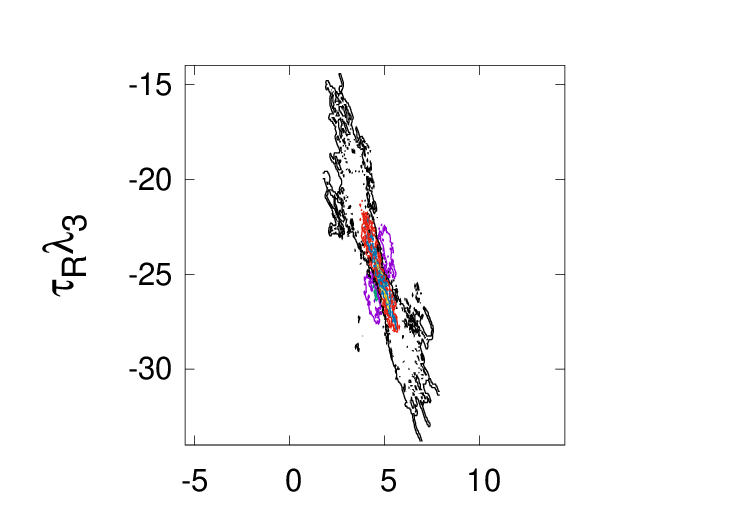}\\
\end{tabular}
\caption{\label{23jntdeltpdf}
        (Left) Isolines of the joint PDF \(p_{23}(\ell_2,\ell_3)\) of the finite-time
        Lyapunov numbers \(L_2(t)=\tau_R\lambda_2(t)\) and
        \(L_3(t)=\tau_R\lambda_3(t)\). (Right) Isolines of the corresponding
        independence residual \(\Delta(\ell_2,\ell_3)\).
}
\end{figure*}
\begin{figure*}
\begin{tabular}{cc}
\hspace{-3mm}\includegraphics[width=0.49\linewidth]{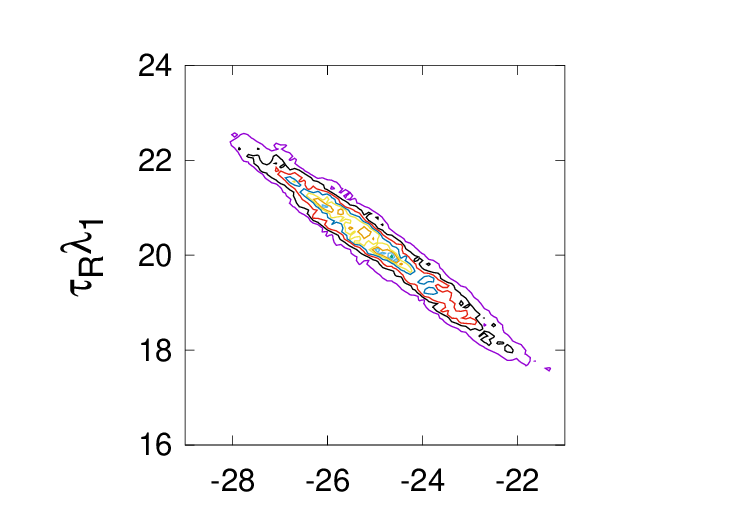}&
\hspace{2mm}\includegraphics[width=0.49\linewidth]{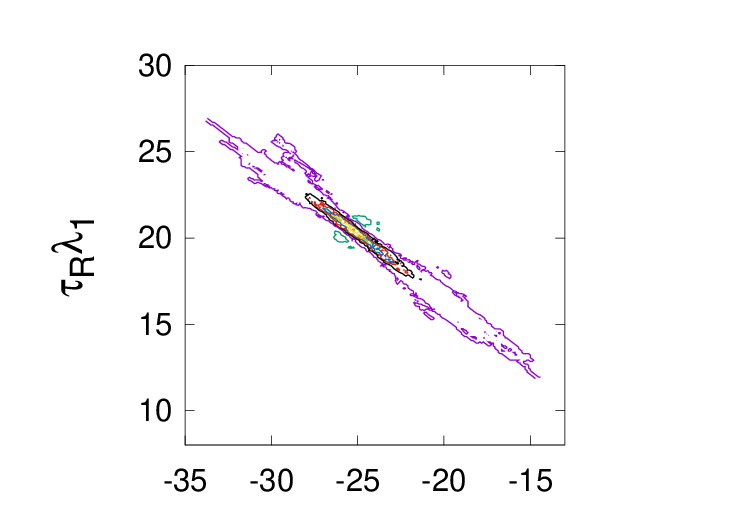}\\
\end{tabular}
\caption{\label{31jntdeltpdf}
        (Left) Isolines of the joint PDF \(p_{31}(\ell_3,\ell_1)\) of the finite-time
        Lyapunov numbers \(L_3(t)=\tau_R\lambda_3(t)\) and
        \(L_1(t)=\tau_R\lambda_1(t)\). (Right) Isolines of the corresponding
        independence residual \(\Delta(\ell_3,\ell_1)\).
}
\end{figure*}

The results in Table~\ref{dependence} give a consistent picture across all
three dependence measures. The pair \((L_3,L_1)\) has the largest values of
\(\mathcal{E}_{ij}\), \(\mathrm{TV}_{ij}\), and \(H_{ij}\), and is therefore
the pair with the strongest departure from independence. This conclusion is
not a direct consequence of incompressibility alone. Incompressibility imposes
the constraint
\[
        L_1(t)+L_2(t)+L_3(t)=0,
\]
so that the compressive member \(L_3(t)\) balances the combined positive
stretching associated with \(L_1(t)\) and \(L_2(t)\), rather than \(L_1(t)\)
alone. The scalar measures therefore provide additional information about how
this constraint is expressed statistically in the pairwise projections of the
finite-time Lyapunov spectrum.

The three measures emphasise different aspects of the same conclusion.
The independence energy \(\mathcal{E}_{ij}\) is largest for \((L_3,L_1)\),
showing that this pair contains the strongest localized departures from the
product-density model. The total variation distance gives the same ordering,
showing that the largest redistribution of probability mass relative to
independence is also associated with \((L_3,L_1)\). The Hellinger distance,
which measures the separation between the square-root densities, again gives
the same ranking. Thus the dominant dependence is not an artefact of a
particular norm or diagnostic. The intermediate finite-time Lyapunov number
\(L_2(t)\) is more strongly dependent on \(L_3(t)\) than on \(L_1(t)\), which
is consistent with the fact that positive intermediate stretching contributes
to the total stretching that must be balanced by the compressive member of the
spectrum.

The orientation of the joint-PDF isolines in
Figs.~\ref{12jntdeltpdf}--\ref{31jntdeltpdf} gives the sign of the
pairwise correlations. In Fig.~\ref{12jntdeltpdf}, the \((L_1,L_2)\)
contours are elongated along a positive-slope ridge, indicating positive
correlation between the largest and intermediate finite-time Lyapunov numbers.
The slope of this ridge is not expected to be unity, because the two variables
have different ranges and variances. By contrast, in Figs.~\ref{23jntdeltpdf}
and \ref{31jntdeltpdf}, the \((L_2,L_3)\) and \((L_3,L_1)\) contours are
elongated along negative-slope ridges, indicating that larger positive
finite-time stretching is statistically associated with stronger negative
compression.

The residual plots in the right panels of
Figs.~\ref{12jntdeltpdf}--\ref{31jntdeltpdf} show that the departures from
independence are localised around the same high-probability ridges identified by
the corresponding joint PDFs in the left panels. Thus the strongest departures
from independence occur in the most frequently sampled parts of the finite-time
Lyapunov spectrum, not only in rare tails. Together with the scalar measures in
Table~\ref{dependence}, these figures show a clear hierarchy of dependence:
the strongest departure from independence occurs for \((L_3,L_1)\), followed by
\((L_2,L_3)\), while \((L_1,L_2)\) is the weakest of the three dependencies
despite being positively correlated.

\section{Concluding Remarks}

We have studied polymer stretching in an ultra-dilute turbulent solution using
a full mesoscopic bead--spring description. The calculation is in a regime where
the polymers do not modify the large-scale turbulent structure. Because the
maximum chain length is smaller than the Kolmogorov scale, the carrier-flow
kinematics relevant to stretching are those of the locally smooth velocity-gradient
field, rather than inertial-range velocity differences. The central aim was to
quantify how nonlinear polymer stretching, trajectory-conditioned strain and
vorticity statistics, and finite-time Lyapunov statistics are organised along
polymer trajectories.

The main conclusions are as follows.

{\renewcommand{\labelenumi}{(\arabic{enumi})}
\begin{enumerate}

\item The polymer end-to-end distance exhibits an apparent power-law scaling
over an intermediate range, while the chains remain strongly stretched at
\(Wi\simeq 80\). Material-line stretching provides a useful leading kinematic
reference for the polymer dynamics, but finite departures occur in the strongly
stretched part of the distribution, where elasticity, finite extensibility,
excluded volume, Brownian forcing, and hydrodynamic interactions enter the
bead--spring force balance.

\item The polymers preferentially sample regions of axisymmetric biaxial extension.
A similar result was reported by Terrapon \emph{et al.}
\cite[][]{terrapon}, who found that highly stretched polymers in turbulent
channel flow had, on average, experienced strong biaxial extensional flow.
Their carrier flow is inhomogeneous and wall bounded, and its
velocity-gradient tensor contains a persistent mean-shear contribution in
addition to the turbulent fluctuations. This mean gradient can produce
sustained stretching and preferential polymer orientation. In the present
homogeneous isotropic turbulence, the mean velocity gradient vanishes and
there are no walls or preferred directions, so polymer stretching is produced
entirely by the fluctuating turbulent velocity gradients. The two studies
therefore concern physically distinct turbulence regimes.

Moreover, the result of Terrapon \emph{et al.} was obtained within a
simplified single-segment dumbbell model. The polymer centre of mass was
prescribed to follow a fluid Lagrangian trajectory, while excluded-volume and
hydrodynamic interactions were neglected and the Brownian forcing was not
coupled through a configuration-dependent multibody mobility tensor. In the
present calculation, all bead positions of the multisegment chain are advanced
directly, with elasticity, finite extensibility, excluded-volume forces,
Brownian forcing, and intrachain and interchain hydrodynamic interactions
determining jointly the polymer conformation and its trajectory. The results
show that the prescribed-Lagrangian-trajectory assumption is not generally
valid in the present regime. In particular, along the polymer-generated
trajectories,
\[
    \frac{E[\lambda_2]}{E[\lambda_1]}
    \simeq
    \frac{4}{17},
\]
whereas the corresponding material-line value is \(1/7\), demonstrating a
measurable difference in the accumulated stretching statistics.

Large extensions and large stretching rates occur in high-strain regions,
while small extensions and relaxation events are concentrated in
high-enstrophy regions. Thus stretching is associated primarily with the
extensional geometry of the local strain-rate tensor, whereas compact or
relaxing conformations are associated with vortical regions.

\item The alignment statistics show a pronounced role for the intermediate
strain-rate eigenvector. The polymer end-to-end vector is preferentially
oriented in the plane spanned by \(s_1\) and \(s_2\), with particularly strong
alignment toward \(s_2\), and tends to avoid the direction \(s_3\) associated
with the most negative strain-rate eigenvalue. Since \(\mu_2\) is often positive
along polymer trajectories, this alignment allows the intermediate strain-rate
direction to contribute significantly to the stretching statistics. Compression
is associated primarily with the eigenvalue \(\mu_3<0\), but its contribution
to the end-to-end stretching rate is weighted by the polymer orientation through
\(\cos^2(s_3,l)\).

\item The vorticity statistics sampled along polymer trajectories differ from
standard Eulerian statistics, from material-trajectory statistics, and from
vortex-filament stretching statistics. In particular, vorticity tends to align
with both the first and second strain-rate eigenvectors along polymer paths.
This shows that polymer-conditioned sampling gives a distinct view of the
turbulent velocity-gradient field, rather than reproducing the usual Eulerian,
material-line, or vortex-filament pictures.

\item We developed and used an SVD-normalised tangent-flow method to compute
finite-time Lyapunov numbers along polymer trajectories in a full mesoscopic
bead--spring calculation. The method avoids finite-precision breakdown by
periodically removing the singular stretches of the deformation-gradient tensor,
while storing their logarithms cumulatively and preserving the rotational
geometric content needed for continued tangent evolution. This provides a stable
algorithm for long-time Lyapunov diagnostics along polymer paths. The computed
finite-time Lyapunov numbers measure the material-line stretching environment
sampled by polymers whose trajectories are generated by the nonlinear
bead--spring dynamics. Their means, standard deviations, and PDFs reach a stable
late-time statistical regime over the computed interval. Together with the
ergodic Lyapunov theory, this statistical convergence makes the late-time
ensemble means useful estimates of the corresponding asymptotic Lyapunov
exponents. The resulting largest Lyapunov number,
\(L_1=\tau_R\Lambda_1\simeq 20.6\), is consistent with the strongly stretched
polymer conformations observed in the nonlinear dynamics.

\item The intermediate direction plays a significant role in both the local
strain geometry and the accumulated finite-time stretching. The polymers
preferentially sample axisymmetric biaxial extension, so the sampled local
strain field often has two extensional eigenvalues. Independently, the polymer
end-to-end vector aligns strongly with the intermediate strain-rate eigenvector
\(s_2\). The finite-time Lyapunov calculation gives the corresponding
accumulated result: the intermediate finite-time Lyapunov exponent is positive
for all sampled trajectories over the reported late-time statistics, not merely
in the mean. These observations are mutually consistent and show that the
intermediate direction is not a passive bystander in polymer stretching. The
finite-time Lyapunov spectrum also displays a clear dependence structure:
\(L_1\) and \(L_2\) are positively correlated, whereas \(L_2\) and \(L_3\), and
\(L_3\) and \(L_1\), are negatively correlated. The strongest departure from
independence occurs for the \((L_3,L_1)\) pair.

\item The finite-time Lyapunov statistics provide a quantitative test of the
usual material-flow stretching reduction. Along polymer-generated trajectories
with full bead--spring dynamics, the present calculation gives
\(E[\lambda_2]/E[\lambda_1]\simeq 4/17\), whereas the corresponding
material-line value is \(1/7\). This difference is not negligible. It shows
that polymer-conditioned Lyapunov statistics contain information that would be
lost if polymer motion were replaced a priori by passive material or tracer
transport.

\end{enumerate}}

These results suggest several directions for further work.

{\renewcommand{\labelenumi}{(\arabic{enumi})}
\begin{enumerate}

\item This investigation could be extended into the dilute regime, where, as
measured by the interaction parameter \(I\), polymer feedback could induce
departures from Navier--Stokes turbulence. The scaling estimate in the present
work indicates that such turbulence-modifying mesoscopic calculations would
require \(O(10^{11})\) beads. Such calculations will require new scalable
mesoscopic algorithms for bead--spring polymers, including efficient treatment
of long-range hydrodynamic coupling and boundary conditions on modern parallel
architectures. Developing such methods is a necessary step toward fully coupled
polymer-resolved computations of turbulence-modifying dilute solutions.

\item In our model, excluded--volume interactions are included, whereas the FENE
elasticity is an ideal--chain approximation. For consistency, a real--chain
elasticity law would be more appropriate in the presence of excluded volume
\cite[][]{kivotides_dense}. This would allow us to test whether such
realism alters polymer dynamics in turbulence, particularly the location
and sharpness of the coil--stretch transition.

\item Given the complexity of full-physics calculations in dilute and semi-dilute
regimes, a less detailed yet informative approach may be desirable.
In particular, one could develop a viscoelastic extension of Navier--Stokes
vortex dynamics \cite[][]{kivotides_lagrangian}.
Such a reduced description could guide and prioritise subsequent, more detailed
computational studies.

\end{enumerate}}


\appendix

\section{Summary of governing equations, numerical method and algorithmic variables}
\label{app:governing-system}

This appendix collects the equations and variables used in the calculation.
The main text discusses their physical meaning, assumptions, limitations and
algorithmic role.

\subsection{Carrier-flow equations and assumptions}
\label{app:carrier-flow}

In this work we consider homogeneous incompressible turbulent carrier flows.
The fluctuating turbulent velocity field \(u_i(x,t)\) satisfies
\begin{equation}
    \frac{\partial u_i}{\partial x_i}=0 ,
    \label{eq:app-incompressibility}
\end{equation}
and
\begin{equation}
    \rho_f\frac{\partial u_i}{\partial t}
    +\rho_f\frac{\partial (u_i u_j)}{\partial x_j}
    +\frac{\partial p}{\partial x_i}
    -\mu_f\frac{\partial^2 u_i}{\partial x_j\partial x_j}
    -\rho_f\,\frac{\epsilon}{\langle u_k u_k\rangle}\,u_i
    =0 .
    \label{eq:app-navier-stokes-lundgren}
\end{equation}
Here \(i\) and \(j\) denote spatial direction indices,
\(\mu_f=\rho_f\nu_f\) is the dynamic viscosity, \(p\) is the fluid pressure,
and
\begin{equation}
    \epsilon
    =
    -\nu_f
    \left\langle
    u_i
    \frac{\partial^2 u_i}{\partial x_j\partial x_j}
    \right\rangle
    \label{eq:app-dissipation-rate}
\end{equation}
is the mean kinetic-energy dissipation rate. The last term in
Eq.~\eqref{eq:app-navier-stokes-lundgren} is Lundgren's linear forcing. It
compensates viscous dissipation and maintains a statistically steady turbulent
state.

The Lundgren forcing is Galilean invariant because it governs the fluctuating
velocity
\[
    u'=u-\overline{u},
\]
rather than the total velocity \(u\). Under a Galilean boost, both \(u\) and
\(\overline{u}\) change by the same constant velocity, so their difference
\(u'\) remains unchanged. The equation appears Aristotelian only if one
incorrectly applies the total-velocity transformation \(u'\mapsto u'-V\) to
the fluctuation field.

The velocity-gradient tensor sampled by the polymers is
\begin{equation}
    L_{ij}(x,t)
    =
    \frac{\partial u_i}{\partial x_j}(x,t).
    \label{eq:app-velocity-gradient}
\end{equation}
Its symmetric and antisymmetric parts are
\begin{equation}
    S_{ij}
    =
    \frac{1}{2}\left(L_{ij}+L_{ji}\right),
    \qquad
    W_{ij}
    =
    \frac{1}{2}\left(L_{ij}-L_{ji}\right),
    \label{eq:app-strain-rotation}
\end{equation}
and the vorticity is
\begin{equation}
    \omega_i
    =
    \epsilon_{ijk}\frac{\partial u_k}{\partial x_j}.
    \label{eq:app-vorticity}
\end{equation}

Since the polymer size remains smaller than the Kolmogorov scale in the regime
considered here, the carrier velocity over a polymer conformation is represented
locally by the Batchelor expansion
\begin{equation}
    u_i(x_c+r,t)
    =
    u_i(x_c,t)
    +
    L_{ij}(x_c,t)r_j
    +
    O(|r|^2/\eta_K^2),
    \label{eq:app-batchelor-expansion}
\end{equation}
where \(x_c\) is the midpoint of the polymer end-to-end vector, \(r\) is the
bead displacement relative to this midpoint, and \(\eta_K\) is the Kolmogorov
length scale.

\subsection{Bead--spring Langevin dynamics and force terms}
\label{app:langevin-forces}

The polymer phase is represented by \(N_c\) coarse-grained bead--spring
chains, each containing \(N_b\) beads and \(N_s=N_b-1\) springs. The total
number of beads is
\begin{equation}
    N=N_cN_b .
    \label{eq:app-total-beads}
\end{equation}
We use \(k,\ell=1,\ldots,N\) for bead labels and \(i,j=1,2,3\) for spatial
components. The position and velocity of bead \(k\) are denoted by
\(r_i^k(t)\) and \(\dot r_i^k(t)\).

The bead-level Langevin force balance used in the main text is
\begin{equation}
    F_i^{{\rm in},k}
    +
    F_i^{{\rm el},k}
    +
    F_i^{{\rm ev},k}
    +
    F_i^{{\rm vd},k}
    +
    F_i^{{\rm th},k}
    =
    0 .
    \label{eq:app-langevin-balance}
\end{equation}
Here \(F_i^{{\rm in},k}\), \(F_i^{{\rm el},k}\),
\(F_i^{{\rm ev},k}\), \(F_i^{{\rm vd},k}\), and
\(F_i^{{\rm th},k}\) are, respectively, the inertial, elastic,
excluded-volume, viscous-drag and thermal-fluctuation forces acting on bead
\(k\). The inertial term is
\begin{equation}
    F_i^{{\rm in},k}
    =
    m_b \ddot r_i^k .
    \label{eq:app-inertial-force}
\end{equation}
In the computations reported in the main text the overdamped limit of
Eq.~\eqref{eq:app-langevin-balance} is used, so that bead motion is governed
by the instantaneous balance of elastic, excluded-volume, viscous-drag and
thermal-fluctuation forces.

\subsubsection{FENE elastic force}
\label{app:fene-force}

Let \(s\) denote a spring. If the spring starts at the bead with global index
\(k\), then it connects bead \(k\) to the next bead \(k+1\) along the same
chain. No spring is defined across the end of one chain and the beginning of
the next. The spring vector is therefore oriented along the bead ordering of
the chain:
\begin{equation}
    Q_i^s
    =
    r_i^{k+1}-r_i^k .
    \label{eq:app-spring-vector}
\end{equation}
The corresponding spring length and unit direction are
\begin{equation}
    \ell_s
    =
    \left(Q_i^s Q_i^s\right)^{1/2},
    \qquad
    \widehat Q_i^s
    =
    \frac{Q_i^s}{\ell_s}.
    \label{eq:app-spring-length-unit}
\end{equation}

The maximum spring length is
\begin{equation}
    \ell_0
    =
    N_{K,s} b ,
    \label{eq:app-maximum-spring-length}
\end{equation}
where \(b\) is the Kuhn length and
\begin{equation}
    N_{K,s}
    =
    \frac{M}{M_K N_s}
    \label{eq:app-kuhn-per-spring}
\end{equation}
is the number of Kuhn monomers represented by one spring. Here \(M\) is the
chain molar mass, \(M_K\) is the Kuhn-monomer molar mass, and \(N_s=N_b-1\)
is the number of springs per chain.

The FENE elastic potential of spring \(s\) is
\begin{equation}
    \Phi_s^{\rm el}
    =
    -\frac{H\ell_0^2}{2}
    \ln\left|
        1-\left(\frac{\ell_s}{\ell_0}\right)^2
    \right|,
    \qquad
    H
    =
    \frac{3k_B T}{N_{K,s}b^2}.
    \label{eq:app-fene-potential}
\end{equation}
The corresponding spring force is
\begin{equation}
    T_i^s
    =
    H
    \frac{Q_i^s}{1-\left(\ell_s/\ell_0\right)^2}
    =
    H
    \frac{\ell_s}{1-\left(\ell_s/\ell_0\right)^2}
    \widehat Q_i^s .
    \label{eq:app-fene-spring-force}
\end{equation}

Since \(Q_i^s=r_i^{k+1}-r_i^k\), the two endpoint positions enter the spring
vector with opposite signs. Consequently, the force exerted by spring \(s\) on
its starting bead \(k\) is \(+T_i^s\), while the force exerted by the same
spring on its end bead \(k+1\) is \(-T_i^s\). The elastic force on a bead is
obtained by adding the force contributions from the springs attached to that
bead. Thus an interior bead receives two spring-force contributions, while a
bead at either end of a chain receives only one spring-force contribution.

Equivalently, the bead elastic force is obtained from the total FENE elastic
potential,
\begin{equation}
    F_i^{{\rm el},k}
    =
    -\frac{\partial \Phi^{\rm el}}{\partial r_i^k},
    \qquad
    \Phi^{\rm el}
    =
    \sum_s \Phi_s^{\rm el}.
    \label{eq:app-elastic-gradient-form}
\end{equation}

\subsubsection{Bead--bead excluded-volume force}
\label{app:excluded-volume-force}

Excluded-volume interactions between beads are represented by a soft repulsive
potential. The excluded-volume potential acting on bead \(k\), generated by
all other beads, is
\begin{equation}
    \Phi^{{\rm ev},k}
    =
    \sum_{\substack{\ell=1\\ \ell\neq k}}^{N}
    \alpha
    \exp\left[
        -
        \frac{
        \left(r_i^\ell-r_i^k\right)
        \left(r_i^\ell-r_i^k\right)
        }{\delta^2}
    \right],
    \label{eq:app-ev-potential-bead}
\end{equation}
where \(N=N_cN_b\) is the total number of beads. The excluded-volume force on
bead \(k\) is
\begin{equation}
    F_i^{{\rm ev},k}
    =
    -\frac{\partial \Phi^{{\rm ev},k}}{\partial r_i^k}.
    \label{eq:app-ev-force-gradient}
\end{equation}
Equivalently,
\begin{equation}
    F_i^{{\rm ev},k}
    =
    -
    \sum_{\substack{\ell=1\\ \ell\neq k}}^{N}
    \frac{2\left(r_i^\ell-r_i^k\right)}{\delta^2}
    \alpha
    \exp\left[
        -
        \frac{
        \left(r_j^\ell-r_j^k\right)
        \left(r_j^\ell-r_j^k\right)
        }{\delta^2}
    \right].
    \label{eq:app-ev-force-explicit}
\end{equation}

The potential range is chosen to be one half of the equilibrium spring size,
\begin{equation}
    \delta
    =
    \frac{R_s}{2}.
    \label{eq:app-ev-range}
\end{equation}
The equilibrium spring size is estimated from Flory mean-field theory and its
renormalization-group refinements as
\begin{equation}
    R_s
    =
    b
    \left(
        \frac{v_{\rm ev}}{b^3}
    \right)^{2\nu-1}
    N_{K,s}^{\nu},
    \label{eq:app-equilibrium-spring-size}
\end{equation}
where \(v_{\rm ev}\) is the Kuhn excluded volume, \(b\) is the Kuhn length,
and \(\nu=0.588\) for a good solvent \((\nu=0.5\) in a theta solvent). The
dimensionless ratio \(v_{\rm ev}/b^3\) measures solvent quality. A common
parametrisation is
\begin{equation}
    v_{\rm ev}
    =
    b^3\frac{T-\theta}{T},
    \label{eq:app-kuhn-excluded-volume}
\end{equation}
where \(T\) is the solution temperature and \(\theta\) is the theta
temperature. Thus \(v_{\rm ev}=0\) at the theta point, while in the ideal
athermal-solvent limit one has \(v_{\rm ev}/b^3=1\).

\subsubsection{Hydrodynamic drag and thermal fluctuation forces}
\label{app:hydrodynamic-brownian}

Hydrodynamic interactions and thermal fluctuations are represented by the
standard bead--spring Brownian-dynamics formulation. Let \(k,\ell=1,\ldots,N\)
denote bead labels and \(i,j=1,2,3\) Cartesian components. The carrier-flow
velocity at bead \(k\) is
\begin{equation}
    u_i^{{\rm c},k}(t)
    =
    u_i(r^k,t).
    \label{eq:app-carrier-velocity-at-bead}
\end{equation}
The velocity of bead \(k\) relative to this carrier velocity is
\begin{equation}
    u_i^{{\rm S},k}
    =
    \dot r_i^k-u_i^{{\rm c},k}.
    \label{eq:app-stokes-relative-velocity}
\end{equation}
This relative velocity is the Stokes velocity associated with the bead motion
relative to the carrier flow.

The viscous-drag force on bead \(k\) is
\begin{equation}
    F_i^{{\rm vd},k}
    =
    -
    \sum_{\ell=1}^{N}
    \sum_{j=1}^{3}
    Z_{ij}^{k\ell}
    u_j^{{\rm S},\ell},
    \label{eq:app-viscous-drag}
\end{equation}
where \(Z_{ij}^{k\ell}\) is the \(3\times3\) bead--bead friction block. 
In terms of the full \(3N\)-dimensional notation used in the earlier
bead--spring formulation, this corresponds to
\[
    F_I^{\rm vd}
    =
    -
    \sum_{J=1}^{3N}
    Z_{IJ}(v_J-u_J),
    \qquad
    I=(k,i),
    \quad
    J=(\ell,j).
\]
Equivalently, since \(v_J-u_J\) is the Stokes velocity \(u_J^{\rm S}\), this
can be written as
\[
    F_I^{\rm vd}
    =
    -
    \sum_{J=1}^{3N}
    Z_{IJ}u_J^{\rm S}.
\]

Let \(M=Z^{-1}\) be the mobility matrix. Then the Stokes velocity can be
written equivalently as the velocity induced at the bead positions by the bead
drag forces:
\begin{equation}
    u_i^{{\rm S},k}
    =
    -
    \sum_{\ell=1}^{N}
    \sum_{j=1}^{3}
    M_{ij}^{k\ell}
    F_j^{{\rm vd},\ell}.
    \label{eq:app-stokes-velocity-mobility}
\end{equation}
The diffusion tensor is related to the mobility by
\begin{equation}
    D_{ij}^{k\ell}
    =
    k_BT\,M_{ij}^{k\ell}.
    \label{eq:app-diffusion-mobility}
\end{equation}

In the unbounded-domain approximation, the diffusion tensor is taken to be the
Rotne--Prager--Yamakawa tensor. Let
\begin{equation}
    R_i^{k\ell}
    =
    r_i^\ell-r_i^k,
    \qquad
    R^{k\ell}
    =
    \left(R_i^{k\ell}R_i^{k\ell}\right)^{1/2},
    \qquad
    \widehat R_i^{k\ell}
    =
    \frac{R_i^{k\ell}}{R^{k\ell}} .
    \label{eq:app-bead-separation}
\end{equation}
The self-diffusion block is
\begin{equation}
    D_{ij}^{kk}
    =
    \frac{k_BT}{6\pi\mu_f a}\delta_{ij},
    \label{eq:app-rpy-self}
\end{equation}
where \(a\) is the effective bead radius. For \(k\neq\ell\) and
\(R^{k\ell}\geq 2a\),
\begin{equation}
    D_{ij}^{k\ell}
    =
    \frac{k_BT}{8\pi\mu_f R^{k\ell}}
    \left[
        \left(
            1+\frac{2a^2}{3(R^{k\ell})^2}
        \right)\delta_{ij}
        +
        \left(
            1-\frac{2a^2}{(R^{k\ell})^2}
        \right)
        \widehat R_i^{k\ell}\widehat R_j^{k\ell}
    \right].
    \label{eq:app-rpy-far}
\end{equation}
For \(k\neq\ell\) and \(R^{k\ell}<2a\),
\begin{equation}
    D_{ij}^{k\ell}
    =
    \frac{k_BT}{6\pi\mu_f a}
    \left[
        \left(
            1-\frac{9R^{k\ell}}{32a}
        \right)\delta_{ij}
        +
        \frac{3R^{k\ell}}{32a}
        \widehat R_i^{k\ell}\widehat R_j^{k\ell}
    \right].
    \label{eq:app-rpy-near}
\end{equation}

Equations~\eqref{eq:app-rpy-self}--\eqref{eq:app-rpy-near} are the
\(3\times3\) bead--bead block form of the RPY diffusion tensor used in the
earlier bead--spring formulation. Together with
Eq.~\eqref{eq:app-diffusion-mobility}, they also specify the thermal
fluctuations consistently with the generalized fluctuation--dissipation
theorem: the same mobility tensor that determines the dissipative response
of the beads determines the covariance of their Brownian displacements.

To generate these displacements, the symmetric positive-definite diffusion
tensor is factorized as
\begin{equation}
    D_{ij}^{k\ell}
    =
    \sum_{m=1}^{N}
    \sum_{q=1}^{3}
    B_{iq}^{km}B_{jq}^{\ell m},
    \label{eq:app-diffusion-factorization}
\end{equation}
which is the bead--block form of \(D=BB^{T}\). The independent Wiener
increments \(dW_i^k\) satisfy
\begin{equation}
    \left\langle dW_i^k\right\rangle=0,
    \qquad
    \left\langle dW_i^k dW_j^\ell\right\rangle
    =
    \delta_{ij}\delta_{k\ell}\,dt .
    \label{eq:app-wiener-increments}
\end{equation}
The thermal contribution to the displacement of bead \(k\) over the time
interval \(dt\) is therefore
\begin{equation}
    d r_i^{{\rm th},k}
    =
    \sqrt{2}
    \sum_{\ell=1}^{N}
    \sum_{j=1}^{3}
    B_{ij}^{k\ell}\,dW_j^\ell ,
    \label{eq:app-thermal-displacement}
\end{equation}
with mean and covariance
\begin{equation}
    \left\langle d r_i^{{\rm th},k}\right\rangle=0,
    \qquad
    \left\langle
        d r_i^{{\rm th},k}
        d r_j^{{\rm th},\ell}
    \right\rangle
    =
    2D_{ij}^{k\ell}\,dt .
    \label{eq:app-thermal-displacement-covariance}
\end{equation}
Thus, both the deterministic hydrodynamic response and the correlated
thermal displacements contain the intrachain and interchain hydrodynamic
interactions represented by the RPY tensor.

\subsubsection{Polymer configuration and tangent-flow diagnostics}
\label{app:polymer-diagnostics}

For each chain \(n=1,\ldots,N_c\), let
\[
    k_s(n)=(n-1)N_b+1,
    \qquad
    k_e(n)=nN_b
\]
denote the first and last bead of the chain. The chain midpoint and end-to-end
vector are
\begin{equation}
    z_i^n(t)
    =
    \frac{1}{2}
    \left(
        r_i^{k_e(n)}(t)
        +
        r_i^{k_s(n)}(t)
    \right),
    \label{eq:app-chain-midpoint}
\end{equation}
and
\begin{equation}
    \Delta z_i^n(t)
    =
    r_i^{k_e(n)}(t)
    -
    r_i^{k_s(n)}(t).
    \label{eq:app-chain-end-to-end}
\end{equation}
The carrier-flow velocity-gradient tensor sampled by chain \(n\) is evaluated
at the midpoint:
\begin{equation}
    L_{ij}^n(t)
    =
    L_{ij}(z^n(t),t)
    =
    \frac{\partial u_i}{\partial x_j}(z^n(t),t).
    \label{eq:app-chain-sampled-gradient}
\end{equation}

The tangent-flow diagnostic evolves an infinitesimal reference vector
\(f_R^n\) along the chain-midpoint trajectory. Its image \(f^n(t)\) is written
as
\begin{equation}
    f_i^n(t)
    =
    F_{ij}^n(t)f_{R,j}^n,
    \label{eq:app-tangent-vector}
\end{equation}
where the deformation-gradient tensor satisfies
\begin{equation}
    \frac{dF_{ij}^n}{dt}
    =
    L_{im}^n(t)F_{mj}^n,
    \qquad
    F_{ij}^n(0)=\delta_{ij}.
    \label{eq:app-deformation-gradient}
\end{equation}
This diagnostic uses the carrier-flow velocity gradient sampled along polymer
trajectories; it is therefore distinct from passive material-line tracking,
because the midpoint trajectories \(z^n(t)\) are generated by the full
bead--spring dynamics.

The singular value decomposition of \(F^n\) is
\begin{equation}
    F^n
    =
    W^n\Sigma^n(V^n)^T,
    \qquad
    \Sigma^n
    =
    \operatorname{diag}
    \left(
        \sigma_1^n,\sigma_2^n,\sigma_3^n
    \right),
    \label{eq:app-svd}
\end{equation}
where \(W^n\) and \(V^n\) are orthogonal matrices and
\(\sigma_i^n>0\) are the singular values. Without SVD normalisation, the
finite-time Lyapunov exponents are
\begin{equation}
    \lambda_i^n(t)
    =
    \frac{1}{t}
    \ln \sigma_i^n(t).
    \label{eq:app-finite-time-lyapunov}
\end{equation}

In the SVD-normalised computation, at each normalisation event \(q\) the
singular values are removed from \(F^n\) and their logarithms are stored:
\begin{equation}
    G_i^n
    =
    \sum_{q=1}^{N_{\rm norm}}
    \ln \sigma_i^{n,(q)} .
    \label{eq:app-svd-stored-stretch}
\end{equation}
The deformation-gradient matrix is then reset according to
\begin{equation}
    F^n
    \leftarrow
    W^n(V^n)^T .
    \label{eq:app-svd-reset}
\end{equation}
At time \(t\), if \(\sigma_i^n(t)\) are the current singular values of the
renormalised \(F^n\), the accumulated finite-time Lyapunov exponents are
\begin{equation}
    \lambda_i^n(t)
    =
    \frac{
        G_i^n+\ln \sigma_i^n(t)
    }{t}.
    \label{eq:app-svd-normalised-lyapunov}
\end{equation}

\subsection{Overdamped stochastic equation and numerical integration}
\label{app:overdamped-numerical-scheme}

Using the mobility--diffusion relation in
Eq.~\eqref{eq:app-diffusion-mobility}, the factorization of the
diffusion tensor in Eq.~\eqref{eq:app-diffusion-factorization}, and
the Wiener-increment statistics in
Eq.~\eqref{eq:app-wiener-increments}, the overdamped bead dynamics
can be written as an It\^o stochastic differential equation. The
equation for bead \(k\) is
\begin{equation}
    d r_i^k
    =
    \left[
        u_i^{{\rm c},k}
        +
        \sum_{\ell=1}^{N}
        \sum_{j=1}^{3}
        \frac{\partial D_{ij}^{k\ell}}
             {\partial r_j^\ell}
        +
        \frac{1}{k_BT}
        \sum_{\ell=1}^{N}
        \sum_{j=1}^{3}
        D_{ij}^{k\ell}
        \left(
            F_j^{{\rm el},\ell}
            +
            F_j^{{\rm ev},\ell}
        \right)
    \right]dt
    +
    \sqrt{2}
    \sum_{\ell=1}^{N}
    \sum_{j=1}^{3}
    B_{ij}^{k\ell}\,dW_j^\ell .
    \label{eq:app-overdamped-sde}
\end{equation}

For the unbounded-domain RPY diffusion tensor defined in
Eqs.~\eqref{eq:app-rpy-self}--\eqref{eq:app-rpy-near}, the It\^o
drift vanishes identically:
\begin{equation}
    \sum_{\ell=1}^{N}
    \sum_{j=1}^{3}
    \frac{\partial D_{ij}^{k\ell}}
         {\partial r_j^\ell}
    =
    0 .
    \label{eq:app-rpy-zero-ito-drift}
\end{equation}
Using \(D_{ij}^{k\ell}=k_BT M_{ij}^{k\ell}\), the equation advanced
in the calculation is therefore
\begin{equation}
    d r_i^k
    =
    \left[
        u_i^{{\rm c},k}
        +
        \sum_{\ell=1}^{N}
        \sum_{j=1}^{3}
        M_{ij}^{k\ell}
        \left(
            F_j^{{\rm el},\ell}
            +
            F_j^{{\rm ev},\ell}
        \right)
    \right]dt
    +
    \sqrt{2}
    \sum_{\ell=1}^{N}
    \sum_{j=1}^{3}
    B_{ij}^{k\ell}\,dW_j^\ell .
    \label{eq:app-implemented-overdamped-sde}
\end{equation}

Equation~\eqref{eq:app-implemented-overdamped-sde} is integrated with
the first-order, It\^o-compatible Ermak--McCammon method. The
first-order semi-implicit predictor--corrector algorithm introduced in
\cite[][]{somasi} and extended to include excluded-volume forces in
\cite[][]{kivotides_stretch} is adapted here to the Ermak--McCammon
scheme.

At each corrector iteration, the new spring vector \(Q_i^s\) is
obtained by solving the corresponding cubic algebraic equation
\begin{equation}
    \ell_s^3
    -
    \left|\mathcal R^s\right|\ell_s^2
    -
    (3C+1)\ell_0^2\ell_s
    +
    \left|\mathcal R^s\right|\ell_0^2
    =
    0,
    \qquad
    \ell_s=\left(Q_i^sQ_i^s\right)^{1/2},
    \label{eq:app-fene-cubic}
\end{equation}
where \(\mathcal R^s\) is the known predictor vector,
\(\ell_0\) is the maximum spring length, and
\begin{equation}
    C
    =
    \frac{k_BT\,\Delta t}
         {3\pi\mu_f a b\ell_0}.
    \label{eq:app-fene-cubic-coefficient}
\end{equation}
Equation~\eqref{eq:app-fene-cubic} has a unique physical solution
satisfying \(0<\ell_s<\ell_0\). The updated spring vector therefore
preserves the maximum-extension restriction exactly, without clipping
or rejecting bead positions. The corrector iterations continue until
the change in the spring vectors between two successive iterations is
smaller than the prescribed convergence tolerance.

\section{Microscopic interpretation of coarse-grained polymer forces}
\label{app:microscopic-forces}

This appendix provides a microscopic and thermodynamic interpretation of the
coarse-grained excluded-volume and elastic forces appearing in the
bead--spring Langevin equation.

\subsection{Microscopic origin of coarse-grained forces}

The Navier--Stokes and Langevin equations have a similar force-balance
structure. For example, both equations include an inertial force. Moreover,
the pressure force in the Navier--Stokes equation and the excluded-volume
force in polymer models have a common microscopic origin: both arise from
coarse-graining microscopic interactions and collisional momentum transfer
between underlying particle degrees of freedom.

In the continuum-fluid limit, this coarse graining produces the isotropic
pressure part of the stress. In the bead--spring description, the same kind
of coarse graining produces an effective excluded-volume interaction between
polymer segments, whose strength and range are renormalised by
polymer--solvent and solvent--solvent interactions in the intervening
solvent. Excluded volume therefore represents a deterministic mean,
solvent-renormalised component of the molecular forcing on the polymer,
distinct from the Brownian, drag, and elastic contributions.

\subsection{Entropic elasticity and solvent momentum balance}

As discussed in the main text, standard bead--spring models neglect the
reaction on the solvent of the Brownian force acting on the polymer beads,
although such a reaction should, in principle, be present. This raises the
related question of whether polymer elasticity implies an additional
accompanying mean force density in the solvent.

To address this question, one must distinguish direct molecular momentum
transfer from the thermodynamic force that appears after coarse graining. Let
\begin{equation}
    R=\{r^k\}_{k=1}^{N_b}
\end{equation}
denote the resolved bead configuration, and let \(Q\) denote a microscopic
torsional or conformational polymer state compatible with a given \(R\). At
the microscopic level, torsion-angle transitions occur within the
intramolecular torsional landscape and are driven by molecular interactions
with the surrounding solvent \cite[][]{rubinstein}. The corresponding
instantaneous microscopic impulses obey the action--reaction principle.

For each fixed bead configuration \(R\), however, the fast torsional
variables are treated as equilibrated over the microscopic states compatible
with \(R\). Within this conditional-equilibrium closure, the direct
torsion-angle-changing molecular impulses form a detailed-balance ensemble of
momentum-transfer events. Averaging over this conditional ensemble produces
no separate accompanying mean force density in the solvent, although
residual equal-and-opposite fluctuations in the polymer and solvent variables
may exist in principle. These residual stochastic impulse pairs are not
retained as independent polymer--solvent momentum-transfer noise terms in
the bead--spring model.

The same microscopic torsional dynamics nevertheless has a nonzero projected
effect on the coarse-grained polymer. The bead--spring model retains \(R\),
rather than the full microscopic conformation \(Q\). The map from microscopic
states \(Q\) to bead configurations \(R\) is many-to-one, and the number and
statistical weights of microscopic states compatible with a given \(R\)
depend on \(R\). The eliminated torsional variables therefore define the
conditional partition function
\begin{equation}
    Z(R)
    =
    \int_{\mathcal{Q}(R)}
    e^{-\beta H(Q;R)}\,dQ,
    \qquad
    \beta=(k_BT)^{-1},
    \label{eq:app-conditional-partition-function}
\end{equation}
where \(\mathcal{Q}(R)\) denotes the set of microscopic conformations
compatible with the bead configuration \(R\). The corresponding
conformational free energy is
\begin{equation}
    A(R)
    =
    -k_BT\log Z(R),
    \label{eq:app-conformational-free-energy}
\end{equation}
and generates the bead-level mean entropic elastic force
\begin{equation}
    F_i^{{\rm el},k}
    =
    -\frac{\partial A}{\partial r_i^k}.
    \label{eq:app-entropic-elastic-force}
\end{equation}

Equations~\eqref{eq:app-conditional-partition-function}--
\eqref{eq:app-entropic-elastic-force} describe the microscopic origin of the
elastic force appearing in Eq.~\eqref{eq:langevin-force-balance}. Its FENE
representation in the coarse-grained bead--spring model is given in
Eqs.~\eqref{eq:app-fene-potential}--
\eqref{eq:app-elastic-gradient-form}. Thus, although the
torsion-angle-changing microscopic impulses produce no separate accompanying
mean force density in the solvent, coarse-graining the associated
torsional and conformational ensemble at fixed bead configuration produces a
genuine thermodynamic force in the polymer equation.

When the solvent stretches the chain through viscous drag on the beads, the
resolved configuration \(R\) changes. Stretched bead configurations are
compatible with fewer microscopic torsional and conformational states and
therefore have a higher conformational free energy. The negative gradient of
this free energy gives the restoring entropic elastic force.

The remaining issue is how this thermodynamic force enters the
polymer--solvent coupling. Viscous drag on the beads is the mechanical
channel through which the solvent performs work on the coarse-grained
polymer. When this drag changes \(R\), the conditional conformational free
energy \(A(R)\) changes correspondingly. Work performed against the restoring
entropic force is stored as conformational free energy, while subsequent
relaxation releases this free energy through bead motion, viscous drag, and
the associated microhydrodynamic response.

Entropic elasticity therefore does not constitute a separate mean elastic
force density in the solvent equation. Its solvent-side mechanical
manifestation occurs through the drag generated by bead motion. Fluctuations
about the projected entropic mean force may also exist in principle, but
standard bead--spring formulations retain only the mean entropic drift of the
coarse-grained polymer configuration.

\end{document}